\documentclass{pasa}%

\usepackage{graphicx}
\usepackage{gensymb}
\usepackage[binary-units]{siunitx}
\usepackage[super]{nth}

\DeclareMathOperator\diag{diag}
\DeclareMathOperator\med{median}
\DeclareSIUnit\jansky{Jy}
\DeclareSIUnit\beam{beam}
\DeclareSIUnit\deg{deg}
\DeclareSIUnit\decibeli{dBi}
\DeclareSIUnit\decibelm{dBm}
\DeclareSIUnit\decibelc{dBc}
\DeclareSIUnit\day{day}
\DeclareSIUnit\year{year}

\title[ASKAP System Description]{Australian Square Kilometre Array Pathfinder: I. System Description}

\author[Hotan et al.]{A.W.~Hotan$^1$, J.D.~Bunton$^{2}$, A.P.~Chippendale$^{2}$, M.~Whiting$^2$, J.~Tuthill$^2$,
V.A.~Moss$^{2}$, D.~McConnell$^{2}$, S.W.~Amy$^{2}$, M.T.~Huynh$^{1}$,
J.R.~Allison$^{11,2}$, C.S.~Anderson$^{2,20}$, K.W.~Bannister$^{2}$, E.~Bastholm$^1$, R.~Beresford$^{2}$, D.C.-J.~Bock$^{2}$, 
R.~Bolton$^{2}$, J.M.~Chapman$^{2}$, K.~Chow$^{2}$, J.D.~Collier$^{5,6}$, F.R.~Cooray$^{2}$, T.J.~Cornwell$^{2,7}$, 
P.J.~Diamond$^{2,8}$, P.G.~Edwards$^{2}$, I.J.~Feain$^{2}$, T.M.O.~Franzen$^{19}$, D.~George$^{2}$, N.~Gupta$^{2,9}$, 
G.A.~Hampson$^{2}$, L.~Harvey-Smith$^{2,10,6}$, D.B.~Hayman$^{2}$, I.~Heywood$^{11,12,2}$, C.~Jacka$^{2}$, C.A.~Jackson$^{2,19}$,
S.~Jackson$^{18}$, K.~Jeganathan$^{2}$, S.~Johnston$^{2}$, M.~Kesteven$^{2}$, D.~Kleiner$^{13}$, B.S.~Koribalski$^{2}$, K.~Lee-Waddell$^{2}$, 
E.~Lenc$^{2}$, E.S.~Lensson$^{14}$, S.~Mackay$^{2}$, E.K.~Mahony$^{2}$, N.M.~McClure-Griffiths$^{15}$, 
R.~McConigley$^{4}$, P.~Mirtschin$^{16}$, A.K.~Ng$^{2}$, R.P.~Norris$^{6,2}$, S.E.~Pearce$^{2}$, C.~Phillips$^{2}$, M.A.~Pilawa$^{2}$,
W.~Raja$^{2}$, J.E.~Reynolds$^{2}$, P.~Roberts$^{2}$, D.N.~Roxby$^{2}$, E.M.~Sadler$^{2,3}$, M.~Shields$^{2}$, A.E.T.~Schinckel$^{2}$, 
P.~Serra$^{2,13}$, R.D.~Shaw$^{2}$, T.~Sweetnam$^{2}$, E.R.~Troup$^{2}$, A.~Tzioumis$^{2}$, M.A.~Voronkov$^{2}$, 
T.~Westmeier$^{17}$
\affil{$^{1}$CSIRO Astronomy and Space Science, PO Box 1130, Bentley, WA 6102, Australia}%
\affil{$^{2}$CSIRO Astronomy and Space Science, PO Box 76, Epping NSW 1710, Australia}
\affil{$^{3}$Sydney Institute for Astronomy, School of Physics A28, University of Sydney, Sydney, NSW 2006, Australia}
\affil{$^{4}$CSIRO Astronomy and Space Science, PO Box 2102, Geraldton, WA 6530, Australia}
\affil{$^{5}$The Inter-University Institute for Data Intensive Astronomy (IDIA), Department of Astronomy, University of Cape Town, Private Bag X3, Rondebosch, 7701, South Africa}
\affil{$^{6}$School of Science, Western Sydney University, Locked Bag 1797, Penrith, NSW 2751, Australia}
\affil{$^{7}$Tim Cornwell Consulting, 17 Elgan Crescent, Sandbach, CW11 1LD, United Kingdom}
\affil{$^{8}$SKA Organisation, Jodrell Bank, Lower Withington, Cheshire, SK11 9FT, United Kingdom}
\affil{$^{9}$Inter-University Centre for Astronomy and Astrophysics, Post Bag 4, Ganeshkhind, Pune 411007, India}
\affil{$^{10}$Deans Unit, Faculty of Science, Dalton Building F12 UNSW Sydney, NSW 2052, Australia}
\affil{$^{11}$Sub-Dept. of Astrophysics, Department of Physics, University of Oxford, Denys Wilkinson Building, Keble Rd., Oxford, OX1 3RH, United Kingdom}
\affil{$^{12}$Rhodes University, PO Box 94, Makhanda (Grahamstown) 6140, Eastern Cape, South Africa}
\affil{$^{13}$INAF - Osservatorio Astronomico di Cagliari, via della Scienza 5, 09047 Selargius, CA, Italy}
\affil{$^{14}$CSIRO Astronomy and Space Science, PO Box 276, Parkes, NSW 2870, Australia}
\affil{$^{15}$Research School of Astronomy \& Astrophysics, The Australian National University, Canberra, ACT 2601, Australia}
\affil{$^{16}$CSIRO Astronomy and Space Science, 1828 Yarrie Lake Rd, Narrabri, NSW 2390, Australia}
\affil{$^{17}$ICRAR, M468, The University of Western Australia, 35 Stirling Highway, Crawley WA 6009, Australia}
\affil{$^{18}$CSIRO Astronomy and Space Science, PO Box 2225, Ellenbrook WA 6069, Australia}
\affil{$^{19}$ASTRON, the Netherlands Institute for Radio Astronomy, Oude Hoogeveensedijk 4, 7991 PD Dwingeloo, the Netherlands}
\affil{$^{20}$Jansky fellow of the National Radio Astronomy Observatory, NRAO, 1003 Lopezville Rd, Socorro, NM 87801 USA}
\affil{\textit{CSIRO's ASKAP telescope was built and brought into operation by a large team working over a number of years. The authorship of this paper and its ordering reflects the effort invested by team members in preparing the paper, in addition to their contributions towards the construction and commissioning of ASKAP itself.}}
}%

\jid{PASA}
\doi{10.1017/pas.\the\year.xxx}
\jyear{\the\year}

\usepackage{aas_macros}
\usepackage{hyperref}
\usepackage{xcolor}

\hypersetup{draft}
\begin{document}

\begin{frontmatter}
\maketitle

\begin{abstract}
In this paper we describe the system design and capabilities of the Australian Square Kilometre Array Pathfinder (ASKAP) radio telescope at the conclusion of its construction project and commencement of science operations. ASKAP is one of the first radio telescopes to deploy phased array feed (PAF) technology on a large scale, giving it an instantaneous field of view that covers \SI{31}{\deg\squared} at \SI{800}{\mega\hertz}.
As a two-dimensional array of 36$\times$12\,m antennas, with baselines ranging from 22\,m to 6\,km, ASKAP also has excellent snapshot imaging capability and 10 arcsecond resolution. This, combined with 288\,MHz of instantaneous bandwidth and a unique third axis of rotation on each antenna, gives ASKAP the capability to create high dynamic range images of large sky areas very quickly. It is an excellent telescope for surveys between \SI{700}{\mega\hertz} and \SI{1800}{\mega\hertz} and is expected to facilitate great advances in our understanding of galaxy formation, cosmology and radio transients while opening new parameter space for discovery of the unknown.
\end{abstract}

\end{frontmatter}

\section{INTRODUCTION}
\label{sec:intro}

\begin{figure}
    \centering
    \includegraphics[width=\columnwidth]{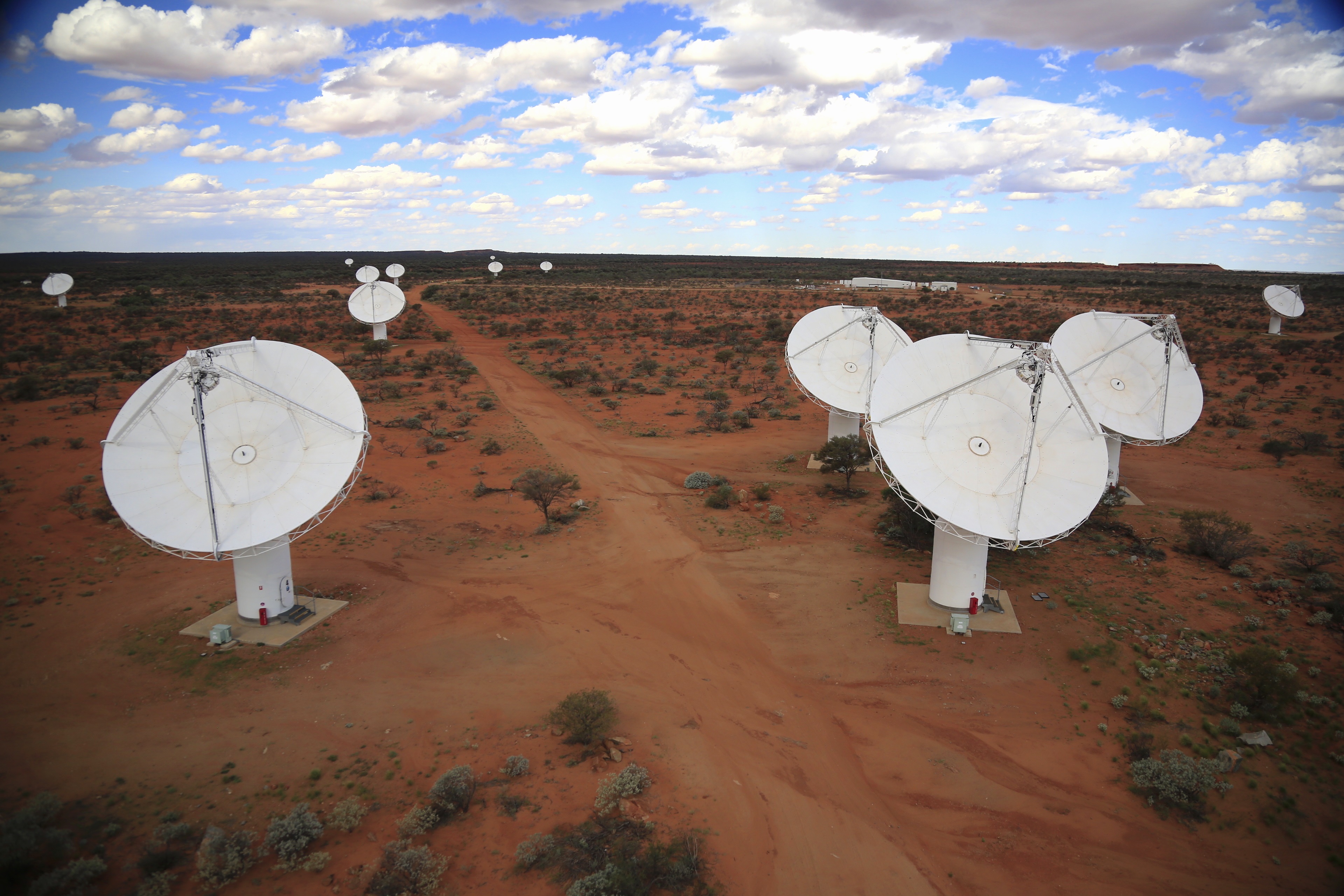}
    \caption{CSIRO's Australian Square Kilometre Array Pathfinder (ASKAP) telescope.}
    \label{fig:askap}
\end{figure}

ASKAP is a Square Kilometre Array (SKA\footnote{\url{https://www.skatelescope.org}}) precursor telescope located at Australia's SKA site, the Murchison Radio-astronomy Observatory (MRO). This radio telescope was developed using new phased array feed (PAF) technology to achieve high survey speed by observing with a wide instantaneous field of view. The design concept is described in more detail by \cite{DeBoer_2009} and the science goals in \cite{2007PASA...24..174J}.  Figure~\ref{fig:askap} shows a photograph of ASKAP's core taken in 2018.

After more than ten years of design, prototyping, construction and commissioning, ASKAP became fully operational in 2019. During construction, we developed a second generation of electronics \citep{Hampson2012} based on lessons learnt from prototype hardware \citep{Schinckel2011, hotan_2014, mcconnell_2016}, along with many iterations of firmware and software. Thousands of components have been installed and connected together to form one of the most complicated and powerful radio astronomy signal processors ever developed.

The first interferometry between PAF-equipped antennas at the MRO was conducted using a prototype system known as the Boolardy engineering test array \citep[BETA,][]{hotan_2014}. This consisted of 6 antennas fitted with first-generation receiver systems \citep{Schinckel2011}. Beyond teaching us how to improve ASKAP's system design, BETA contributed to astrophysical research \citep[e.g.,][]{Serra2015,2016MNRAS.460.2180H,Hobbs2016,Heywood2016,2017MNRAS.465.4450A,2017MNRAS.471.2952M} including the discovery of neutral hydrogen in a young radio galaxy at redshift $z=0.44$ through an absorption-line search \citep{10.1093/mnras/stv1532}. BETA also demonstrated real-time spatial radio-frequency interference (RFI) mitigation \citep{Hellbourg2016}.

A diverse early science program conducted on a subset of 12 to 18 antennas fitted with the second-generation PAF systems also produced a wide range of results, showcasing the utility of a wide-field radio telescope at gigahertz frequencies. See Section~\ref{sec:ops} for details. 

In 2019 we commenced a program of pilot surveys with the full array and an all-sky survey known as the Rapid ASKAP Continuum Survey (RACS\footnote{\url{https://www.atnf.csiro.au/content/racs}}), demonstrating the full science capability of the telescope for the first time.

With up to 36 beams per antenna and 36 antennas, ASKAP produces a torrent of raw data (approximately \SI{100}{\tera\bit\per\second}).  This is correlated and averaged at the observatory, producing an output visibility data stream of up to \SI{2.4}{\giga\byte\per\second} that is sent to the Pawsey Supercomputing Centre\footnote{\url{https://pawsey.org.au}} in Perth, some 600\,km south of the telescope, for image processing. This gives the research community a hint of the challenges to come in the era of the SKA.

This paper describes the technical details of ASKAP and documents key performance metrics based on commissioning data. Future papers in the series will describe the image processing software and performance metrics in more detail. We describe ASKAP's design (Section~\ref{sec:surveytelescope}) followed by information about the observatory site (Section~\ref{sec:mro}), an overview of the system and its components (Section~\ref{sec:system}), and more detailed information about key subsystems (Sections~\ref{sec:antenna} to~\ref{sec:CASDA}). We report on measured performance metrics in Section~\ref{sec:performance}, the site radio frequency environment in Section~\ref{sec:rfi}, the telescope operations model in Section~\ref{sec:ops} and future upgrade options in Section~\ref{sec:future}.

\subsection{Designing a radio telescope for surveys}
\label{sec:surveytelescope}

The primary design goal of ASKAP was to maximise survey speed, the rate at which the telescope can observe a given area of sky to a certain sensitivity limit \citep{2007PASA...24..174J}. Compared to existing telescopes, it was clear that high survey speed could be achieved by building an array of many small antennas to keep the primary beam size large, provide sensitivity over multiple spatial scales, and achieve good surface brightness sensitivity. Such an array would have many baselines (antenna pairs) and would therefore produce a very large amount of data and cause the computational cost of imaging to dominate array design. 

ASKAP uses PAF receivers to achieve a \SI{31}{\deg\squared} instantaneous field of view at \SI{800}{\mega\hertz}.  PAFs capture more information from each antenna and can be a more cost effective way to increase survey speed than using a much larger number of smaller antennas with cryogenic single-pixel feeds \citep{Chippendale2007}.    

The survey speed of a radio telescope array also depends on bandwidth \citep{Bunton2003} and the precise nature of the survey parameters \citep{Johnston2006, 2007PASA...24..174J}, but can be broadly represented by the figure of merit \citep{Bunton2003, DeBoer_2009, 5651120}
\begin{equation}
\label{eq:ssfom}
\text{SS} = \left(\frac{A_\text{e}}{T_\text{sys}}\right)^2\Omega_\text{FoV}
\end{equation}
where $A_\text{e}$ is the total effective area of all antennas, $T_\text{sys}$ is the system equivalent noise temperature, and $\Omega_\text{FoV}$ is the instantaneous (processed) field of view.  Effective area is related to the physical area $A$ of the antennas via $A_\text{e}=\eta A$ and $\eta$ is the antenna efficiency.

Table~\ref{tab:specs} summarises key indicative performance parameters of the ASKAP array using \eqref{eq:ssfom} to measure survey speed and $A_\text{e}/T_\text{sys}$ to measure sensitivity. A nominal measured value of 75\,K is used for the effective system temperature $T_\text{sys}/\eta$. With 36 dual-polarisation beams, ASKAP achieves a \SI{31}{\deg\squared} field of view at \SI{800}{\mega\hertz} but this reduces to \SI{15}{\deg\squared} at \SI{1700}{\mega\hertz} \citep{McConnell_2017_Survey}. Field of view and survey speed may be increased at the high end of the band in the future by processing more than 36 beams.  See Section~\ref{sec:performance} for more details on the definition and measurement of sensitivity, field of view, and the resulting survey speed.

\begin{table}
\begin{center}
\caption{Key parameters of the ASKAP telescope.}
\label{tab:specs}
\begin{tabular}{ll}
\hline 
Number of antennas & \num{36} \\
Antenna diameter & \SI{12}{\meter} \\
Focal ratio $f/D$ & \num{0.5} \\
Total collecting area & \SI{4071.5}{\meter\squared} \\
Maximum baseline & \SI{6}{\kilo\meter} \\
Angular resolution & \SI{10}{\arcsecond} at \SI{1}{\giga\hertz} \\
Observing frequency & \num{0.7} to \SI{1.8}{\giga\hertz} \\
Processed bandwidth & \SI{288}{\mega\hertz} \\
Frequency channels & \num{15552} \\
Frequency resolution & \SI{18.5}{\kilo\hertz} to \SI{0.58}{\kilo\hertz}\\
Effective system temperature & \SI{75}{\kelvin} \\
Sensitivity & \SI{54}{\meter\squared\per\kelvin} \\
Dual-polarisation beams & 36 \\
Field of view$^\text{a}$ (\SI{800}{\mega\hertz}) & \SI{31}{\deg\squared} \\
Field of view$^\text{a}$ (\SI{1700}{\mega\hertz}) & \SI{15}{\deg\squared} \\
Survey speed$^\text{b}$  (\SI{800}{\mega\hertz})& \SI[per-mode=reciprocal]{91400}{\meter\tothe{4}\deg\squared\per\kelvin\squared} \\
Survey speed$^\text{b}$  (\SI{1700}{\mega\hertz}) & \SI[per-mode=reciprocal]{44200}{\meter\tothe{4}\deg\squared\per\kelvin\squared} \\
\hline
\end{tabular}
\end{center}
$^\text{a}$Calculated via \eqref{eq:foveff} \\
$^\text{b}$Calculated via \eqref{eq:ss}
\end{table}

ASKAP's large field of view and increased survey speed came at the expense of sensitivity since cooling each PAF to cryogenic temperatures was not economical on the scale required at the time of design.  This resulted in a beamformed $T_\text{sys}/\eta$ in the range \SIrange{60}{80}{\kelvin} over most of the frequency range, or a system equivalent flux density (SEFD) of approximately \SI{1800}{\jansky} for a single antenna (see Section~\ref{sec:performance}).  

ASKAP's $T_\text{sys}$ is dominated by low-noise amplifier (LNA) noise, in part because the LNA uses a transistor (ATF-35143, circa 2006) that is now quite old.  Sensitivity could be improved significantly in the future by updating the room-temperature LNA design with new transistors \citep{Shaw2015, Weinreb2019} or by scaling up the manufacturability and affordability of cryogenic PAF technology like that under development for the Parkes 64\,m telescope \citep{Dunning2016, Dunning2019}.

Pilot survey observations (see Section \ref{sec:ops}) have tested many different observing modes, providing direct experience of the sensitivity that can be achieved within practical constraints. In many cases, we approach the thermal noise limit. However, for some observing modes (particularly with short integration times) other factors such as deconvolution errors contribute to an elevated noise floor. The values given in Table \ref{tab:tsys} should provide a realistic basis for planning observations, and may improve as updates are made to the telescope. The data used to compile Table \ref{tab:tsys} were obtained using different beam arrangements and processing strategies, so some variation is expected on top of the intrinsic spectral behaviour shown in Figure \ref{fig:sefdSpectrum}. 

Importantly, the measured noise can be increased above the theoretical value by weighting the data differently to natural weighting. ASKAP uses a preconditioning approach with robust weighting. The robustness parameter has a similar effect to the description in \citet{briggs1995}, ranging from -2.0 (uniform weighting) to 2.0 (natural weighting). The robustness values typically used for pilot survey processing were: 0.0 for continuum imaging at low frequencies, -0.5 for continuum imaging in the mid-band and 0.5 for spectral imaging in the low and mid-bands. The noise increases by factors of approximately 2.5, 1.5 and 1.2 for robustness values of -0.5, 0.0 and 0.5 respectively.

\begin{table}
\begin{center}
\caption{RMS noise measured in pilot survey phase I data. RMS noise per beam is given as the minimum and average over all observations in CASDA, then the minimum scaled to a standard \SI{1}{\hour} and \SI{288}{\mega\hertz} (for continuum, above the line) or \SI{18.5}{\kilo\hertz} (for spectral line, below the line). }
\label{tab:tsys}
\begin{tabular}{lllll|l}
\hline 
$\nu$ & $\Delta\nu$ & $t_{int}$ & {Min} & {Avg} & {Scaled} \\
{(\si{\mega\hertz})} & {(\si{\mega\hertz})} & & (\si{\micro\jansky}) & (\si{\micro\jansky}) & (\si{\micro\jansky}) \\
\hline
944 & 288 & 10\,h & 24 & 37 & 74 \\
864 & 288 & 12\,m & 202 & 420 & 90 \\
1368 & 144 & 8\,h & 40 & 70 & 80 \\
1665 & 9 & 8\,h & 186 & 582 & 95 \\
\hline
856 & \SI{18.5}{\kilo\hertz} & \SI{2}{\hour} & 4860 & 5700 & 6870 \\
1368 & \SI{18.5}{\kilo\hertz} & \SI{8}{\hour} & 2000 & 2500 & 5660 \\
\hline
\end{tabular}
\medskip\\
\end{center}
\end{table}

Most of ASKAP's observing time will be dedicated to large-scale survey projects and the observatory will provide science-ready data products via a public archive (see Section \ref{sec:CASDA}). ASKAP's deep survey projects are expected to discover millions of new radio sources in the southern sky \citep{2007PASA...24..174J}. The large instantaneous field of view also opens up new parameter space for the study of transient sources.  ASKAP will excel at wide-area, high cadence surveys for slow transients; wide-area surveys for spectral line emission and absorption; rapid, wide-field searches for gravitational-wave counterparts; and the discovery and localisation of fast transients via 1\,ms cadence autocorrelations and a deep voltage buffer (up to \SI{14}{\second}).

\section{The Murchison Radio-astronomy Observatory}
\label{sec:mro}
ASKAP is located on a remote site in Western Australia, specifically established as the Australian radio quiet zone (WA) for the SKA and its precursors \citep{Wilson2013, Wilson2016}.  The total power in RFI signals over the ASKAP band at this site is typically more than an order of magnitude less than ASKAP's system noise power \citep{Chippendale2013}. Legislation regulates the use of radio transmitters within \SI{260}{\kilo\metre} of the site \citep{acma2007}, helping to ensure that the environment will remain as favourable as possible for radio astronomy into the future. Further protection is afforded by carefully testing and, if necessary, shielding all electronic equipment installed at the site \citep{Beresford2013}.

\begin{figure}
    \centering
    \includegraphics[width=\columnwidth]{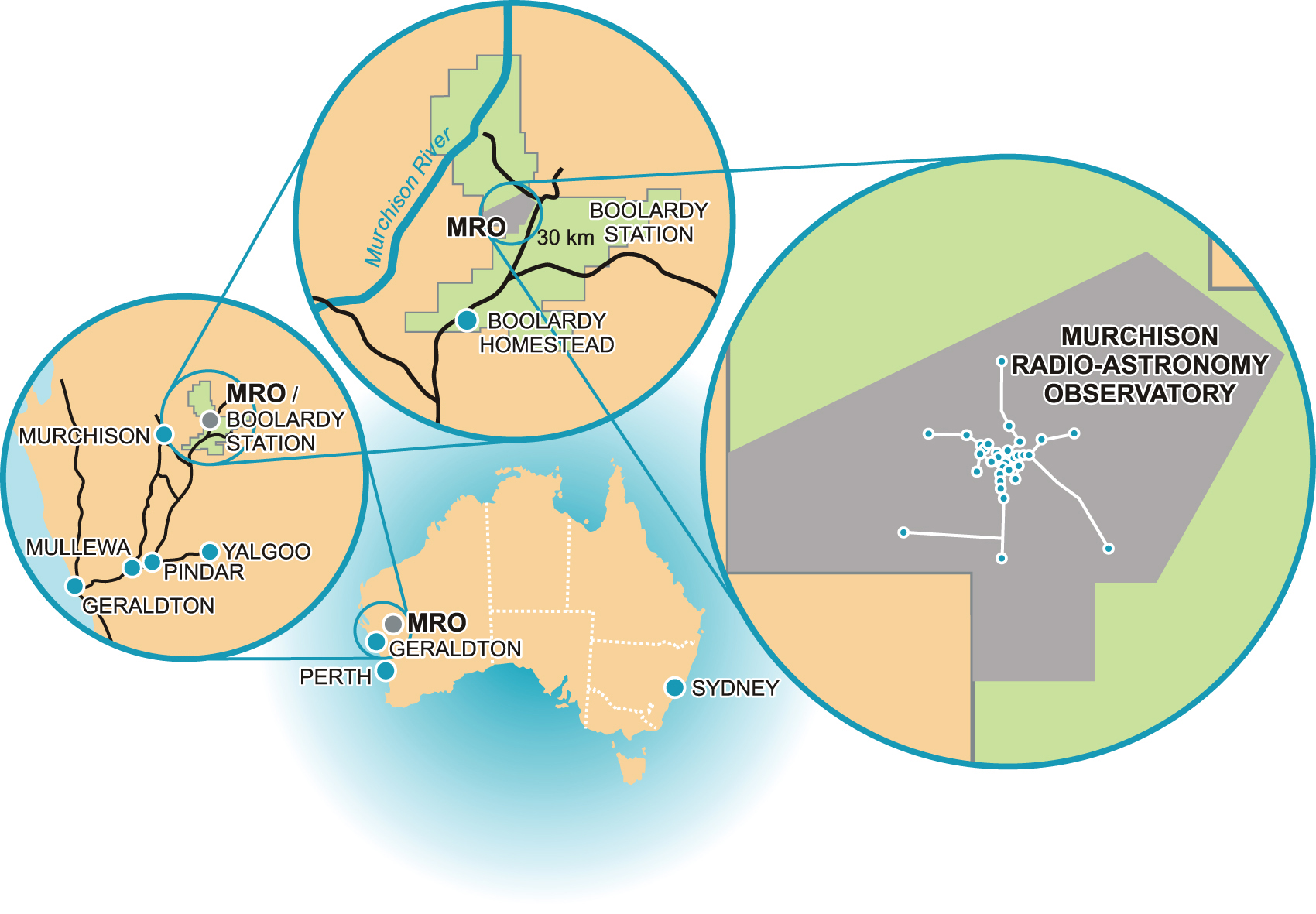}
    \caption{Diagram showing the relative location and size of the Murchison Radio-astronomy Observatory, with the final inset showing ASKAP antennas as blue dots and service tracks as white lines.}
    \label{fig:MRO_map}
\end{figure}

The observatory site (see Figure~\ref{fig:MRO_map}) is roughly 315\,km north-east of Geraldton, the largest town in the region. It is operated by the Australian Commonwealth Scientific and Industrial Research Organisation (CSIRO) and hosts the Murchison Widefield Array (MWA) \citep{2013PASA...30....7T,2018PASA...35...33W} and EDGES \citep{2018Natur.555...67B} telescopes in addition to ASKAP. In future, the low-frequency component of the SKA will be established nearby.

In order to prevent ASKAP's infrastructure from creating RFI, all processing hardware is contained within a dual-shielded central control building \citep{Abeywickrema2013} that attenuates any radio emission by 160\,dB. Signal processing and computational systems are housed inside the second layer of shielding, while office and workshop space are inside the first layer. The building provides working space for maintenance crews but no on-site accommodation. The number of people on site is kept to a minimum and crews are housed in the Boolardy station accommodation complex, roughly 40\,km from the telescope.

\section{SYSTEM OVERVIEW}
\label{sec:system}
\begin{figure*}
\begin{center}
\includegraphics[width=\textwidth]{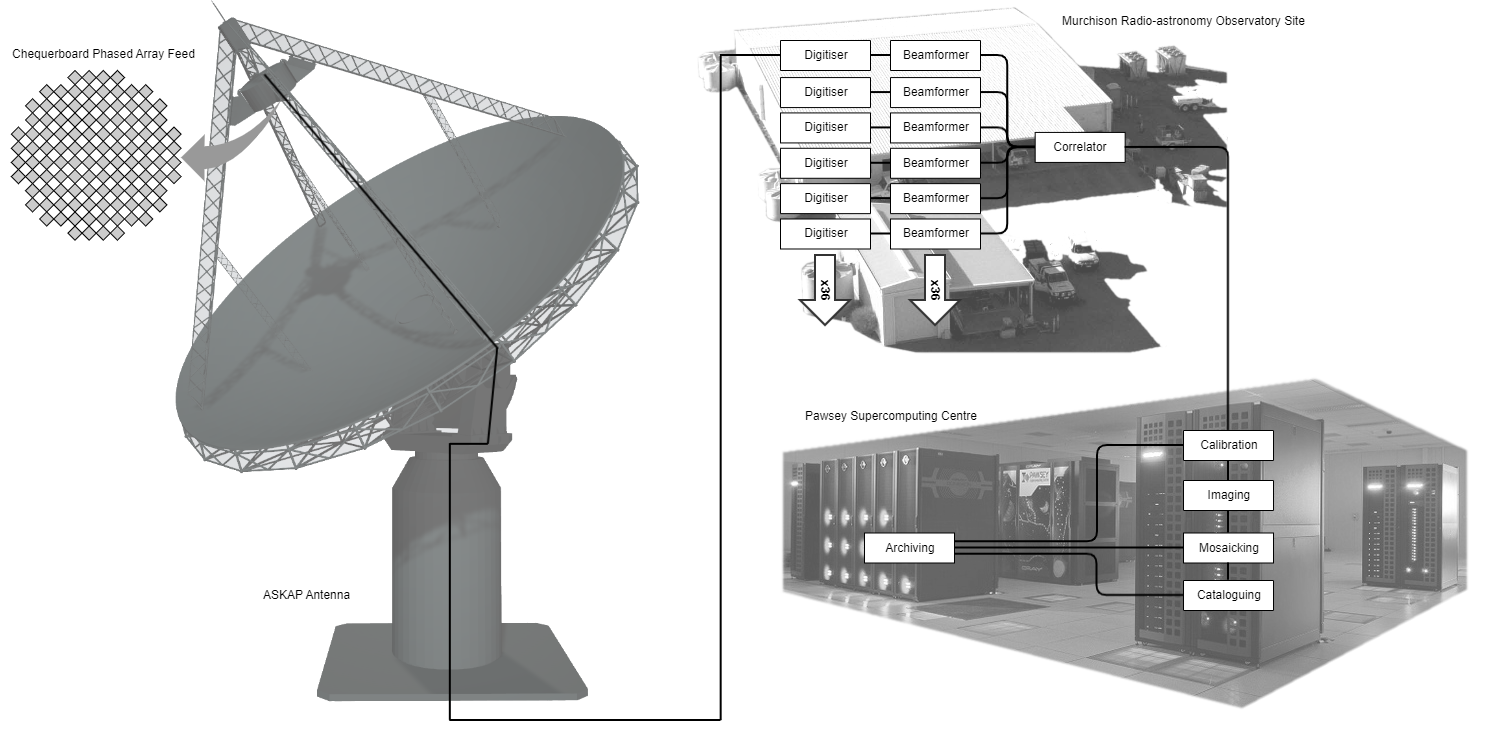}
\caption{Overview of key ASKAP subsystems and data flow.}
\label{fig:system}
\end{center}
\end{figure*}

Figure~\ref{fig:system} shows the major components of the telescope and highlights how data flows from the antennas through to the image archive. Each subsystem is described in more detail in the sections below.

ASKAP consists of 36 paraboloidal reflector antennas, each 12\,m in diameter with a chequerboard PAF at the primary focus.  Within each PAF, the radio frequency signals received by 188 active feed elements (94 per polarisation) are converted into analogue optical signals.  The optical signal from each element is transmitted over its own dedicated optical fibre back to the central control building. See Section~\ref{sec:PAF} for details.

Inside the control building, each PAF element signal is digitised and then passed through a coarse oversampled polyphase filter bank with 1\,MHz channel spacing. This generates 640 or 768 channels depending on which of three frequency bands is being used. The digital receiver system selects 384 of these channels and sends them via digital optical communications links and a passive optical cross-connect to the beamformers.

The beamformers compute the weighted sum (see Section \ref{sec:beamforming}) across PAF elements for each frequency channel and antenna, forming 36 dual-polarisation beams over \SI{336}{\mega\hertz}. The beamformed signals then pass into a fine filter bank that divides each 1\,MHz channel into 54 channels of 18.5\,kHz width, for a total of \num{18144} fine channels. Bandwidth can be exchanged for higher frequency resolution by up to a factor of 32 using zoom modes. See Section~\ref{sec:DSP} for details. The beamformer also generates 1\,ms power averages of the signal for each 1\,MHz channel of each beam. This is processed on site to search for fast transients.

The fine channels are sent via digital optical links and an optical cross-connect to the correlator, which computes and accumulates the visibilities for each baseline.  The correlator has the capacity to process \num{15552} fine channels for a maximum bandwidth of \SI{288}{\mega\hertz}. These visibilities are sent directly to the Pawsey Supercomputing Centre in the city of Perth via a long-haul underground optical fibre network, with no buffering at the observatory. Three petabytes of disk buffer is available at Pawsey, where raw visibilities are formatted and written to a number of Common Astronomy Software Applications (CASA) {\tt MeasurementSet} files \citep[version 2, ][]{MSv2} in parallel.

The visibility data are imaged offline in batch processing mode using a custom software package called ASKAPsoft \citep{2019ascl.soft12003G}. The Pawsey Supercomputing Centre currently maintains a dedicated compute cluster called \textit{Galaxy} for the purpose of imaging ASKAP science data. This system is due to be replaced with a more powerful supercomputer in 2021. See Section~\ref{sec:SDP} for details.

Data products are made available to astronomers via a public archive, described in Section~\ref{sec:CASDA}. As shown in Figure~\ref{fig:system}, the archive can store calibrated visibility data (averaged in frequency to reduce disk space usage), mosaicked image cubes (but typically not individual beam images) and source catalogues.

\subsection{Control and monitoring}

ASKAP's control system is built on the Experimental Physics Instrument Control System (EPICS) library\footnote{\url{https://epics.anl.gov}}. Each major hardware subsystem has an associated EPICS input/output controller (IOC) which is a software application that issues commands and gathers monitoring information via network transactions. Several servers located at the MRO run these IOCs and there is a dedicated monitoring and control network in addition to the office network and astronomical data network. The design of the control system is described in \cite{2010SPIE.7740E..1JG}.

The digital signal processing hardware was designed with a common command and control interface that translates instructions from the network into firmware registers on field-programmable gate arrays (FPGAs) throughout the system. A common software library was written to support the hardware layer and this library is used by the EPICS IOCs. Various other commodity electronics devices (such as drive controllers and power supplies) are also used, requiring different network protocols at the IOC level.

\subsection{Power infrastructure}

Due to its remote location, the MRO is not connected to a mains power grid. The site is supplied by a hybrid solar-diesel power station purpose-built with sufficient RFI shielding to protect the observatory from stray radio emissions. The power station is located approximately 6\,km from the core of the telescope to further aid in RFI mitigation. The solar component of the power station can supply 1.6\,MW (enough to run the entire site during the day) and is linked to a battery bank with 2.5\,MWh capacity. Several redundant diesel generators are the main power source at night. 

Power is transmitted via underground high-voltage cable to the central control building.  Distribution to the antennas is on six independently-switched underground tracks that roughly follow the surface roads and power several antennas each. A transformer at each antenna pad steps the high-voltage input down to three-phase 415\,V for the equipment in the antenna pedestal.

\subsection{Cooling infrastructure}

ASKAP's digital signal processing hardware in the central control building consumes 280\,kW of power and requires a commensurate amount of cooling. This is achieved by circulating chilled water to each electronics cabinet in the central building, where internal heat exchange units extract heat from the air. The cooling system vents this waste heat into a geothermal exchange system. When the air temperature is low a water-to-air heat exchanger can be put in series, extending the life of the geothermal system. The entire cooling system was designed to minimise radio frequency emissions, as described in \cite{Abeywickrema2013}.

\subsection{Network infrastructure}

The network between Perth and the MRO consists of approximately 900\,km of optical fibre using carrier-grade transmission equipment. The network was designed and is managed jointly by CSIRO and AARNet (the Australian Academic and Research Network). Using dense wavelength division multiplexing, the system can support up to 80 channels on a single fibre pair.  Over this distance, optical amplification is required every \SIrange{80}{100}{\kilo\metre}. Only two of these channels are used for production services, each lit by a 100\,Gbit/s transponder split out into ${10 \times \SI{10}{\giga\bit\per\second}}$ services. The first runs direct to the Pawsey Supercomputing Centre in Perth with $4 \times \SI{10}{\giga\bit\per\second}$ used for ASKAP data and one 10\,Gbit/s service to carry MWA data. The second \SI{100}{\giga\bit\per\second} link provides services between Perth, Geraldton and the MRO such as CSIRO internal network connectivity, IP telephony, and monitor and control traffic.

There are additional transponders to support SKA development activities at Curtin University and for long-haul network performance testing undertaken by AARNet and CSIRO. Currently, there is 500\,Gbit/s of lit capacity between Perth and the MRO using four of the 80 available channels. There is also a parallel network consisting of two 1\,Gbit/s links directly patched to fibre pairs between Geraldton and the MRO. This provides network services at the various locations en route and acts as a backup network should the transmission system fail, however it does not protect against a fibre cable cut.

The network at the MRO itself is a dual-core design. One core handles all the telescope data and connects the digitisers and beamformers. The second core handles the monitoring and control of the telescope as well as the general purpose network at the MRO. The dual-core allows for separation of the different systems providing better control and resilience. The data network is large, the core router has over 2400 network ports. 

The second core makes extensive use of virtual local area networks (VLANs) to partition traffic. This avoids telescope monitor and control commands being mixed with unrelated network traffic. VLANs provide different network connections at relevant locations across the MRO.

The conditions at the MRO present some challenges to network equipment; the antenna pedestals are not cooled, so industrial switches are used, providing greater reliability and enabling connectivity for the varied range of equipment that has been installed at the observatory. The flexible network design allows for services to be provided to several clients located at the observatory, helping to broaden the research taking place at the site.

\section{ANTENNA DESIGN}
\label{sec:antenna}
ASKAP's antennas were designed, constructed and installed by the 54th Research Institute of China Electronics Technology Group Corporation (CETC54) to CSIRO's specification \citep{Jackson2008}. The antennas are \SI{12}{\metre} diameter unshaped prime-focus paraboloidal reflectors with a focal ratio $f/D$ of 0.5.  They have a solid surface, specified up to \SI{10}{\giga\hertz}, and are mounted on a novel ``sky-mount'' that can roll the entire reflector, quadrupod and feed structure about the optical axis \citep{DeBoer_2009}.  Slew rates of \SI{1}{\deg\per\second} in elevation and \SI{3}{\deg\per\second} in azimuth and roll mean that ASKAP can slew to and track an arbitrary position on the sky within one minute of request. 

Continuous rise-to-set tracking of most sources is possible above the antenna elevation limit of \ang{15}. From the observatory's position at \ang{26.7}\,S latitude, the telescope can observe sources between \ang{-90} and \ang[retain-explicit-plus]{+48} declination, although with decreasing time above the horizon for sources closer to the northern limit. The PAF's wide field of view extends this reach by another \ang{2.5}.

\subsection{Mechanical design and axes of motion}

The antenna structure consists of a steel pedestal and support frame, topped with solid panels made of non-metallic honeycomb sandwiched between aluminium sheets. The feed is located at the prime focus.

Due to the small size of the reflectors and the comparatively large size of the PAF, special consideration was given to the strength and rigidity of the prime-focus support structure. The robust feed legs, specified to support a \SI{200}{\kilo\gram} PAF receiver, block \SI{4}{\percent} of the total collecting area with the symmetric antenna design used. However, the symmetric design produces more uniform offset beams, in both total intensity and polarisation.

The use of a PAF with offset beams also makes it necessary to track parallactic angle on the sky during an observation. This is typically done with an equatorial mount, which creates extra mechanical complexity. For ASKAP, a unique hybrid approach was developed. CSIRO designed a simple azimuth-elevation mount that included a third axis of rotation for the reflector itself.\footnote{The third antenna axis was first suggested by Dr. Peter Dewdney and colleagues from the Dominion Radio Astrophysical Observatory in Canada.}

While it would be possible to track parallactic angle by continuously updating the electronic beamformer weights or rotating just the feed, the third axis exchanges a small amount of mechanical complexity for greatly reduced computational complexity at the time of calibration and imaging, since it maintains the angular relationship of the feed elements with respect to the support structure.

Unlike the azimuth axis, the antenna polarisation axis (also known as the roll axis) does not have overlapping range limits and can only be driven slightly less than \ang{\pm180} from its neutral position. This means it is not possible to continuously track northern sources without unwrapping at transit or beginning the track at an offset roll angle.

The antennas were specified to operate up to \SI{10}{\giga\hertz}, providing the flexibility to consider higher-frequency receiver upgrades in future. A single ASKAP antenna was successfully used for several years with a single-pixel feed for VLBI observations at \SI{8.4}{\giga\hertz} \citep{Kadler2016}. Preparation for a large-scale receiver system upgrade to higher frequencies should include direct measurements of atmospheric water vapour to assess the suitability of the site itself. Photogrammetry was used to measure both the location of the PAF at the focus and the reflector accuracy after installation. The rms deviation from the expected parabolic shape is typically less than \SI{1}{\milli\metre}. This is consistent with measurements of the first antenna during test assembly at the factory \citep{Feng2010}.

\subsection{Wind limits and storm monitoring}

The ASKAP antennas were specified to operate under conditions similar to other radio telescopes. The antennas will stow if the peak wind speed exceeds \SI{45}{\kilo\meter\per\hour} during the last \SI{30}{\minute}. We find that very little observing time is lost as a result.

Strong winds at the MRO are usually associated with oncoming storm fronts. These can cause the wind speed to rise extremely rapidly, sometimes faster than the time it takes the antennas to stow. On a few occasions, this has led to drive errors during the stow procedure as wind loading exceeded the safe operating range of the motors. To prevent circumstances such as this, we have implemented a storm stow system that uses satellite meteorological data and lightning detectors to sense approaching storm fronts and stow the antennas in advance of their arrival  \citep{2018SPIE10704E..1WI}.

Not all nearby storms lead to high winds at the telescope, but associated lightning activity usually produces extensive radio frequency interference. False positive stow triggers are therefore not a major concern as astronomy data would be impacted by a storm in any case.

\subsection{Array configuration}
\begin{figure}
\begin{center}
\includegraphics[width=\columnwidth]{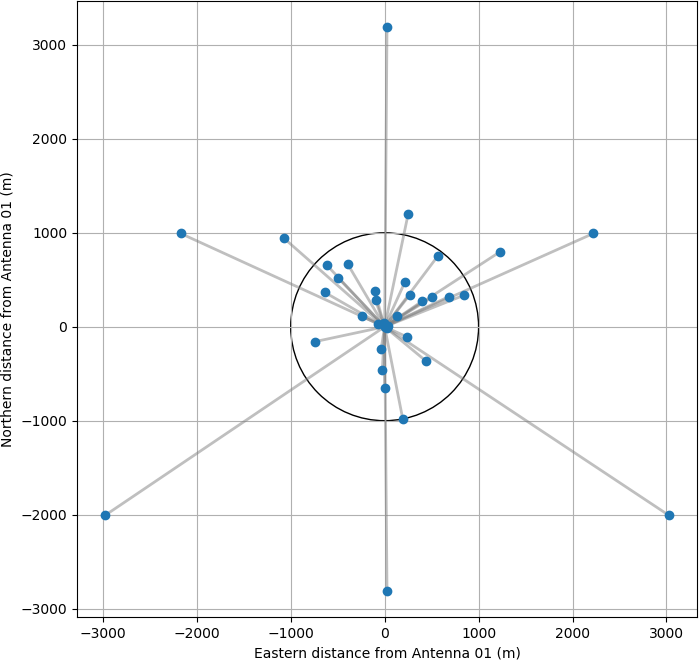}
\caption{Location of each ASKAP antenna plotted relative to Antenna 1. A circle of radius 1\,km is drawn for scale. The outer six antennas form a Reuleaux triangle and provide approximately 10$''$ resolution. The dense cluster of antennas in the core provides excellent surface brightness sensitivity.}
\label{fig:FullArrayMap}
\end{center}
\end{figure}

Of the 36 ASKAP antennas, 27 were placed to provide a Gaussian distribution of spatial scales with a point spread function of 30$''$ \citep{ASKAPconfig}. Three additional antennas were added to the core of the array to increase surface brightness sensitivity and another six antennas were added on longer baselines (up to 6\,km) in a Reuleaux triangle (see Figure~\ref{fig:FullArrayMap}) to provide improved resolution (approximately 10$''$) for compact sources.  Antennas are plotted relative to ASKAP antenna 1 which is located at \ang{26.6970007225}~S, \ang{116.6314242861}~E and an elevation of \SI{361}{\metre}.

ASKAP has excellent instantaneous ($u,v$) coverage \citep{ASKAPconfig}, making snapshot surveys and equatorial imaging possible.

\section{ASKAP PHASED ARRAY FEED}
\label{sec:PAF}


\subsection{Design background}

The first production prototype PAFs for ASKAP used copper coaxial cables for signal transport, with digitisation inside the antenna pedestal \citep{hotan_2014}. 6 of these units were constructed for assessment in the field. Aperture array measurements \citep{Schinckel2011, 2014PASA...31...19C}, tests at the focus of a 12\,m antenna located at the Parkes observatory \citep{DeBoer_2009, 5613298, Sarkissian2017} and the BETA array at the MRO \citep{hotan_2014} revealed several issues with the Mk I PAF. These included effective system temperatures in excess of \SI{150}{\kelvin} (more than twice the design requirement) in the upper half of the frequency band \citep{mcconnell_2016} and the need to fully remove the PAF from the focus of the antenna to perform internal maintenance.

In order to better assess the performance of prototype PAF designs, \cite{2014PASA...31...19C, PAF_T, Chippendale2016} developed an aperture-array method of determining the noise temperature of a PAF beam using measurements of a microwave absorber, the radio sky, and broadband noise transmitted from a reference antenna. This was a valuable step in the testing process as it helped validate electromagnetic modelling of the PAF that in turn enabled improvements in sensitivity \citep{Shaw2012}. Aperture-array testing was also practical as it allowed rapid testing of PAF prototypes on the ground instead of at the focus of an antenna.

During these early tests, research into RF over Fibre showed that the PAF signal could be transported to the central site for processing rather than digitising the signals at the antenna. A major redesign was initiated to incorporate experience from the Mk I tests and take advantage of new technological developments. The ASKAP Design Enhancements (ADE) project developed a Mk II system with lower overall cost, improved maintainability and effective system temperature less than \SI{80}{\kelvin} across most of the band. LNAs could be changed without removing the PAF from the dish and sensitive electronics were relocated from the antenna pedestal to the climate-controlled central building. All 36 ASKAP antennas are now fitted with Mk~II PAFs and digital systems.

\begin{figure}
\begin{center}
\includegraphics[width=\columnwidth]{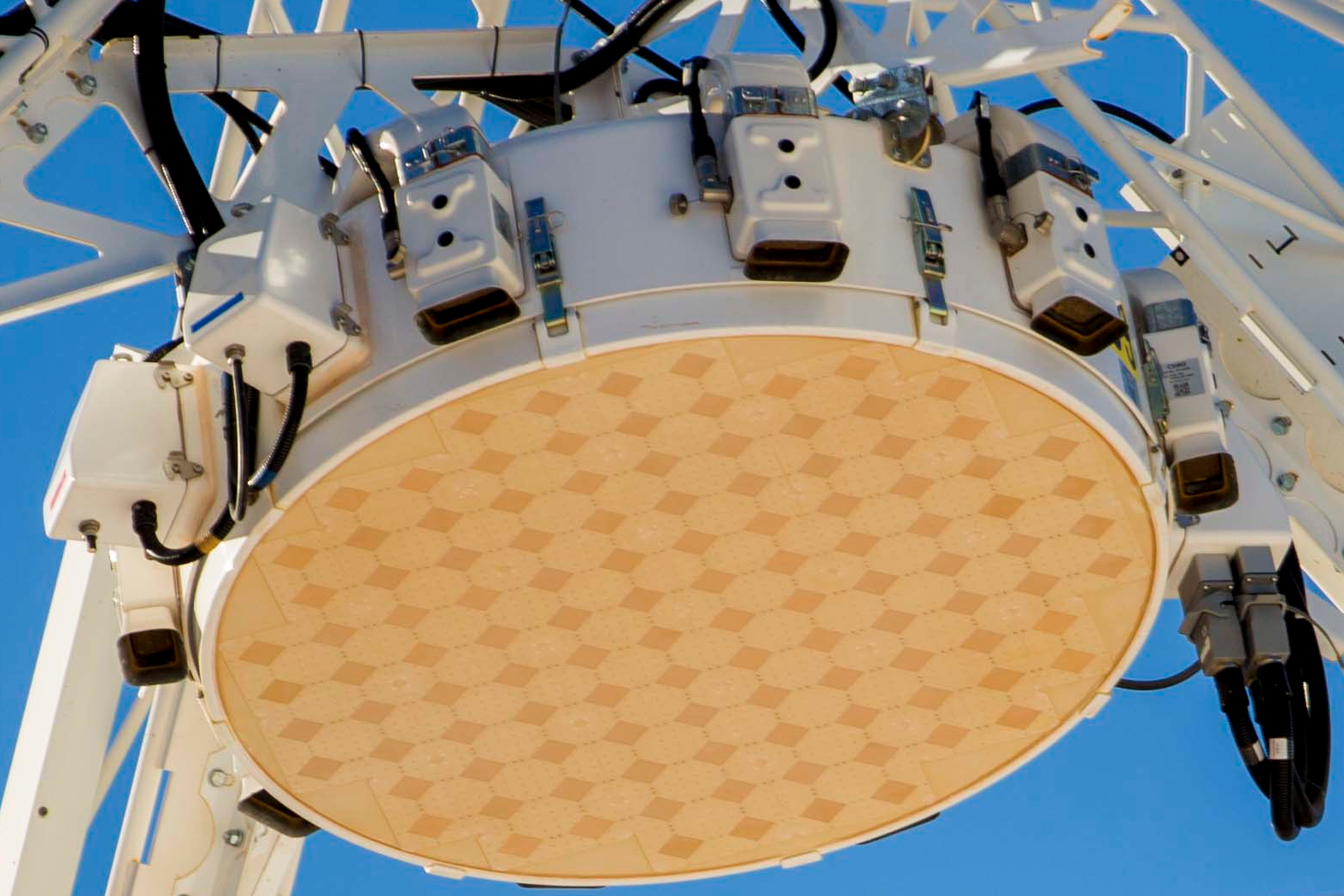}
\caption{Photograph of a Mk~II PAF installed on one of the ASKAP antennas at the MRO. The chequerboard surface is visible, along with the composite case and air vents for the cooling system. Power and optical cables attach via several bulkheads on the side of the case.}
\label{fig:paf_photo}
\end{center}
\end{figure}

\subsection{Planar chequerboard connected-array}

The ASKAP PAFs fill a region of the antenna's focal plane with small receptors that consist of flat, square conductive patches (see Figure~\ref{fig:paf_photo}) printed on a circuit-board substrate to form a planar connected-array antenna \citep{Hay2007, Hay2008}. The distance between each patch is 90\,mm. A row of patches can be considered as a row of bow-tie antennas that are connected edge to edge, forming a linearly connected array. The feed point at the centre of each ``bow-tie'' is differential and both orthogonal linear polarisations are available due to the two-dimensional nature of the grid. Each element has a radiation pattern that would over-illuminate the dish surface, but the beamforming process combines these elements to create an efficient illumination pattern for a given direction on the sky.

The precise geometry of the chequerboard surface was determined through electromagnetic simulation \citep{Hay2011}. Matching the simulation results to experimental data from the Mk~I system was a key step to ensure that design modifications could be assessed prior to manufacturing. Co-design of the chequerboard array and its LNAs was critical for achieving high sensitivity. An initial impedance target for the LNA and matching network design was derived by optimising for maximum beamformed sensitivity over the field of view \citep{Hay2010}.   Slight modifications were made to the shape of the surface elements on the Mk~II design to improve the impedance match with the LNAs in the upper observing band (\SIrange{1400}{1800}{\mega\hertz}) and therefore reduce the system temperature.  

The surface panels are bonded to a ground plane via several centimetres of non-conductive honeycomb backing which provides structural rigidity as well as insulation. Twin-wire feed lines are used to connect the corners of each ``bow-tie'' element to LNAs housed underneath the ground plane. A conformal coating is applied to the outward-facing surface for protection from dust and weather.

\subsection{Composite enclosure}

The density of electronics inside the PAF and the sometimes harsh environmental conditions at the focus of the antenna makes the design of the enclosure extremely challenging. As well as keeping the internal systems dry, free of dust and thermally isolated from the external environment, the enclosure must provide more than 30\,dB of extra RF shielding \citep{PAF_shielding}. Experience with the BETA array showed that poorly-shielded enclosures led to significant leakage of noise that could correlate, including between PAF elements and nearby antennas.

The Mk~II enclosure is a moulded composite structure that is bonded to the metal ground plane and shielded with carbon fibre. Four access hatches at the top are fitted with RF gaskets to reduce radio frequency leakage.  The access hatches allow replacement of faulty electronics modules, even when a PAF is installed at the focus of an antenna.

\subsection{Dominoes and monitoring}
\label{Domino}

Each pair of LNAs is housed in a self-contained RF shielded enclosure known as a domino.  The domino design was developed for the Mk~II PAF to improve shielding of the components with the highest gain, and also to improve modularity and maintainability.  The dominos are bolted directly to the ground plane to provide conductive heat transfer and can be removed individually for maintenance. 

\begin{figure}[htb]
    \centering
    \includegraphics[width=\columnwidth]{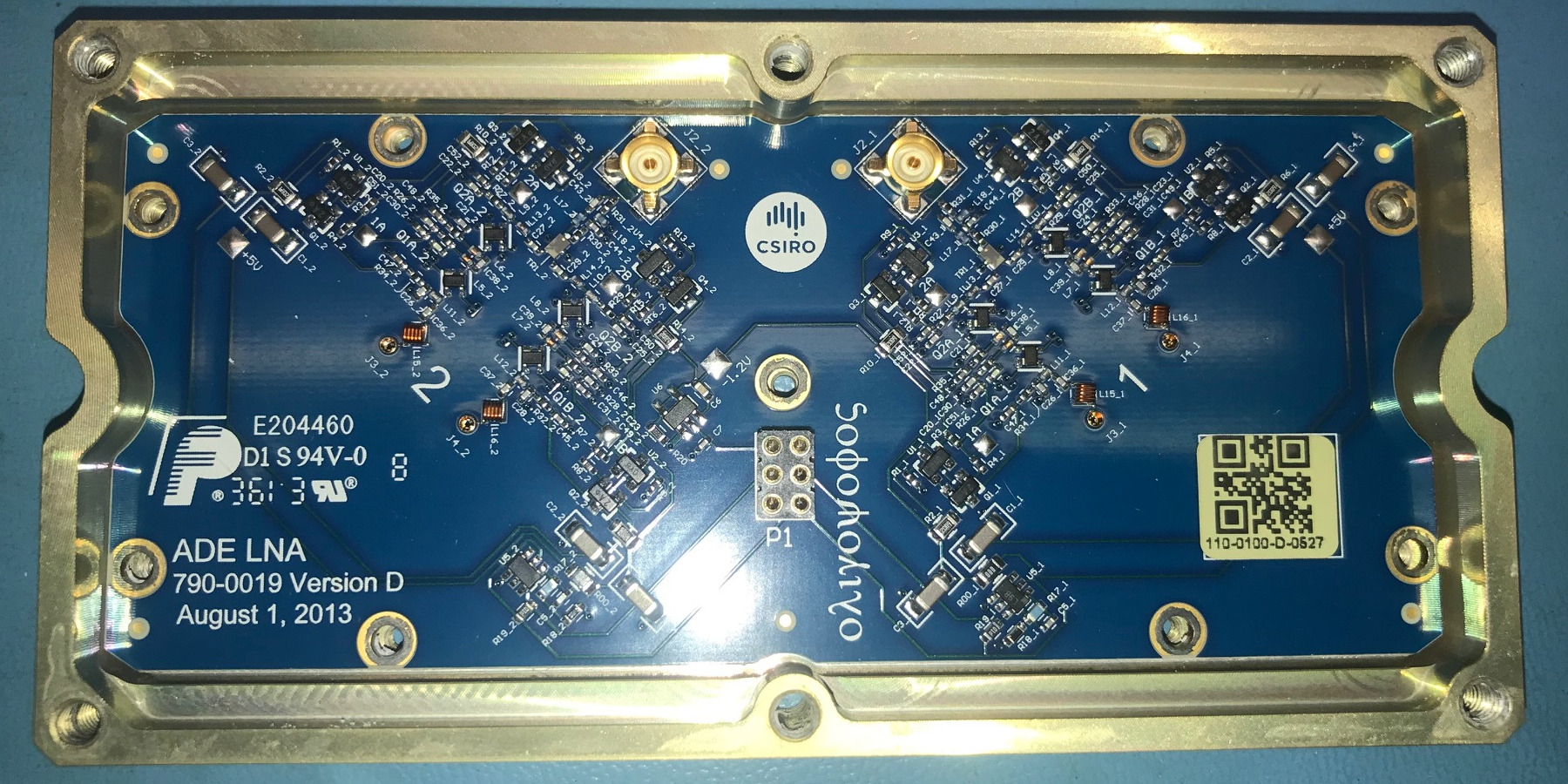}
    \caption{Photograph of the low-noise amplifier layer in the domino, referred to elsewhere as LNA.}
    \label{fig:DominoLNA}
\end{figure}

\begin{figure}[htb]
    \centering
    \includegraphics[width=\columnwidth]{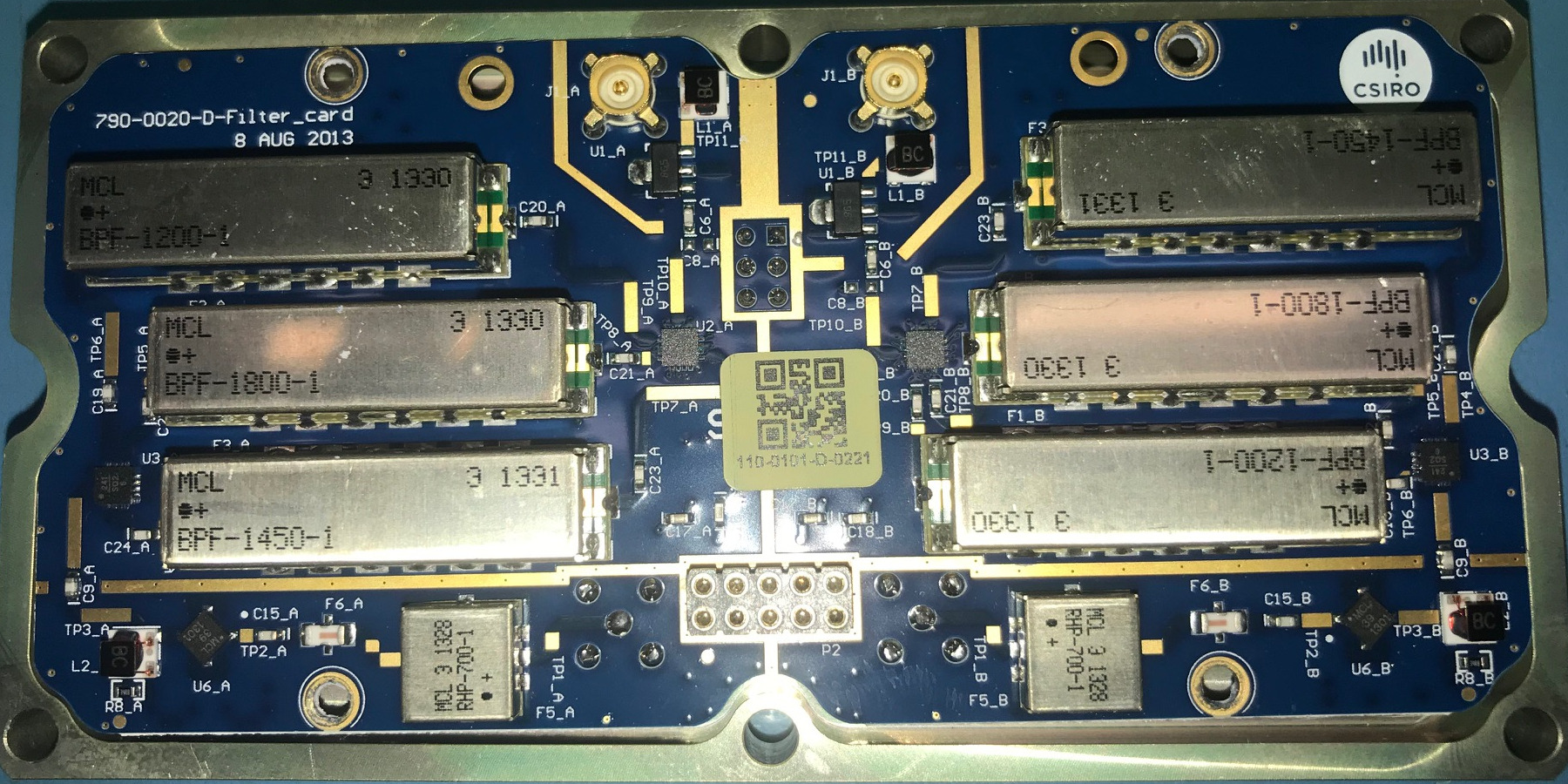}
    \caption{Photograph of the filter layer in the domino, referred to elsewhere as FILTER.}
    \label{fig:DominoFIL}
\end{figure}

\begin{figure}[htb]
    \centering
    \includegraphics[width=\columnwidth]{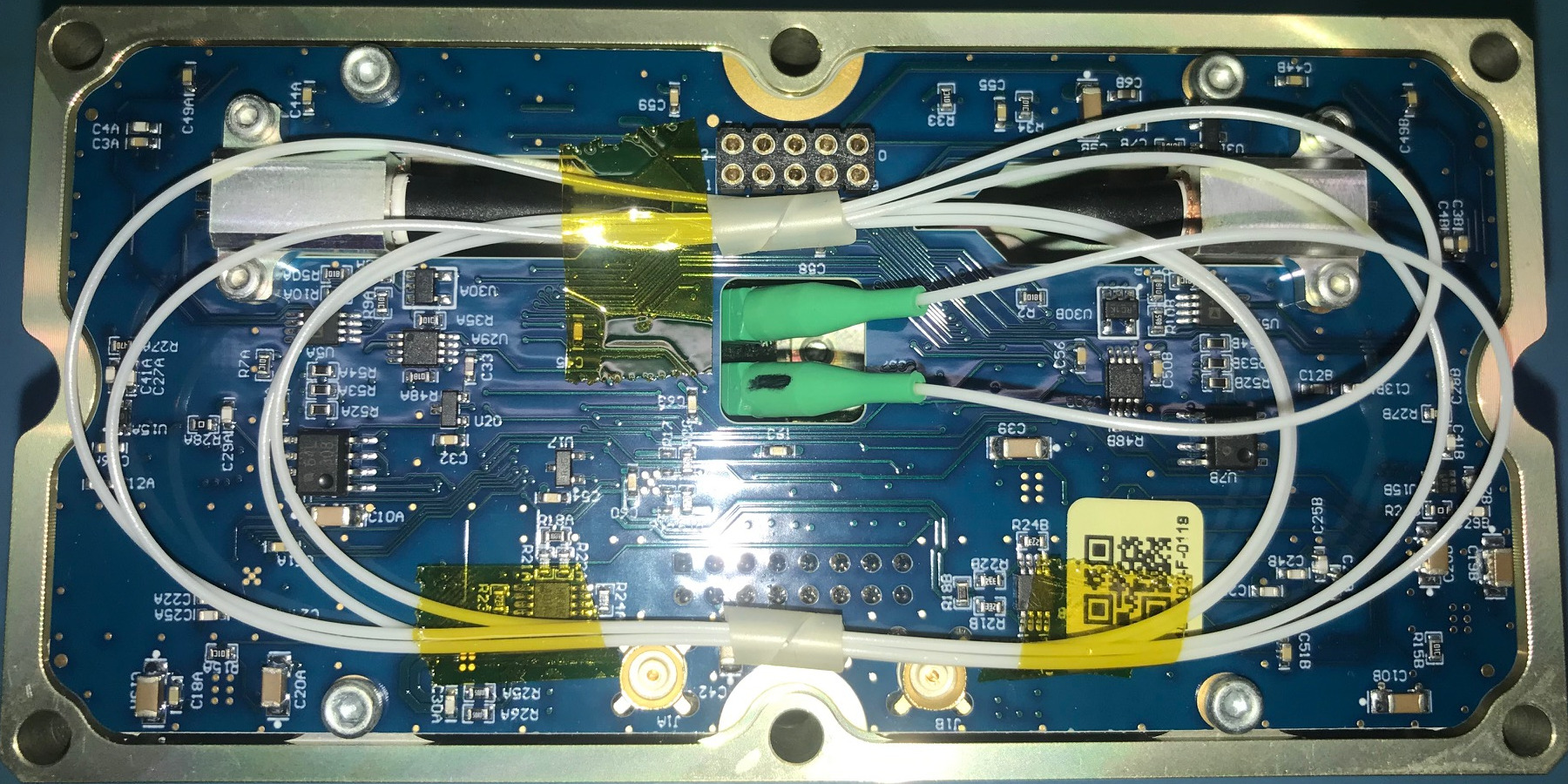}
    \caption{Photograph of the optical transmitter layer in the domino, referred to elsewhere as TRANSMITTER.}
    \label{fig:DominoOPT}
\end{figure}

\begin{figure}[htb]
    \centering
    \includegraphics[width=\columnwidth]{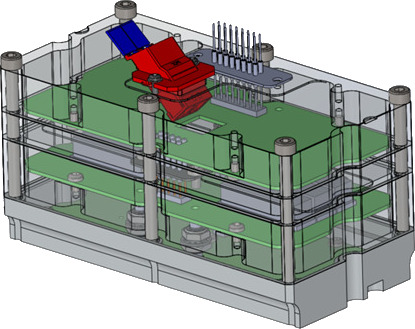}
    \caption{Diagram showing the assembled domino with transparent walls. The LNA card is closest to the bottom of the stack, followed by the filter card in the middle and the optical transmitter at the top.}
    \label{fig:DominoSTA}
\end{figure}

Each domino has three layers that contain differential LNAs, filters and electro-optical converters stacked in order of distance from the ground plane. A photograph of the interior of each layer is shown in Figures~\ref{fig:DominoLNA},~\ref{fig:DominoFIL} and~\ref{fig:DominoOPT}, and a transparent view of the assembled stack is shown in Figure~\ref{fig:DominoSTA}. The signal path through the domino layers is shown in the upper-left of Figure~\ref{fig:analogue_path} in the boxes marked LNA, FILTER and TRANSMITTER.

The top of each domino provides connections for two fibre optic cables that carry the analogue RF output, and for a copper ribbon cable that supplies DC power and supports monitoring and control within the PAF. Due to the proximity of the LNAs, front-end monitoring has a very high risk of generating RFI. ASKAP's PAF monitoring system was designed to be low-impact, with no continuous clock signal and RFI filters on all connections to wires within the enclosure. The monitoring system is based on the Serial Peripheral Interface (SPI) bus and is shown in the lower-right part of Figure~\ref{fig:analogue_path}.  Transmission of data packets may generate RFI, but most monitoring is disabled during an observation, with short updates between scheduling blocks to maintain a record of the system health. Control of each PAF is managed by a front-end controller (FEC) unit that converts instructions from an EPICS IOC into optical signals for the PAF and returns monitoring data to the software layer.

\subsection{Ground plane and cooling}

The Mk~I PAFs employed water cooling inside their enclosures, using a heat exchange unit in the PAF and fans to circulate cool air throughout the electronics. Although this provided good cooling capacity, it required housing a water chiller at each antenna pedestal (with associated RFI issues) and pumping water up to the focus of the antenna. In addition, the possibility of condensation around the chilled water inlet or water leakage inside the PAF enclosure was a significant concern.

The Mk~II PAFs use heat-pipes embedded in the ground plane to transport heat to the edge of the enclosure.  Further heat pipes connect to thermo-electric coolers (TECs, shown in Figure~\ref{fig:analogue_path} and seen around the edge of the PAF in Figure~\ref{fig:paf_photo}) that transport the heat to the outside of the case where a ducted fan blows ambient air over heat-sink fins. This system has proven effective at keeping the internal temperature as much as \SI{15}{\degreeCelsius} below ambient. It can be regulated via the voltage applied to each TEC. A servo loop keeps the internal temperature at a steady \SI{25}{\degreeCelsius} set point (which can be changed in software if necessary).

The PAFs were designed to operate in ambient temperatures up to \SI{45}{\degreeCelsius}. This threshold is rarely exceeded except during the most extreme summer conditions. The cooling system is capable of maintaining safe internal operating temperatures in all conditions experienced to date. However, it is not always possible to keep the internal temperature fixed.  The optimal internal temperature of \SI{25}{\degreeCelsius} is exceeded when ambient temperatures rise above \SI{35}{\degreeCelsius}, which is quite common during summer. The PAF is fitted with an automatic safety shutdown that is triggered when the internal air temperature  exceeds \SI{50}{\degreeCelsius}.

The exact cooling capacity is slightly different for each unit. When the cooling capacity is exceeded, the internal temperature of the PAF begins to track the external ambient temperature, with a constant offset of approximately \SI{10}{\degreeCelsius}. This can cause amplitude calibration errors during astronomical data processing as the gain of the analogue electronics depends on their operating temperature. The impact of this on image quality has yet to be determined and it may be possible to correct using time-dependent self-calibration, a temperature-based lookup table, or the on-dish calibration (ODC) system described in the next section.

\subsection{On-dish calibration}
\label{sec:odc}
For a dish with a single feed, amplitude calibration is often obtained by placing a switched noise source in the feed that adds a known amount of power to both polarisations when activated.  To determine the system amplitude gain, the power induced by the noise source is synchronously demodulated and measured with respect to the background.

Individually injecting a calibration signal into each PAF element is nontrivial due to the number of active elements. Instead, a common noise signal is radiated from the dish vertex to the PAF, illuminating all elements as uniformly as possible. The calibration signal is typically set to inject around \SI{15}{\kelvin} equivalent noise temperature into the PAF, but this can be adjusted within the range \SI{100}{\milli\kelvin} to \SI{200}{\kelvin} as required.  We expect to reduce the calibration signal level as we gain experience in its use. The calibration noise is currently on continuously during all observations, but it is largely cancelled by beamforming with maximum signal-to-noise ratio beamformer weights resulting in less than \SI{1}{\percent} increase to beamformed system temperature \citep{Chippendale_2018_ODC}.  

Using a radiated calibration signal has the disadvantage that it can leak into neighbouring antennas, but the signal power is such that the correlated contribution in a nearby antenna is less than that due to leakage of amplifier noise.  Such common noise contributions could be further mitigated via Walsh modulation of the phase switches incorporated into each domino module, however this has not been found necessary.

\begin{figure}
    \centering
    \includegraphics[width=\columnwidth]{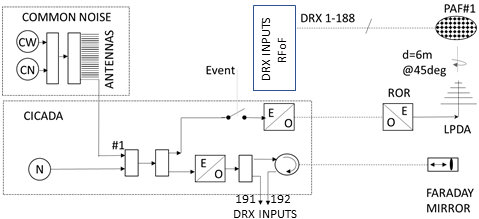}
    \caption{Diagram showing key components and signal paths for ASKAP's on-dish calibration (ODC) system}
    \label{fig:ODC}
\end{figure}

Processing one or two calibration signals is a small engineering overhead compared to the 188 signals generated by each PAF.  This allows the injected calibration signal to be digitised and correlated with the signals from all PAF elements, yielding an amplitude and phase calibration for each port.  Corrections are then made to the beamformer weights to maintain stable beam patterns and sensitivity as described in Section~\ref{sec:odc_use}.

The on-dish calibration system consists of several modules shown in Figure~\ref{fig:ODC}. There is one Cicada module per antenna located in the central building. During normal operation, it generates an independent noise signal that is directly modulated onto an optical carrier and sent to the antenna on single-mode fibre.  At the antenna, the optical signal is detected and the resulting electrical signal is sent to a small log-periodic dipole array antenna (LDPA, $G=\SI{4}{\decibeli}$) located at the vertex of the \SI{12}{\metre} dish, which points at the PAF and is oriented so that its linear polarisation is at \ang{45} to that of the PAF elements.  The illumination of the PAF is close to uniform but there is a path length difference of up to \SI{21}{\milli\metre} from the central elements to the edge. To a first approximation the signal injected into all PAF elements is the same.
 
The Cicada also generates an optical copy of the noise signal.  Half the power is sent directly to the corresponding antenna's digital receiver and half is sent to the antenna on the same multi-core fibre cable as the PAF signals. This copy is reflected by a Faraday mirror in the antenna pedestal and returns to the Cicada, then continues into the digital receiver.  In the digital receiver both optical signals are processed in the same way as the PAF signals, becoming inputs to the array covariance matrix and a higher-sensitivity calibration correlator that forms all correlation products between a nominated reference signal and the PAF signals. 

The direct signal from the Cicada gives the best estimates of gain and phase variation.  The signal returned from the Faraday mirror allows the optical path to the LPDA to be calibrated via a round-trip phase measurement.  This allows the phase variation of the full analogue optical path to be characterised.  The system also provides a way to inject a common calibration signal into the PAFs on all dishes, though this is not routinely used.

Although the main purpose of the ODC system is to measure complex gain and provide updates to the beamformer weights, it can also be used to improve system diagnostics in areas of RF continuity, RF distortion assessment and filter-bank frequency integrity. The main contribution of the ODC at present is to stop beams from being completely destroyed by phase slopes corresponding to unpredictable delay changes that occur with each full reset of the analog-to-digital converters (ADCs). With further research it is hoped to use the ODC to temporally stabilise PAF beam (voltage) patterns to better than \SI{1}{\percent} at the half power points \citep{Hayman2010}.

\section{ANALOGUE SIGNAL PROCESSING}
\label{sec:domino}

\subsection{Design background}

In the BETA version of the ASKAP PAF \citep{Schinckel2011, hotan_2014} all analogue signal processing and digitisation occurred in the antenna pedestal. At the time of the BETA design, photonics for radio frequency over fibre (RFoF) signal transport appeared expensive and sufficiently unproven to deploy on the scale of a \num{6768} element array for radio astronomy. 

The BETA system comprised two shielded cabinets packed full of equipment and occupying most of the space in each antenna pedestal, along with a chiller plant located on the concrete antenna foundation pad to keep both the cabinets and PAF cooled. The chillers required continuous maintenance and the pedestal electronics were not easily accessible due to the confined space. Containment of radiated emissions was less than satisfactory due to the single layer of shielding and the large number of filtered conductive connections to the outside world. Even so, BETA served well to verify basic PAF performance and signal processing algorithms and was a pivotal step to the more advanced Mk~II design. 

For the Mk~II version of ASKAP described here, the price of high linearity uncooled distributed feedback (DFB) laser diodes had fallen. Investigations of performance of low-cost RFoF analogue signal transport using these DFB components were undertaken with fibre spans up to \SI{6}{\kilo\metre}. It was found that, with suitable precautions, a directly modulated DFB laser solution could meet astronomy requirements \citep{Beresford2017_low_cost}. The cost of the optical transmitter was reduced to one hundred dollars rather than several hundred dollars for cooled laser devices or thousands of dollars for high quality externally modulated lasers traditionally used in broadband applications.  

This allowed all analogue electronics, except the PAF itself, to be moved to the central building, which eliminated self RFI problems, lowered cost and improved accessibility and maintainability of the analogue signal processing hardware.


\begin{figure*}[t]
\begin{center}
\includegraphics[width=\textwidth]{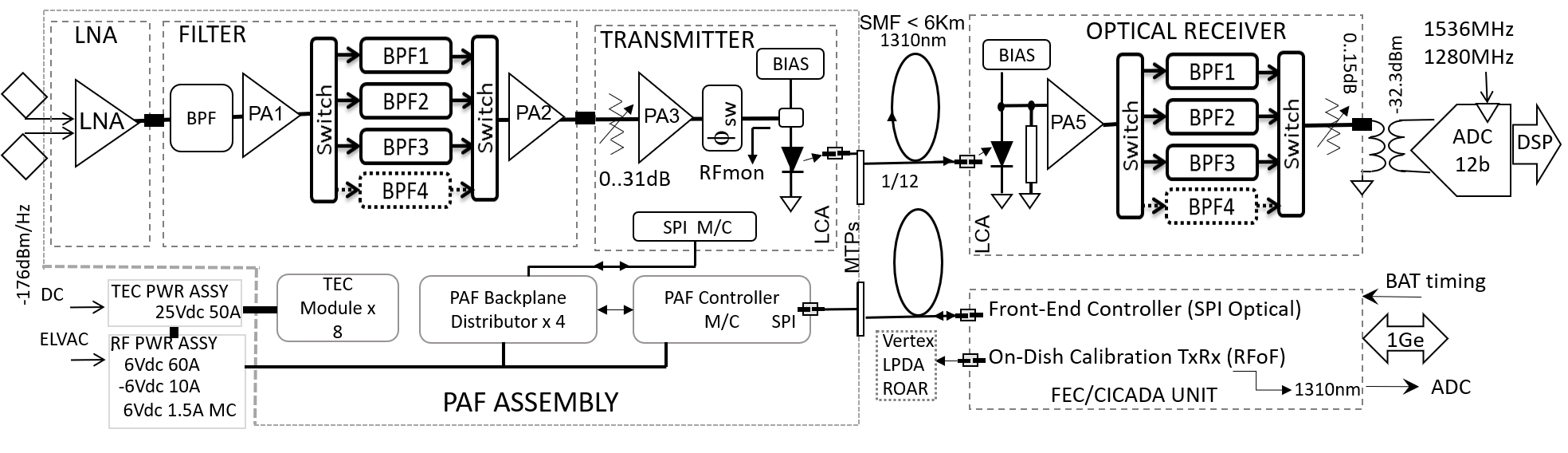}
\caption{The analogue signal path for a single PAF element from the feed to the ADC in the digital receiver. After the chequerboard feed elements, the signal encounters a low noise amplifier (LNA), followed by switchable bandpass filters (BPF) and power amplifiers (PA). The transmitter module provides an option to introduce phase switching ($\phi_{\textrm{SW}}$), after which the signal is used to modulate a laser diode for transmission over single mode optical fibre (SMF) from the antennas to the control building. After analogue to digital conversion (ADC), the digital signal processing (DSP) stage begins. PAF monitoring and control (M$/$C) is distributed to all elements via a system based on the serial peripheral interface (SPI) standard. Cooling inside the PAF is provided by a series of 8 thermoelectric cooler (TEC) modules. Direct current (DC) and extremely low voltage alternating current (ELVAC) powers the various subsystems inside the PAF. The on-dish calibration system consists of a log-periodic dipole antenna (LPDA) mounted at the dish vertex and connected to an optical receive module (ROAR) fed by a noise source (CICADA) located in the front end control (FEC) module. This can be driven synchronously using binary atomic time (BAT) events loaded via a gigabit ethernet (1Ge) interface.}
\label{fig:analogue_path}
\end{center}
\end{figure*}

\subsection{LNA}

Each domino houses the low-noise amplifiers (LNAs) for a pair of PAF elements (Figure~\ref{fig:DominoLNA}) and there are 94 dominoes in each PAF. ASKAP's LNA design forms an active balun by connecting the twin-wire feed line of one connected-array element on the PAF surface directly into two single-ended LNAs with a passive output-side balun \citep{Shaw2012}.  A passive input-side balun would have introduced too much loss.  The differential input complicates the measurement of LNA noise temperature, so an innovative method was developed by \citet{Shaw2012}. The measured LNA minimum noise temperature $T_\text{min}$ of the current design is \SI{30}{\kelvin} in the middle of the frequency range \citep{Shaw2015}.

\subsection{FILTER}

After the LNA, the signal encounters selectable anti-aliasing filters \citep{Beresford2017} as shown in Figure~\ref{fig:DominoFIL}. These filters define four observing bands summarised in Table~\ref{tab:drxbands} and Figure~\ref{fig:sampbands}.  The fourth band was included for assessment of the low-frequency performance of the PAF and has not been used for astronomy.  

\begin{table}[b]
\caption{ASKAP receiver bands.}\label{tab:drxbands}
\sisetup{parse-numbers = true,input-decimal-markers={-},output-decimal-marker = \text{--}}
\centering
\begin{tabular}{c S c c}
\hline
  {Band} & {Sampled} & {Sample clock} & {\SI{1}{\mega\hertz}}  \\ 
    {}   & {RF band} & {frequency}    &  {channels}\\
    {}  &  {(MHz)}   & {(MHz)}        &  {}\\ 
\hline
  1 & 700-1200 & 1280  & 640 \\
  2 & 840-1440 & 1536  & 768 \\
  3 & 1400-1800 & 1280 &  640 \\
  4 & 600-700  & 1536  & 768 \\
\hline
\end{tabular}
\end{table}

\subsection{TRANSMITTER}

After filtering, the signal is converted from electrical to optical for transmission back to the central building. The optical RFoF transmitter \citep{RFoF, Beresford2017, Hampson2012} is shown in Figure~\ref{fig:DominoOPT}. The RF signal is directly added to the laser diode DC bias current and modulates the optical power. An average power control loop ensures constant optical level, compensating for component ageing. The low loss in the single-mode fibre allows the signal to be transported several kilometres. Each antenna connects to the central processing building via underground fibre-optic cables containing 216 cores per antenna, 188 of which carry RFoF PAF signals. Three additional fibres are used for the RFoF ODC system described in Section~\ref{sec:odc}.

\subsection{Optical link design considerations}

\begin{figure}
\begin{center}
\includegraphics[width=\columnwidth]{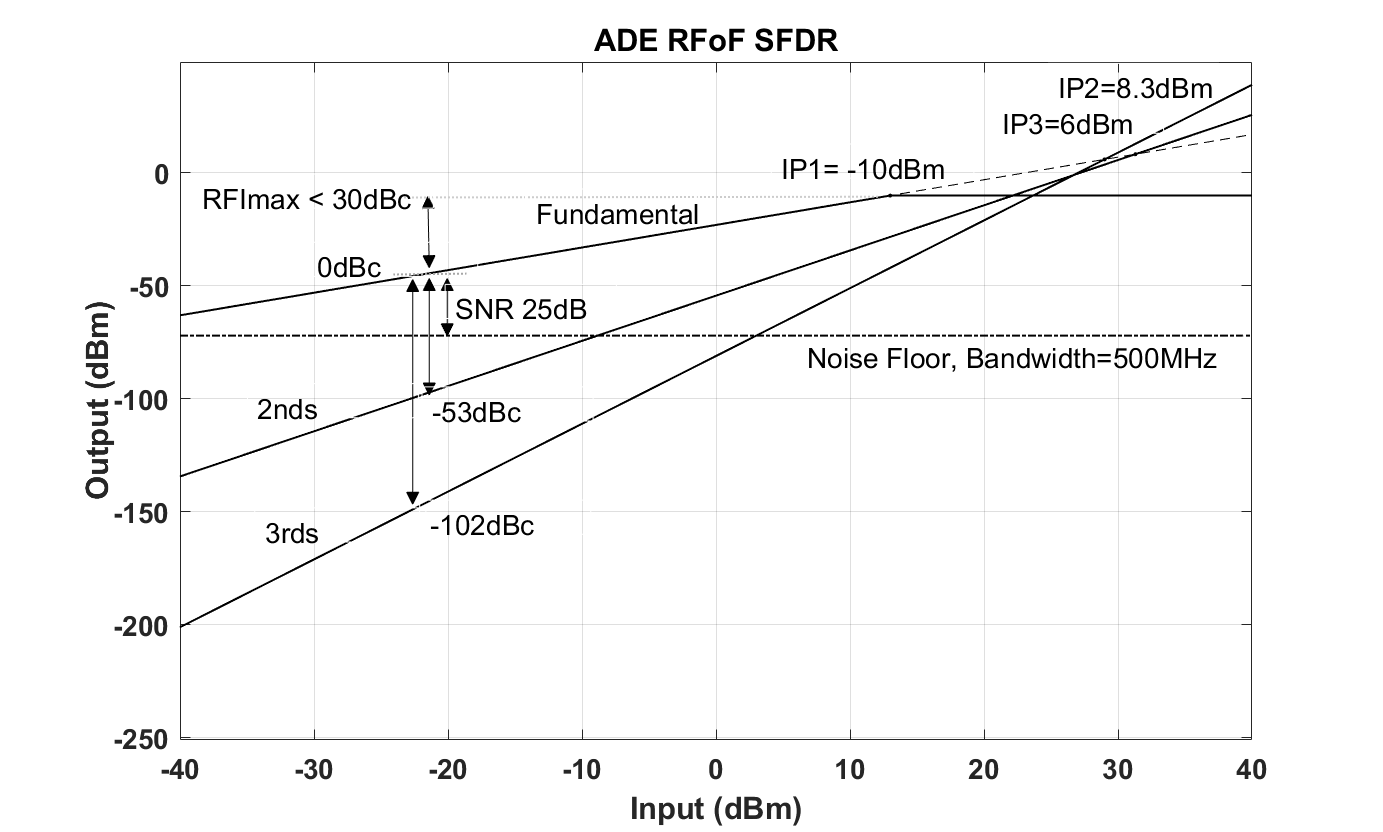}
\caption{Distortion performance, measured as the spurious-free dynamic range (SFDR), of the RF over fibre link.}
\label{fig:dynamic_range}
\end{center}
\end{figure}

All observing bands are suboctave as they are in the second or third Nyquist band of the analogue-to-digital converter (ADC). This mitigates the second-order harmonic distortion of the RFoF link. Radio astronomy signal paths are designed to respond linearly to their input and this was an important consideration for the optical link as well. For systems operating close to linearity, small departures are often characterised using a measure known as the input or output intercept point (IIP or OIP) for various polynomial orders. The ASKAP RFoF link has an OIP1 of \SI{-10}{\decibelm}, an OIP2 of \SI{8.3}{\decibelm}, and an OIP3 of \SI{6}{\decibelm} \citep{Beresford2017_low_cost} as shown in Figure~\ref{fig:dynamic_range}. This marginally degrades as the fibre span increases \citep{RFoF}. The nominal RFI-free RFoF link output signal level is \SI{-47}{\decibelm} (\SI{0}{\decibelc}), relative intensity noise (RIN) and shot noise define the output noise floor at \SI{-72}{\decibelm} (\SI{-25}{\decibelc}).

For narrow-band RFI equal in power to the system noise at the RFoF output (\SI{0}{\decibelc}) the second-order distortion is \SI{-53}{\decibelc} and the third is \SI{-102}{\decibelc}. The strongest narrow-band RFI is due to aviation Automatic Dependent Surveillance-Broadcast (ADS-B) transmitters, located on aircraft flying over the observatory at high altitude. This can be as high as \SI{29}{\decibelc} for short periods of time (minutes) if the aircraft passes overhead.

Placement of half the anti-aliasing filter after the RFoF link eliminates the severe in-band second-order response from the RFoF link as well as out of band relative intensity noise and shot noise \citep{RFoF}.  This allows the system to handle an extra \SI{17}{\decibel} of RFI (third-order intermodulation at \SI{-53}{\decibelc}, see Figure~\ref{fig:dynamic_range}). The other half of the anti-aliasing filter preceding the RFoF link removes out of band RFI that could corrupt the link. Adjustment of the attenuator in the domino transmitter (see Figure~\ref{fig:analogue_path}) can reduce the signal level on the RFoF link to increase the compression headroom, but this comes at the cost of increased additive noise. With nominal settings, the RFoF link adds \SI{1.5}{\kelvin} to the receiver noise temperature. Decreasing the power on the RFoF link by \SI{3}{\decibel} gives \SI{3.7}{\decibel} extra headroom for RFI but adds \SI{1.9}{\kelvin} to the receiver noise temperature for the most distant antennas.

\subsection{Optical receiver}

The conversion from optical back to electrical signals occurs in the digital receiver (see Section~\ref{sec:DigitalRX}). Further amplification increases the electrical signals to the level needed to drive the analog-to-digital converter (ADC) in the digital receiver (\SI{-35}{\decibelm}). The ADC has a specified spurious-free dynamic range (SFDR) of approximately \SI{61}{\decibel}. The RF signal chain has an SFDR of approximately \SI{48}{\decibel}. Marginal degradation of SFDR occurs for longer fibre spans due to the optical loss of \SI{0.35}{\decibel\per\kilo\metre}. The longest fibre span in the array is \SI{6}{\kilo\metre}.

\section{DIGITAL SIGNAL PROCESSING}
\label{sec:DSP}

The next stage in the ASKAP signal chain consists of a distributed, custom digital signal processing (DSP) system built on field-programmable gate array (FPGA) technology \citep{Hampson2012}. This system is responsible for the following key processing steps:

\begin{enumerate}
  \item sampling the PAF analogue RF signals;
  \item channelising the band with \SI{1}{\mega\hertz} coarse resolution;
  \item selecting channels for further processing;
  \item beamforming on the selected coarse channels;
  \item applying time-varying coarse delays to align wavefronts with an error of at most half a sample at \SI{1}{\mega\hertz}; 
  \item further channelising the beamformed output to the final frequency resolution of the correlator;
  \item applying a time-varying phase slope across \SI{1}{\mega\hertz} channels to provide fine delay control (fraction of coarse delay step); and
  \item for each beam, cross-correlating the beam voltages.
 \end{enumerate}
 
Processing is implemented in three stages, first the digital receiver, then the beamformer and finally the correlator.  The result is the cross-correlation of beam voltages for each fine channel across all antenna pairs and polarisation products (the raw visibilities) for the specified phase centre.  Each beam is correlated with itself (autocorrelation) and with each corresponding beam that points in the same direction from every other ASKAP antenna.  Correlations between beams pointing in different directions are not calculated.

\subsection{Digital receiver}
\label{sec:DigitalRX}

\begin{figure}
\begin{center}
\includegraphics[width=\columnwidth]{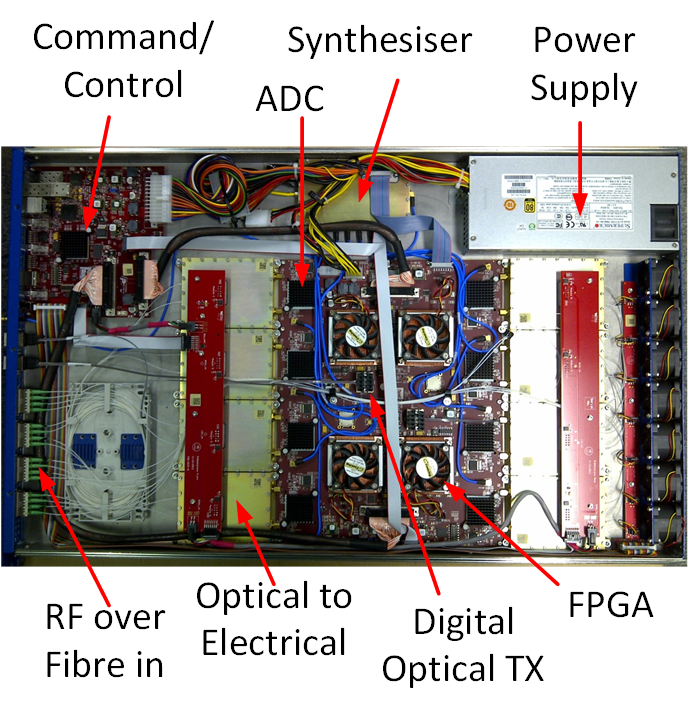}
\caption{The Dragonfly digital receiver module, designed by CSIRO specifically for ASKAP. There are 12 per antenna.}
\label{fig:dragonfly}
\end{center}
\end{figure}

ASKAP's digital receiver \citep{Dragonfly} is implemented with 12 Dragonfly modules (see Figure~\ref{fig:dragonfly}) for each antenna, each processing 16 optical signals. The 192 signals include 188 from each PAF, two calibration signals, and two spares for future applications such as RFI mitigation.

The ADC devices used for ASKAP are National Semiconductor ADC12D1600 parts that have 12-bit resolution with an effective number of bits (ENOB) of 9 and a full-power analogue bandwidth of \SI{2.8}{\giga\hertz}. The sample clock is generated in the Dragonfly synthesiser module as described in section~\ref{sec:timing}.

The ASKAP digital receiver has a bandpass sampling architecture.  It directly samples the PAF signals at RF (Figure~\ref{fig:analogue_path}) without the need for analogue frequency conversion.  The full \SIrange{700}{1800}{\mega\hertz} frequency range of the PAF is covered by three selectable and overlapping sampling bands (see Table~\ref{tab:drxbands} in Section~\ref{sec:domino}).  

\begin{figure*}
\begin{center}
\includegraphics[width=\textwidth]{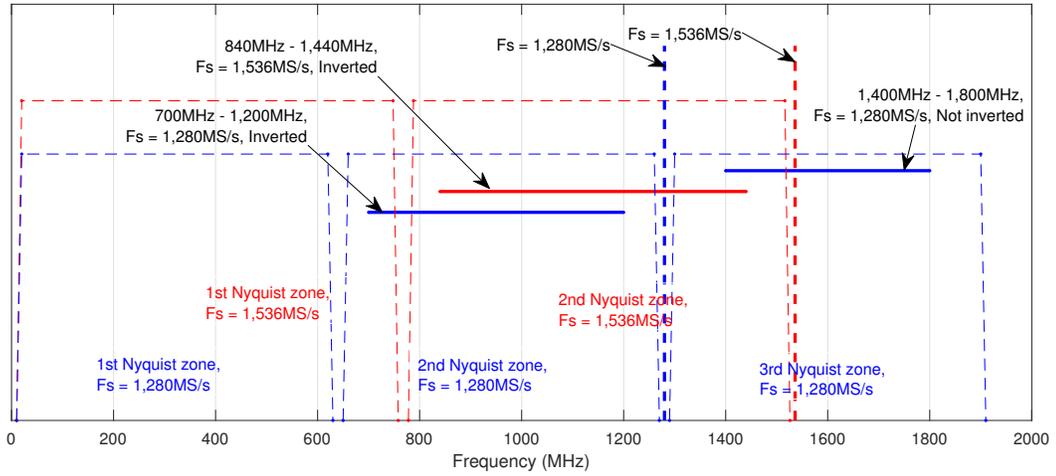}
\caption{ASKAP digital system sampling bands. For convenience, we refer to these as bands 1, 2 and 3 in left-to-right order elsewhere in this document. The labels indicate whether the frequency channel order is inverted or not, with respect to the natural ascending order.}
\label{fig:sampbands}
\end{center}
\end{figure*}

Each sampling band is designed to match a Nyquist zone of the ADC. Two sampling frequencies, \SI{1280}{\mega\hertz} and \SI{1536}{\mega\hertz}, are used to sample at RF with an instantaneous bandwidth of \SI{640}{\mega\hertz} or \SI{768}{\mega\hertz}. Figure~\ref{fig:sampbands} illustrates how the sampling bands correspond to sampling frequency, how they overlap, and how much usable RF band is left after filtering.  The lower and upper observing bands are located in the second and third Nyquist zones respectively of the ADC running at \SI{1280}{\mega\hertz}, while band 2 is located in the second Nyquist zone of the ADC running at \SI{1536}{\mega\hertz} which overlaps the adjacent bands at the other sampling frequency. The sampling frequency and corresponding FPGA firmware are configured at the start of an observation.

\begin{figure*}
\begin{center}
\includegraphics[width=\textwidth]{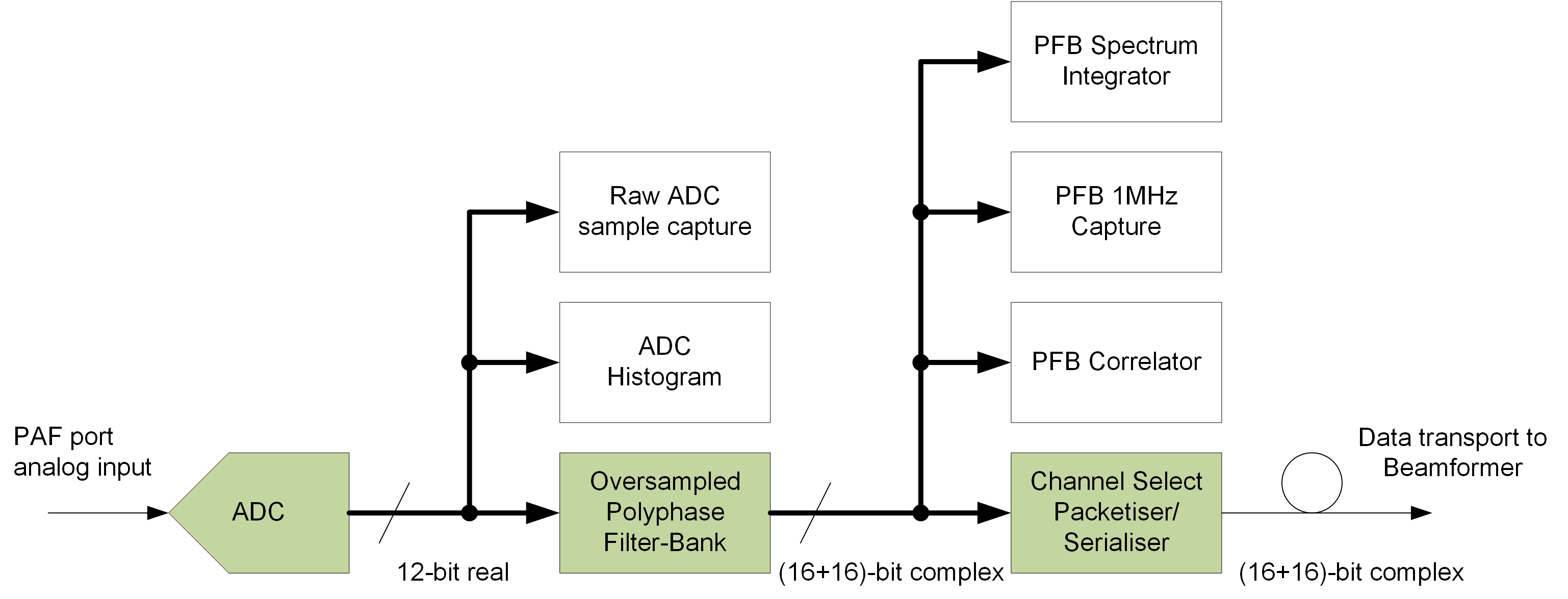}
\caption{The signal path for a single PAF element through the ASKAP digital receiver ADC and PFB.}
\label{fig:askap_drx}
\end{center}
\end{figure*}

An overview of the signal path following the ADC, for a single PAF element, is shown in Figure~\ref{fig:askap_drx}. Also shown, in unshaded boxes, are a number of other ancillary signal statistics and monitoring modules. In operation, the ADC histogram is particularly useful as it shows whether the signal is Gaussian and helps to set the signal level into the ADC.

The sampled PAF element voltages feed directly into a polyphase filter bank (PFB) with an oversampling ratio of 32/27 \citep{Tuthill2012}. The PFB forms channels with a sample rate of \SI{1.185}{\mega\hertz} and \SI{1}{\mega\hertz} spacing. As the channels have a noise bandwidth of \SI{1}{\mega\hertz} and a channel spacing of \SI{1}{\mega\hertz} they will be referred to as \SI{1}{\mega\hertz} channels. The oversampling provides substantial gains in sub-band fidelity and downstream processing capabilities with only a modest increase in system complexity \citep{2003_Bunton_ALMAmemo447,Tuthill2012}. The channel-dependent frequency rotation within each output channel that is introduced by oversampling is compensated for within the PFB implementation so that the output 1\,MHz channels are all correctly DC-centred before being forwarded on to the beamformer subsystem \citep{Tuthill2015}.

The PFB produces either 768 or 640 \SI{1}{\mega\hertz} channels across the selected observing band (Figure~\ref{fig:sampbands}), depending on the sampling frequency. In both cases the channel sample rate is the same and the channel centre frequency is an integer multiple of \SI{1}{\mega\hertz}. The output of the PFB for each channel is the complex-valued analytic envelope of the full-band signal component within that sub-band, i.e. the single-sided spectrum in that sub-band, translated to DC \citep{1982_Sig_Env_Rice}. The equivalent operations to generate the output of the PFB for each channel are a multiplication of the input by $e^{2 \pi ift}$, where $f$ is the channel centre frequency, followed by low pass filtering and decimation to the output sample rate.

Of the total number of output channels from the PFB, only 384 channels are selected for transport to the beamformer subsystem for further processing. These are transported as eight data streams each with 48 channels. However, cost constraints and computing limitations mean that only seven of these data streams are beamformed and only six are correlated.  This limits the total bandwidth available for imaging (see Section~\ref{sec:SDP}) to \SI{288}{\mega\hertz}.

The channel selection process is very flexible and permits contiguous or non-contiguous groups of channels to be selected and also multiple copies of the same channels to be selected and transported to different downstream end-points. The complex task of managing channel selection is done by the Telescope Operating System (TOS) software. At present, only contiguous bands are implemented to simplify visibility data storage and imaging. In future, we plan to support two or more split bands for purposes such as avoiding satellite interference and observing widely separated spectral lines.

Bit growth is permitted through the PFB to ensure that the DSP system meets dynamic range and noise floor requirements, so the output number representation for the coarse \SI{1}{\mega\hertz} channels is 16-bit real and 16-bit imaginary two's complement (signed) integers. The resulting aggregate data rate at the output of the ASKAP digital receiver and PFB is approximately \SI{100}{\tera\bit\per\second}.  This is efficiently transported to the beamformer using a custom point-to-point packetised streaming protocol that has very low overheads. The physical transport medium consists of \num{10368} multi-mode optical fibres in 864 12-fibre ribbons.  All fibres in a ribbon route to different locations and the fibre reordering is achieved by using custom passive optical cross-connects.  

\subsection{Beamformer}
\label{subsec:bmf}

\begin{figure}
    \centering
    \includegraphics[width=\columnwidth]{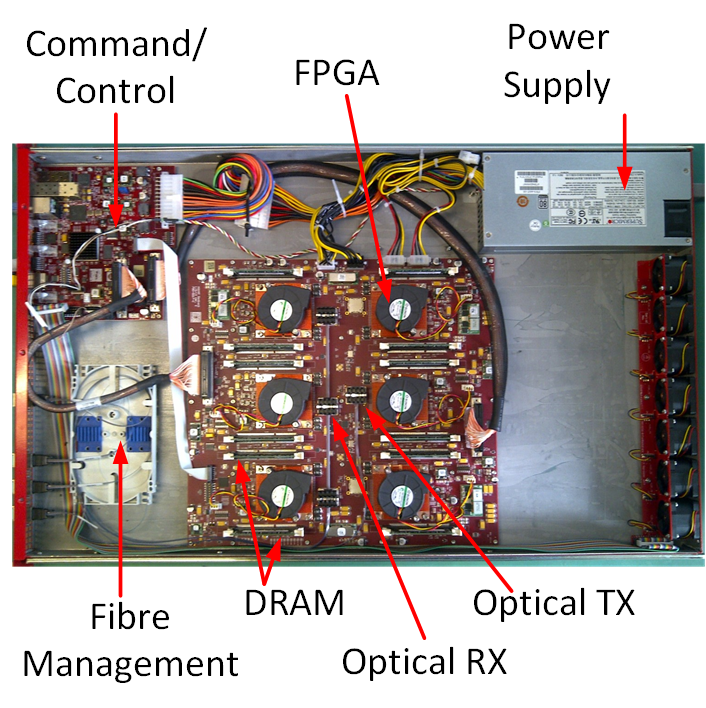}
    \caption{Redback module used for beamformer and correlator. Field programmable gate arrays (FPGAs) used for digital signal processing are hidden beneath cooling fans. Banks of dynamic random access memory (DRAM) are visible beside each FPGA.}
    \label{fig:redback}
\end{figure}

Output from the digital receiver for a single antenna is processed by seven Redback modules \citep{Hampson_2014_ASKAP} shown in Figure~\ref{fig:redback}. Each receives \SI{48}{\mega\hertz}  of bandwidth from all PAF elements for a single antenna and can generate 36 independent dual-polarisation beams. With seven modules, 336\,MHz out of \SI{384}{\mega\hertz} available from the digital receiver is processed. Wiring for the eighth module is in place but cost constraints meant it was not installed. Each Redback has six processing FPGAs, which are fully interconnected electrically.  The interconnect is used to distribute data so that each FPGA has \SI{8}{\mega\hertz} of data from all 192 digitised ports (188 PAF elements + two calibration signals + two spare ports).

\begin{figure*}
\begin{center}
\includegraphics[width=\textwidth]{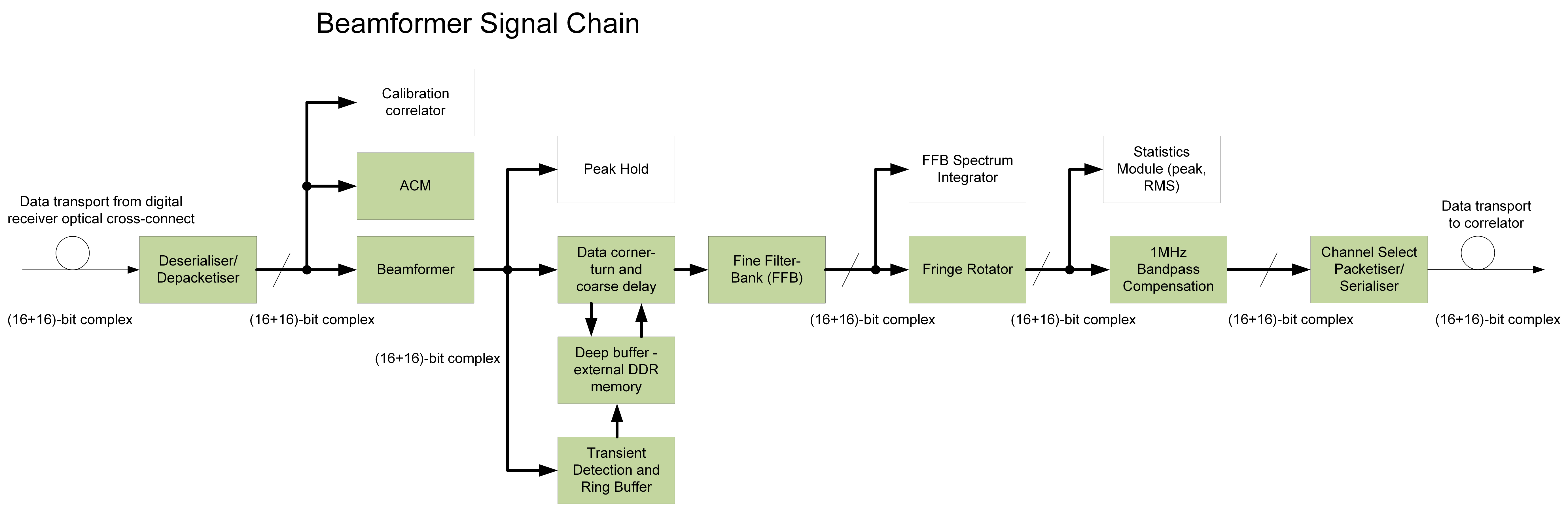}
\caption{Major components of the signal path through the ASKAP Beamformer}
\label{fig:askap_bmf}
\end{center}
\end{figure*}

The signal path through the beamformer is shown in  Figure~\ref{fig:askap_bmf}. The beamformer engine produces up to 72 independent beams on the sky, each a linear combination of $M$ digitised port voltages ${\mathbf{x}=\left[ {{x}_{1}},{{x}_{2}},...,{{x}_{M}} \right]^{T}}$.  The beamforming operation is defined by
\begin{equation}
y(n) = \mathbf{w}^H\mathbf{x}(n)
\end{equation}
where $\mathbf{w}^H$ is the conjugate transpose of the beamformer weights and $\mathbf{x}(n)$ is the vector of port voltages for a single frequency channel at time sample $n$.  The beamformer weights are complex valued and may be varied independently for each \SI{1}{\mega\hertz} channel of each beam. The weights are represented as 14-bit real and 14-bit imaginary two's complement (signed) integers. 

The number of ports $M$ that are weighted into each beam can be as many as all 192 ports from the digital receiver for one ASKAP antenna, but it reduces to 60 ports as the number of beams increases to the full complement of 72 single polarisation beams. The reduction of input ports with beams is summarised in Table~\ref{tab:ports_per_beam} and is required due to hardware resource limitations in the beamformer.  The selection of $M$ ports from 192 can be specified arbitrarily for each \SI{1}{\mega\hertz} channel of each beam.

\begin{table}
  \caption{Number of ports per polarisation per beam for various numbers of dual-polarisation beams.}
  \label{tab:ports_per_beam}

  \begin{center}
    \begin{tabular}{cc}
    \hline
      \textbf{Dual-polarisation beams} & \textbf{Ports per beam $M$}\\
      \hline%
      \numrange{1}{9} & 192\\
      \numrange{10}{18} & 126\\
      \numrange{19}{27} & 82\\
      \numrange{28}{36} & 60\\
      \hline
    \end{tabular}
  \end{center}
\end{table}

The system is typically configured to produce 36 dual-polarised beams.  The beamformed voltages for each \SI{1}{\mega\hertz} channel pass into a second polyphase filter bank, which performs a final fine frequency channelisation. For normal operation the frequency resolution is \SI{18.5}{\kilo\hertz}.

A key advantage of beamforming on 1\,MHz channels is that the beamformer can be most efficiently implemented mid-stage as 336 separate narrow-band beamformers \citep{1988_Beamforming_VanVeen}. This allows beamforming with complex weights as opposed to frequency dependent weighting with FIR filters, which would be needed if the beamforming occurred on the ADC voltages. Furthermore, since the fine filter bank (FFB) is operating on beam voltages rather than raw PAF element voltages, there is a reduction in the FFB data throughput by a factor of approximately three (188 PAF-elements/72-beams).  This reduction also means that although three 12-fibre ribbons are needed to input data to a Redback module only one is needed at the output. That is, two fibres per FPGA each carrying \SI{4}{\mega\hertz} of beamformed data.

A key data product for many beamforming algorithms is the array covariance matrix (ACM), defined as
\begin{eqnarray}
\label{acm}
	\mathbf{R}=\left< \left( \mathbf{x}-{{\mu }_{\mathbf{x}}} \right){{\left( \mathbf{x}-{{\mu }_{\mathbf{x}}} \right)}^{H}} \right>
\end{eqnarray}
where ${\mathbf{x}=\left[ {{x}_{1}},{{x}_{2}},...,{{x}_{K}} \right]^{T}}$ is a vector of $K$ array element voltages assumed to be stationary discrete-time stochastic processes, ${{\mu }_{\mathbf{x}}}$ is the element-wise mean value of $\mathbf{x}$, and $\left<\cdot\right>$ denotes expectation. For the ASKAP signal model the $x_i$ are the frequency-channelised PAF element voltages, which are complex-valued random time-series whose statistics are assumed to be stationary with zero mean over the beam calibration interval. Under these conditions, the elements of $\mathbf{R}$ in \eqref{acm} are adequately approximated for each \SI{1}{\mega\hertz} channel by the time-average
\begin{eqnarray}
\label{acm_est}
	{{\hat{{r}}}_{i,j}} = \frac{1}{N}\sum\limits_{n=1}^{N}{{{{x}}_{i}}\left( n \right){x}_{j}^{*}\left( n \right)}
\end{eqnarray}
where ${x}_{i}\left( n \right)$ is the $n\text{th}$ voltage sample of the $i\text{th}$ input port of $K=192$ and $x^*$ is the complex conjugate of $x$.  With this definition, the ACM estimate $\hat{\mathbf{R}}$ is a ${192\times 192}$ matrix where each element $\hat{r}_{i,j}$ is the correlation between $x_i$ and $x_j$.  This is computed for all \SI{1}{\mega\hertz} frequency channels.  

The full computation of the ACM has a higher compute load than the final correlator for each antenna. This load is reduced by using every fourth time sample and computing ACMs for ${42 \times \SI{1}{\mega\hertz}}$ frequency channels concurrently for each PAF.  The ACM estimation process must be run eight times to cover the entire $8\times\SI{42}{\mega\hertz}=\SI{336}{\mega\hertz}$ of processed bandwidth. Raw ACM data products are transported to conventional compute hardware where they are used to calculate the beamformer weights offline as described in Section~\ref{sec:beamforming}. The beamformer weights are then uploaded back to the beamformer block, which is shown below the ACM block in Figure~\ref{fig:askap_bmf}. 

Depicted above the ACM in Figure~\ref{fig:askap_bmf} is the calibration correlator. This calculates the correlation between a reference calibration signal and all PAF element outputs. This is mathematically equivalent to computing one row of the ACM, but in the calibration correlator it is done for a specific reference input without time or frequency decimation. This achieves the best possible SNR for tracking PAF element gains. The calibration system is described in more detail in Section~\ref{sec:odc}.

The beam voltages output from the beamformer block are fed into two separate processing modules:

\begin{enumerate}
  \item a deep ``corner-turn'' buffer; and
  \item a transient ring buffer.
\end{enumerate}

Both of these modules share a large double data rate (DDR) memory resource for temporarily buffering large amounts of beam voltage data. The transient module implements a ring buffer that can be frozen by an external trigger and is described in Section~\ref{sec:fast_transient}.

The primary signal flow for ASKAP, however, is the corner-turn buffer that implements a streaming data transpose, supplying large blocks of sample data for the same beam and frequency channel to the FFB, which is operated in block-mode for efficiency. The FFB is a second-stage frequency channeliser implemented as a critically-sampled polyphase filter bank on the \SI{1}{\mega\hertz} channels which have a sample rate of \SI{1.185}{\mega\hertz}. This module can be configured at the start of an observation to provide a standard observing frequency resolution, or one of five  ``zoom'' resolutions, according to Table~\ref{tab:askap_zooms}.

\begin{table}
  \caption{ASKAP frequency zoom modes. In all cases the number of frequency channels correlated is \num{15552} for the current hardware deployment.} 
  \label{tab:askap_zooms}
  
  \begin{center}
    \begin{tabular}{c S[table-format=4] S[table-format=2.4] S[table-format=3]}
    \hline
      Mode & {FFB} & {Channel} & {Correlated}\\      
      number & {output} & {width} & {Bandwidth}\\
             & {channels} & {(kHz)}& {(MHz)}\\
      \hline%
      1 & 64 & 18.5185 & 288\\
      2 & 128 & 9.2593 & 144\\
      3 & 256 & 4.6296 & 72\\
      4 & 512 & 2.3148 & 36\\
      5 & 1024 & 1.1574 & 18\\
      6 & 2048 & 0.5787 & 9\\
      \hline
    \end{tabular}
  \end{center}
\end{table}

Note that since each of the 42 FPGAs in a single PAF beamformer sub-system processes the $8 \times \SI{1}{\mega\hertz}$ channels that are routed to it from the digital receiver, different zoom modes can be applied to different or even the same parts of the total processed bandwidth by loading the appropriate zoom mode firmware on each FPGA and routing the corresponding 1\,MHz channels to it. Control software to implement this is not available currently. Each FPGA always generates 423 frequency channels independent of zoom mode.  Thus the total bandwidth decreases as frequency resolution improves as shown in Table~\ref{tab:askap_zooms}. Processing of a full 1\,MHz channel, for zoom modes, requires multiple correlator FPGAs to process the same 1\,MHz channel.

\subsection{Fast transient detection}
\label{sec:fast_transient}
As shown in Figure~\ref{fig:askap_bmf}, a secondary beamformer output is sent to the fast-transient detector. This operates with the 1\,MHz channels and has an adjustable time resolution that is typically set to \SI{864}{\micro\second} or \SI{1.7}{\milli\second}. Power averages are sent over the MRO network to a dedicated computer that searches for dispersed pulses in real time \citep{2019ascl.soft06003B}. In the search algorithm these data are currently added incoherently across all antennas. 

If a pulse is detected, a trigger is sent to the transient ring buffer in the beamformer, telling it to freeze the contents of the ring buffer (which stores an adjustable amount of raw voltage data from all beams, corresponding to the last \SIrange{0.9}{14.2}{\second} depending on data precision selected) and download to disk. The voltage data are then correlated offline using the DiFX software correlator \citep{2011PASP..123..275D} and imaged in order to localise the position of the event with sub-arcsecond resolution and provide high sensitivity data on the pulse \citep{2019Sci...365..565B}.

\subsection{Delay tracking and fringe stopping}

Distant astronomical sources remain effectively stationary in the sky, while the Earth (and therefore the telescope) rotates once per day. This rotation means that the projected baseline vector between any pair of antennas in the array is constantly changing. Interference fringes, therefore, move during an observation, which would decorrelate the signal from the interferometer if not compensated for. The process of adjusting the delay and phase of each antenna to keep interference fringes stationary is known as delay tracking and fringe stopping.

Since ASKAP is a multi-beam instrument, we independently stop fringes in the nominal pointing direction of each beam. This is straightforward when using the antenna polarisation axis to keep beams pointing at a fixed position on the sky. We also offer the option to stop fringes in the boresight direction of the antenna for all beams, which can be useful for experiments such as holography where polarisation axis tracking is not used.

The ASKAP digital hardware implements:
\begin{enumerate}
  \item a pure time delay to a precision of \SI{0.84}{\micro\second} applied across all beams and 1\,MHz coarse frequency channels; and
  \item a time-dependent and frequency-dependent phase correction applied to each fine frequency channel.
\end{enumerate}

A pure ``coarse'' time delay (with precision of \SI{0.84}{\micro\second}) is implemented between the beamformer and the FFB by offsetting the read pointer into the corner-turn sample buffer for the FFB frame input. The coarse delay is the same for all beams and 1\,MHz coarse frequency channels for a single PAF system, but is typically different between antennas in the array to allow for a coarse compensation of the difference in arrival time (up to \SI{20}{\micro\second}) of a wavefront at different antennas. During an observation, the required coarse delay is computed repeatedly and set in the hardware at times synchronised with signals from the timing and synchronisation subsystem (see Section~\ref{sec:timing}). Changes occur at the start of each data frame and do not introduce any discontinuities.

The second ``fine'' stage of fringe stopping occurs in the \textit{fringe rotator} (Figure~\ref{fig:askap_bmf}), which uses complex multipliers to rotate the phase of each fine frequency channel time-series as they come out of the FFB. The rotation phase is calculated from coefficients loaded via the control system software and permits a unique phase correction to be applied at each of the four levels:

\begin{enumerate}
  \item coarse 1\,MHz channels;
  \item dual-polarised beams;
  \item fine frequency channels; and
  \item particular time samples.
\end{enumerate}

The fringe rotator module exploits the fact that the rotation phase, for any beam, is linearly dependent on frequency. Furthermore, the module parameters are updated often enough that the time-variation of phase is adequately approximated with a piecewise-linear function. Under these conditions, the rotator can calculate the required phase, for a beam, from a unique starting phase for each coarse frequency channel and two increment parameters. These parameters are the change in phase with time increment and change in phase with frequency increment.

For each coarse channel and beam combination, the phase correction applied for an individual time sample $\Delta t$ units after the update epoch and $\Delta f$ fine frequency channels from a reference channel, is given by
\begin{eqnarray}
\label{fr_update}
   {{{\phi }\left(\Delta t,\Delta f\right)}={{\phi }_{O}}+{{\phi }_{T}}\Delta t+{{\phi }_{F}}\Delta f}
\end{eqnarray}
where ${{\phi }_{O}}$ is an initial phase offset value and ${{\phi }_{T}}$  and ${{\phi }_{F}}$ are the time slope and frequency slope values respectively.

The array-wide synchronous event that updates the coarse time delay has an appropriate processing delay applied before being used to update new fringe rotator parameters ${{\phi }_{O}}$, ${{\phi }_{T}}$  and ${{\phi }_{F}}$ so that the coarse delay and fringe parameters are consistent for each update interval. 

The smallest incremental change that can be made to the fringe stopping parameters at the fine control stage (corresponding to 18.5\,kHz channel bandwidth) is \SI{0.206}{\nano\second} in delay, \SI{0.044}{\deg} in phase and \SI{0.0248}{\deg\per\second} in phase rate. The fine-level delay slope across the band is implemented as a combination of delay steps and matching phase offsets with the increments given above. 

At the output of the FFB, the beamformer subsystem produces ${{2}^{M-1}}\times 64$ fine frequency channels for each 1\,MHz channel of each PAF beam, where $M$ is the zoom mode number in Table~\ref{tab:askap_zooms} and $336 \times \SI{1}{\mega\hertz}$ channels are processed in the beamformer. As a result of the oversampling ratio of ${32}/{27}$ in the coarse filter bank, each 1\,MHz of processed bandwidth is contained within the centre ${{2}^{M-1}}\times 64\times {27}/{32={{2}^{M-1}}\times 54}\;$ fine channels. The redundant ${{2}^{M-1}}\times (64-54)=10\times {{2}^{M-1}}$ fine channels at the band edges of each 1\,MHz sub-band are discarded before further processing. The number of channels processed is independent of zoom mode.

\subsection{Correlator}

The correlator subsystem is the final stage in the FPGA-based signal processing chain at the MRO central site. It takes in fine frequency channelised beam voltage time-domain data, streaming from the 36 antenna beamformer subsystems, and generates the raw visibilities (cross-correlations) for the final science data processing stage. The data are carried from the beamformer to the correlator via multimode optical fibre and are transported using a custom, unidirectional streaming protocol for maximum link efficiency. 

The outputs from the beamformer Redback modules connect to the correlator subsystem via several 12-fibre ribbon cables. A single correlator subsystem (also known as a block) processes 48\,MHz of bandwidth and currently six of a possible eight blocks are installed. Each correlator block consists of 12 Redback modules (programmed for correlator functionality) and passive optical cross-distribution networks. The numbers in this section represent the installed capacity.

The accumulation operation in computing the visibilities results in a large reduction in the data rate such that the correlator output can be sent as Ethernet data over conventional network infrastructure to the remote high-performance computing facility in Perth, which is approximately 600\,km from the telescope site.

The aggregate data arriving at the correlator input consists of \num{15552} fine channels for 72 beams from each of the 36 antennas. To provide the capacity for polarimetry, these are configured as 36 dual-polarisation beams (with a linear basis). For each fine channel of each dual-polarisation beam, the correlator computes the correlations between all 36 antennas for both polarisations, resulting in $(2\times 36 + 1)\times (2\times 36)/2=2,628$ products.  This is full Stokes cross-correlation for each of 630 baselines and full Stokes autocorrelation for each of 36 antennas. The total correlator output in one integration period, for all 36 dual-polarisation beams and all fine channels, is $\num{2628}\times 36\times \num{15552} = \num{1471343616}$ complex-valued visibilities in the current 6-block configuration with \num{15552} fine channels.

The complex-valued visibilities are streamed from the correlator as pairs (real/imaginary) of 32-bit floating-point numbers. The data streaming cycle from the correlator is locked to the integrate-and-dump cycle which, in turn, is controlled through the timing event subsystem (see Section~\ref{sec:timing}). For the minimum allowed integration interval of \SI{5}{\second}, the total output data rate from the correlator is $\num{1471343616}\times 32\times 2/(\SI{8}{\bit}/\text{Byte}\times \SI{5}{\second}) \approx \SI{2.4}{\giga\byte\per\second}$. See Section \ref{sec:deployment} for more information on system-level constraints.

Data from the correlator are transmitted as user datagram protocol (UDP) packets, each containing up to 657 complex floating-point numbers corresponding to a single spectral channel, beam and polarisation. Partitioning the data in this way makes it easy to scale up the size of the telescope, but means the data must be reordered before further processing. Each UDP packet contains a total of \num{5256} bytes of raw data, to which an additional 48 bytes of metadata are added, including the reference time of the accumulation cycle. The metadata expands the total amount of data leaving the correlator hardware by approximately \SI{1}{\percent}.

\subsection{Timing and synchronisation}
\label{sec:timing}
\begin{figure*}
\begin{center}
\includegraphics[width=\textwidth]{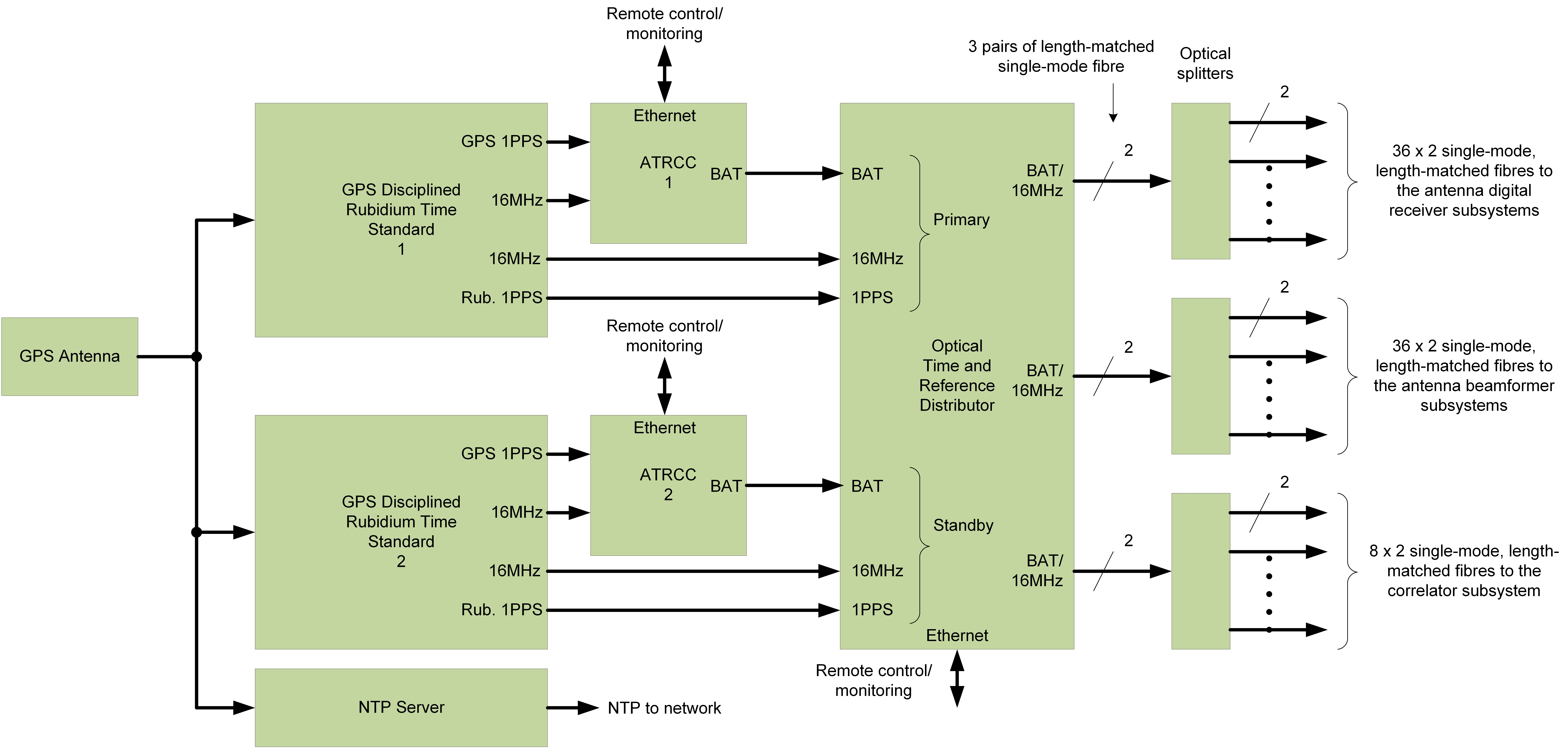}
\caption{Station reference: ASKAP's master clock generation and event timing system. The ATRCC (ASKAP Timing Reference Control Computer) block is an industrial PC with a custom PCIe card that generates a serial encoded timestamp referred to as Binary Atomic Time (BAT) that provides a precise, absolute time reference with a resolution of \SI{1}{\micro\second}.}
\label{fig:station_ref}
\end{center}
\end{figure*}

For a synthesis array radio telescope, the accurate distribution of a stable clock signal to multiple endpoints is a critical part of the instrument design. ASKAP is an array of arrays and this aspect of the system architecture is particularly important and challenging, where signal processing is distributed across some \num{4000} separate FPGA devices and \num{8000} separate ADCs. The station reference and timing system (Figure~\ref{fig:station_ref}) must ensure synchronous sampling and processing of the data across the entire array and throughout the whole DSP chain. 

ASKAP's timing system is based on a general architecture used at CSIRO's observatories \citep{Hoyle2015}.  It distributes a 16\,MHz reference clock to phase reference the signals from the antennas and a serial-encoded timestamp to enable absolute timing of synchronous events across the array. 

The 16\,MHz reference clock is provided by a GPS (global positioning system)-disciplined rubidium time standard. The rubidium clock provides the necessary short-term stability while the GPS corrections ensure absolute long-term stability of the reference.  Also present on site is a hydrogen maser that can serve as an external reference input to the rubidium sources. This is not used for typical ASKAP operation, but is used for VLBI (very long baseline interferometry) experiments \citep[e.g.,][]{Kadler2016}.

The rubidium time standard provides a one pulse per second (1PPS) signal as well as the 16\,MHz reference clock to the ASKAP timing reference control computer (ATRCC), which generates a serial-encoded timestamp known as binary atomic time (BAT). The BAT serial data stream is phase-locked to the 16\,MHz reference clock and encodes a 64-bit integer representing the count of microseconds since the epoch MJD 0.0, which is 00:00\,UT on 17 November 1858.

The BAT serial stream and the 16\,MHz reference clock are distributed together in a star topology from the ATRCC by the optical time and reference distributor, Figure~\ref{fig:station_ref}, to the rest of the ASKAP DSP hardware subsystems (digital receivers, beamformers and correlators). The transmission is by single-mode, length-matched optical fibres using optical splitters. The BAT signal is transported using Manchester encoding to ensure no DC bias for compatibility with optical transmission.

As shown in Figure~\ref{fig:station_ref}, the system has dual-parallel redundancy, and is remotely configurable via the Optical Time and Reference Distributor. Either of the redundant time standards can be selected to provide the primary 16\,MHz and BAT references for distribution to ASKAP DSP endpoints. The final stage in the time and reference distribution occurs at the digital receiver, beamformer or correlator subsystems where the optically distributed 16\,MHz and BAT signals are used to generate all the timing signals needed for the subsystem.
 
\begin{figure*}
\begin{center}
\includegraphics[width=\textwidth]{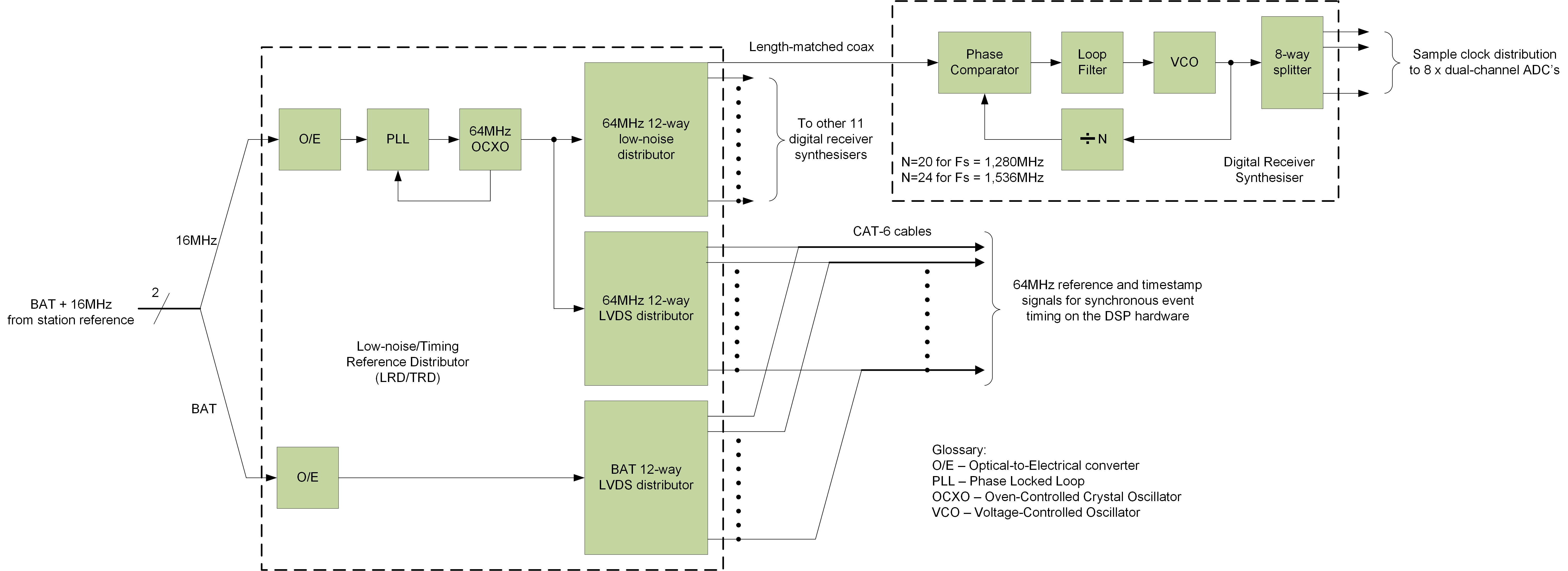}
\caption{End-point of the reference/timing distribution system. The low-noise reference distributor (LRD) generates and distributes a 64\,MHz reference clock for the digital receiver sample clocks plus ${12 \times (\SI{64}{\mega\hertz} + \text{BAT})}$ signals for synchronous event timing. The timing reference distributor (TRD) is the same hardware as the LRD without the separate 12-way 64\,MHz reference distributor. The TRD is used for synchronous timing of events in the beamformer and correlator subsystems.}
\label{fig:lrd_trd}
\end{center}
\end{figure*}

In each beamformer and correlator subsystem the timing signals are received by the Timing Reference Distributor (TRD). The digital receiver subsystems use the Low-noise Reference Distributor (LRD).  The architecture of the TRD and LRD are  shown in Figure~\ref{fig:lrd_trd}. In the LRD and TRD the 16\,MHz clock is multiplied up to 64\,MHz and the 64\,MHz and BAT signal are distributed by CAT-6 cables to all modules in the subsystem. These signals are used to generate synchronous processing events in all Redback and Dragonfly modules.  Each TRD can supply signals to eight Redback modules.

The LRD has an additional low noise distributor for the 64\,MHz clock, which uses length-matched coaxial cable to distribute a low phase-noise copy of the 64\,MHz signal to the 12 digital receiver Dragonfly modules. This generates a high quality ADC sample clock with low jitter.  

\section{OPERATIONAL MODES}
\label{sec:deployment}

ASKAP's digital system is very modular and was designed to exceed the nominal bandwidth of 300\,MHz for a full hardware deployment. For reasons outlined here, the existing system does not have a full complement of hardware and does not exercise all possible configuration flexibility, though more may be unlocked in future upgrades.

\subsection{Digital system hardware considerations}

The digital receiver outputs 384 channels of 1\,MHz bandwidth each. Processing all of these would require 8 beamformers per antenna, but only the first 7 were installed. Wiring for the final beamformer is in place, but the construction budget did not allow the additional boards to be manufactured and they were not needed to meet the nominal bandwidth specification. Similarly, only 6 of the full 8 sets of correlator redbacks were installed. Since the correlator works in blocks of 48 MHz, meeting the nominal 300 MHz specification would have required installation and configuration of a partial block. While possible, this would have increased the complexity of control system logic and configuration management, for only a few percent improvement in continuum sensitivity. Using all available test hardware and reducing the stock of spare parts, it may be possible to install a 7th correlator block. However, the data ingest and image processing system was only designed to work with up to 300\,MHz of bandwidth, so expanding to 336\,MHz will be left for a future upgrade once operational experience shows how much spare capacity exists in the post-correlator stages.

\subsection{Correlator output modes}

The flexibility of ASKAP's digital processing system allows each beamformer FPGA to operate in a different frequency resolution mode with arbitrary channels from the digital receiver. However, the complex channel routing logic required to implement a generalised configuration manager is prohibitive, and the control system is currently restricted to a single frequency resolution mode with contiguous bandwidth across all hardware. A summary of the available modes and the resulting bandwidth available for imaging is shown in Table \ref{tab:askap_zooms}. Regardless of frequency resolution, the 6-block correlator always outputs 15552 spectral channels.

To reduce the amount of storage space required for initial capture of data intended for continuum science, the ingest pipeline software can perform online averaging down to \SI{1}{\mega\hertz} frequency resolution. This reduces the storage footprint by a factor of 54 and can be used to balance resource requirements when scheduling observations. This averaging mode is only intended for use with the full bandwidth, not zoom modes.

At the time of writing a \SI{10}{\second} integration interval is used in the correlator accumulation stage, reducing the data volume by a factor of two, for a total of \SI{1.2}{\giga\byte\per\second} leaving the observatory. If pilot surveys show that this accumulation interval does not introduce any significant smearing, it will be used for large-scale surveys as well. We have tested cycle times down to roughly 7 seconds successfully with the existing system, so there is some room for improvement. Attempts to record with the nominal 5-second cadence lead to data loss that will require more investigation to understand. An upgrade to the ingest computing hardware at Pawsey scheduled for November 2020 should also improve performance.

\subsection{Data processing assumptions}

At the time of design, ASKAP presented an extreme computational challenge. Short-term storage of the correlator output was deemed impossible (and still presents challenges, though advances in technology have made temporary storage possible) and efforts were made to minimise resource-intensive aspects of downstream processing. One such limitation is that the outer six antennas were not expected to be used in spectral line mode, since the size of the resulting cube would be too large. Advances in visualisation and analysis software make it desirable to consider creating high-resolution sub-cubes, at least for specific regions of interest.

\section{DIGITAL BEAMFORMING}
\label{sec:beamforming}
The use of digital beamforming allows ASKAP to alter the illumination of the reflector antennas (and therefore the primary beam shape and direction) by changing numerical coefficients. This level of flexibility is of great benefit, but also significantly increases the complexity of the system. Unlike a mechanical feed, ASKAP's primary beam shape and receiver noise can change due to drifts in the complex gains of the elements that combine to form a beam.

Radio astronomy imaging algorithms typically assume that the primary beam is fixed, symmetric, and identical from one antenna to the next. For ASKAP, extra calibration steps are needed to ensure that these assumptions are met. The on-dish calibration system (ODC), described in Section~\ref{sec:odc}, is used to monitor the complex gain of each PAF element. Although this does not provide a direct measure of the beam shape, it can be used to compensate for individual-element gain changes so that beams are stable over time.

ASKAP uses measurements of an astronomical reference source to determine the numerical weights used by the digital beamformers. A strong source is placed in the desired beam centre for each of 36 offset positions and the array covariance matrix $\hat{\mathbf{R}}_{n+s}$ is recorded. These observations are compared with a noise field $\hat{\mathbf{R}}_{n}$ that does not contain any strong sources.

\subsection{Maximum sensitivity weights}
We compute the maximum sensitivity (maxSNR) beamformer weights following the procedure described by \citet{hotan_2014} and \citet{mcconnell_2016}, using the Sun as our reference source. The estimated array response $\hat{\mathbf{a}}$ of the PAF to a far-field signal incident from the direction of the desired beam is given by $\hat{\mathbf{a}}=\mathbf{u}_1$ where $\mathbf{u}_1$ is the dominant eigenvector of $(\hat{\mathbf{R}}_{n+s} - \hat{\mathbf{R}}_{n})$ \citep{Jeffs2008_Signal}.  The maxSNR weights \citep{Lo1966, Applebaum1976} are then computed as
\begin{eqnarray}
\label{wgts}
\mathbf{w} & = & \hat{\mathbf{R}}_n^{-1} {\hat{\mathbf{a}}}.
\end{eqnarray} 

The \emph{noise-plus-signal} covariance estimate $\hat{\mathbf{R}}_{n+s}$ is measured via \eqref{acm_est} with the antenna pointed so that the calibration source (Sun) is at the same position within the field of view as the desired beam pointing.  The \emph{noise} covariance estimate $\hat{\mathbf{R}}_{n}$ is measured with the antenna pointing at an `empty' piece of sky, typically \ang{15} south of the calibration source.   

To ensure there is only one dominant eigenvector, we use only X-polarisation ports to make X-polarisation beams and only Y-polarisation ports to make Y-polarisation beams.  One eigenvalue equation is solved for the sub-array of X-polarisation ports and then a second eigenvalue equation is solved for the near-independent sub-array of Y-polarisation ports.  This means that the polarisation states and angles of the X and Y beams will be tied to the inherently linear and orthogonal polarisations of the X and Y elements of the chequerboard array.  Commissioning measurements have shown that this technique results in beams with low and stable polarisation leakages \citep{Sault_2014_Initial,Sault_2015_Widefield,Deng2017} and that ASKAP already has excellent polarisation imaging capability \citep{Anderson2018}.

Weights are calculated independently for each \SI{1}{\mega\hertz} channel of each beam, which can lead to a randomised phase response.  As described in \citet{mcconnell_2016}, we adjust the phase of the weights at each \SI{1}{\mega\hertz} channel to ensure a smooth phase response with frequency for each beam.  

Digital beamforming with \SI{1}{\mega\hertz} resolution introduces quasi-periodic bandpass errors with a \SI{1}{\mega\hertz} period.  To work around this, for spectral-line observations, we sometimes reduce the beamformer resolution by fixing the beamformer weights to be identical over sub-bands of $N \times \SI{1}{\mega\hertz}$ where $N$ is an odd integer.  This increases the magnitude of the bandpass errors, but ensures their smoothness over the larger $N \times \SI{1}{\mega\hertz}$ sub-bands.  This is advantageous when looking for astronomical features at $\SI{1}{\mega\hertz}$ resolution.  It also reduces the probability of an astronomical feature falling on a sub-band boundary.  

  Since \citet{mcconnell_2016}, we have extended our maxSNR calibration algorithm to: 
\begin{itemize}
\item calibrate the XY phase of beams to simplify polarisation observations; 
  \item accept the limit on the number of ports that our hardware can weight into each beam;
  \item avoid bias due to ports with abnormally high or low signal levels;
  \item correct weights for gain, phase and delay changes in the receiver electronics; and 
    \item avoid bias due to RFI by interpolating weights with frequency. 
\end{itemize}   

\subsection{Calibrating {XY} phase}
Some of these changes were required to improve the robustness and reliability of the beamforming process for operational deployment. However, one great advantage of digital beamforming is that it allows control over aspects of the primary beam that would normally be fixed by construction. One such example is the relative phase of the two orthogonal polarisations. 

We use the ODC system and the technique of \citet{Ch_An_2019_XYphase} to adjust the beamformer weights so that the XY phase of each dual-polarisation beam pair is near zero by design.  Calibrating XY phase up-front in the beamforming allows polarisation calibration and imaging to take place in the standard ASKAP  software pipeline, so that polarisation studies can be made commensally with all ASKAP observations.   
\subsection{Beamforming with limited port selection}
ASKAP's beamformer (see Section~\ref{subsec:bmf}) can simultaneously form 36 dual-polarisation beams compared to BETA's nine.  However, as the number of processed beams increases towards the full 36, the number of ports that can be weighted into each beam reduces towards 60 according to Table~\ref{tab:ports_per_beam}.  This is due to hardware resource limitations in the beamformer.

If all ports were functional and had the same gain and phase response, we could just select the weights with the largest amplitude when selecting the finite number of weights $M$ that can be processed by the beamformer.  However, the inversion of the noise covariance in \eqref{wgts} can lead to very high numeric weights for ports that have low gain.  A first step towards avoiding these problems is to use RF attenuators along the signal-chain of each port to normalise the signal level that is presented to the digital backend from each PAF element.  

\subsection{Robust port selection}
In the case of a malfunctioning port, with near zero gain even after attenuator adjustment, the covariance matrices are poorly conditioned and can further complicate the solution for maxSNR beamformer weights. To avoid this problem, we use a diagonal loading process to avoid singularities in the inversion of the full covariance matrix.  This adds a small amount of white noise to the noise covariance matrix before its inversion and application in \eqref{wgts} to calculate the weights 
\begin{eqnarray}
\label{wgts_diag}
\mathbf{w} & = & \left(\hat{\mathbf{R}}_n+\alpha\mathbf{I}\right)^{-1} {\hat{\mathbf{a}}}.
\end{eqnarray}
The diagonal loading places an upper limit on the weight that will be ascribed to any malfunctioning port with near-zero gain.

We use diagonal loading in a two-step process to robustly exclude malfunctioning ports from the beamformer weights.  First, we select $M$ ports with the largest signal contribution to the beam defined by \eqref{wgts_diag}.  The signal contribution of each port is estimated with the aid of the ODC (see Section~\ref{sec:focal_field}).  Second, we recalculate the basic maxSNR weights using \eqref{wgts} on a reduced $M\times M$ covariance matrix of the selected ports. 

The first step of the process tries to select the ports that are important to the beam, using diagonal loading to exclude malfunctioning ports.  The second step applies the basic maxSNR algorithm, with no diagonal loading and therefore no bias, to the selected ports.  

There is a sizeable literature on the appropriate selection of $\alpha$ in the diagonal loading process.  We use an empirical technique guaranteed only to make improvements upon the maxSNR solution.  We set
\begin{equation}
\alpha = \beta \med(\diag(\hat{\mathbf{R}}_n))
\end{equation}
and iterate the above solution over diagonal loading values of $\beta = [0, 0.1, 0.2, 0.3, 0.4]$, calculating the Y-factor (signal plus noise to noise ratio) at each step.  Only solutions that improve Y-factor upon previous trials are retained.  The first trial, with $\beta=0$, is equivalent to the maxSNR solution with no diagonal loading, so the iterative process is guaranteed to achieve or improve upon the sensitivity of the basic maxSNR solution.
  
Comparison of SEFDs achieved by this two-step process to those achieved by the raw maxSNR algorithm show that it improves sensitivity over the raw maxSNR algorithm in almost all scenarios.  The improvement is most noticeable when there are low-gain ports with significant energy falling on them due to the focal plane field of the desired PAF beam.

\subsection{Focal plane field estimation}  
\label{sec:focal_field}
We use the technique of \citet{Chippendale2016} to decouple the PAF element gains from the beamformer weights and estimate the focal plane field $\mathbf{s}$ matched by beamformer weights $\mathbf{w}$

\begin{equation}
\label{eq:weightcal}
\mathbf{s} = \mathbf{d^*_\text{cal}}\circ\mathbf{w}.
\end{equation}
Here $\circ$ is the Hadamard (element-wise) product and $\mathbf{d^*_\text{cal}}$ is the complex conjugate of the PAF's response to a plane wave.  The plane-wave response is estimated by correlating each PAF element with a copy of the broadband calibration noise that is radiated into the PAF
\begin{equation}
\mathbf{d_\text{cal}} = \frac{\left<\mathbf{x}_\text{paf}x^*_\text{rtn}\right>}{\left<|x_\text{rtn}|\right>}.
\end{equation}
Here $\mathbf{x}_\text{paf}$ is a vector of the PAF element voltages sampled at the digital receiver and $x_\text{rtn}$ is the sampled copy of the calibration noise that has been sent out to the antenna and reflected back to the digital receiver via a Faraday mirror as shown in Figure~\ref{fig:ODC}.  We commonly refer to $\mathbf{d}_\text{cal}$ as the ODC response.  In fact $\mathbf{d}_\text{cal}$ are the conjugate field match beamformer weights that would form an aperture array beam matched to the ODC source.

Ideally, the calibration source would present a plane wave to the PAF, illuminating all ports equally and in phase.  In this case, $\mathbf{s}$ is a good estimate of the focal plane field excited when the telescope observes a point source in the centre of the beam created with weights $\mathbf{w}$.  The ports with the largest component amplitudes in $\mathbf{s}$ are the ports with most significant focal plane field excitation by the desired signal.  They are the ports with the largest signal contribution to the beam.

\subsection{Correcting PAF element gain drifts}
\label{sec:odc_use}
We infrequently observe the Sun to make new beamformer weight solutions; typically after several months have passed or when significant changes have been made to the hardware. Between these solution intervals we use the ODC system to correct existing beamformer weights for changes in the complex gains of the PAF elements.  This is effected by the Hadamard product of the original weights by the Hadamard (element-wise) division $\oslash$ of ODC responses at the beamforming (subscript 0) and update (subscript 1) epochs  

\begin{equation}
\label{eq:weight_update}
\mathbf{w}_1 = \mathbf{d^*_\text{cal,0}} \oslash \mathbf{d^*_\text{cal,1}}  \circ \mathbf{w}_0.
\end{equation}
The phase component of this correction is applied independently for each \SI{1}{\mega\hertz} channel, but the median value over the \SI{336}{\mega\hertz} beamformer bandwidth is used in amplitude.  This forces the correction to be frequency independent in amplitude despite the ODC source having some undesired variation in spectral shape with dish antenna motion.  We will revise the need for this smoothing in amplitude when the stability of the noise source is improved.  The problem appears to be related to mechanical stability of the ODC vertex antenna and its RF connection to the noise source.    

Since the ODC system is constantly injecting a low-level broadband reference signal into the feed \citep{Chippendale_2018_ODC}, the array covariance measurements used for beamforming also capture a measure of the complex response of each element to the ODC signal.  When we wish to update the beamformer weights, we record a smaller data product from part of the beamformer known as the calibration correlator (see Sections~\ref{sec:odc} and~\ref{subsec:bmf}). Updated weights are calculated via \eqref{eq:weight_update} and uploaded to the beamformer.

This process requires only a small amount of data and since the beamformer weights are loaded via a double-buffered mechanism in the beamformer firmware, it is possible to update the weights without interrupting an observation in progress. The minimum update timescale is set by the time taken to compute, upload and switch to a new set of weights (a few seconds). However, the weights are currently left fixed during scheduling blocks.  We perform an update before each observing session and leave the weights fixed until it is necessary to restart the digital receivers, typically when changing sampling bands. 

Currently, the ODC correction in \eqref{eq:weight_update} stops beams from being completely destroyed by unpredictable delay changes that occur with each full reset of the ADCs.  In the future we hope to use the ODC to stabilise PAF beam (voltage) patterns to better than \SI{1}{\percent} at the half power points \citep{Hayman2010}. 

\subsection{Avoiding RFI bias of beam direction}
Basic maxSNR weights can be biased at frequencies with strong RFI.  If there is RFI stronger than the beamformer calibration source in the measurement of $\hat{\mathbf{R}}_{n+s}$, the resulting beam will try to point in the direction of the RFI instead of the calibration source.  These weights will have a low gain in the desired beam direction at the frequencies with strong RFI.  The astronomical signal in the intended beam direction cannot be recovered when using these weights, even with careful flagging of the RFI-impacted data in the time domain. 

We avoid this problem by interpolating valid beamformer weights for ASKAP at interference-affected channels following the technique in \citet{Chippendale2017}.  This uses iterative fitting of low-order polynomials to the weight amplitude spectrum of each individual port, taking advantage of the inherent smoothness of maxSNR weight amplitudes with frequency. 

\subsection{Using maxSNR beams for astronomy}
Optimisation for sensitivity tends to create stronger primary beam side lobes than a mechanical feed and allows the beam shape to take on coma distortion off axis. Both of these can introduce undesirable effects in the imaging process. 

Holography observations show that the total intensity beam shape is consistent with a circular Gaussian model to within roughly \SI{10}{\percent}, according to measured eccentricities \citep{McConnell_2016_Field} for a beam spacing of \ang{0.9}. However, the level of coma distortion is a strong function of distance from the field centre, which means that the consequent  variation in beam shape  needs to be well characterised  to form an accurate linear mosaic.

Over time we intend to develop other optimisations that may improve the symmetry, polarisation and other properties of the beams.  A starting point will be linearly constrained minimum variance (LCMV) beamforming \citep{frost1972}, which can be used to optimise sensitivity with additional constraints on main beam shape.  See \citet{Elmer2014} for an example of constrained beamforming with a PAF. 

\subsection{Measurement of beams using holography}
\label{sec:holography}

To assist with beam optimisation research, we have developed a raster grid holography method that provides a measure of the beam shapes over a large fraction of the field of view \citep{Hotan_2016_Holography}. Conducting a full measurement requires roughly six hours of observing time. This can be used to measure the stability and reliability of the beamforming process and assess the impact of changes to the constraints used when computing beamformer weights.

One of the next planned improvements to the imaging software will allow primary beam correction in the mosaicking stage using beam shapes measured via holography, instead of the circular Gaussian model currently applied.

\subsection{Beam footprints for science observations}
Digital beamforming provides the freedom to place primary beams anywhere within the field of view of the antenna. However, in practice it is important to maximise the packing efficiency of the beams and develop standard arrangements that minimise sensitivity fluctuation across the field of view and also tessellate well when covering larger areas with multiple pointings.

\begin{figure}
\begin{center}$
\begin{array}{cc}
\includegraphics[width=1.5in]{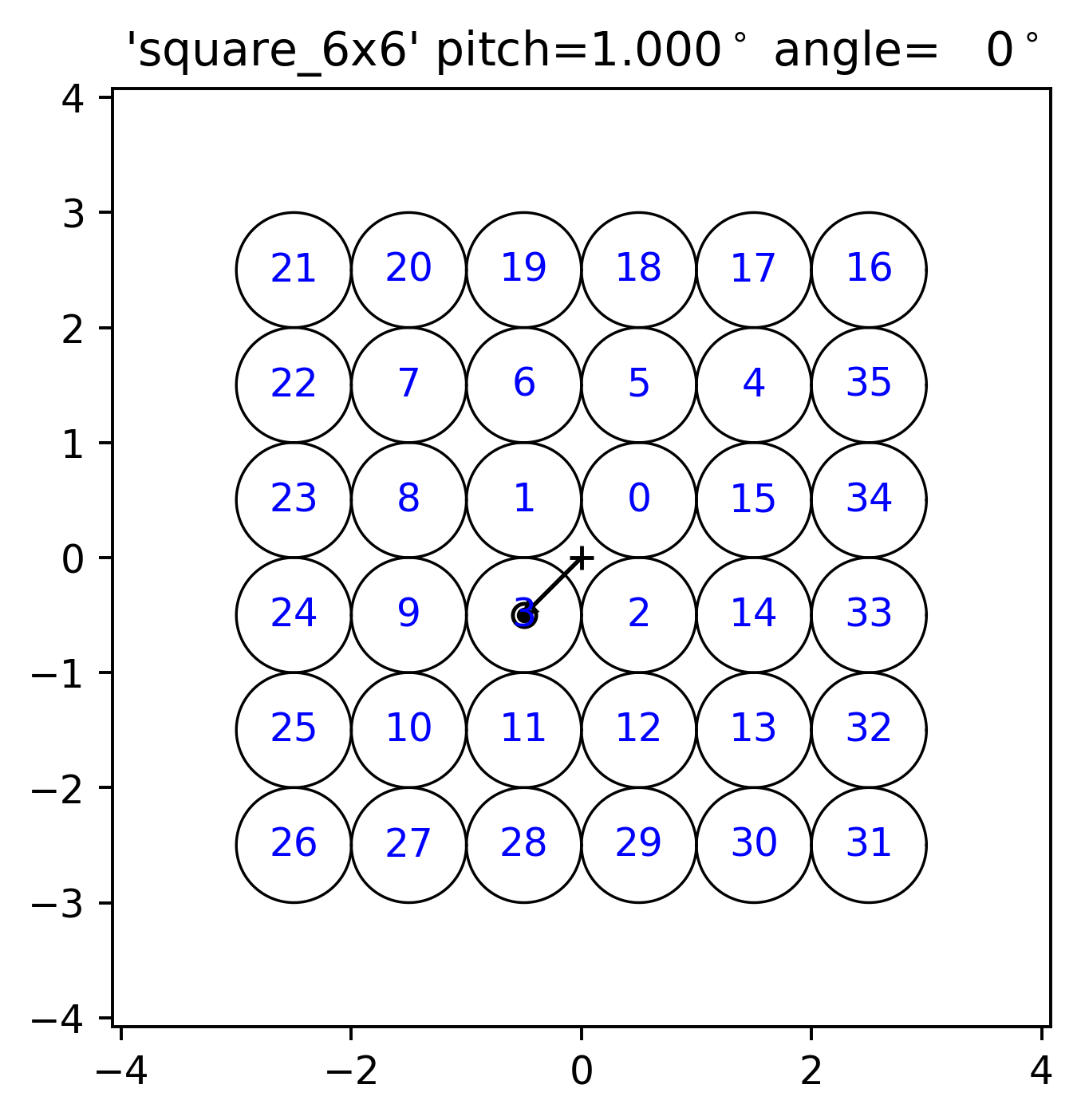} &
\includegraphics[width=1.5in]{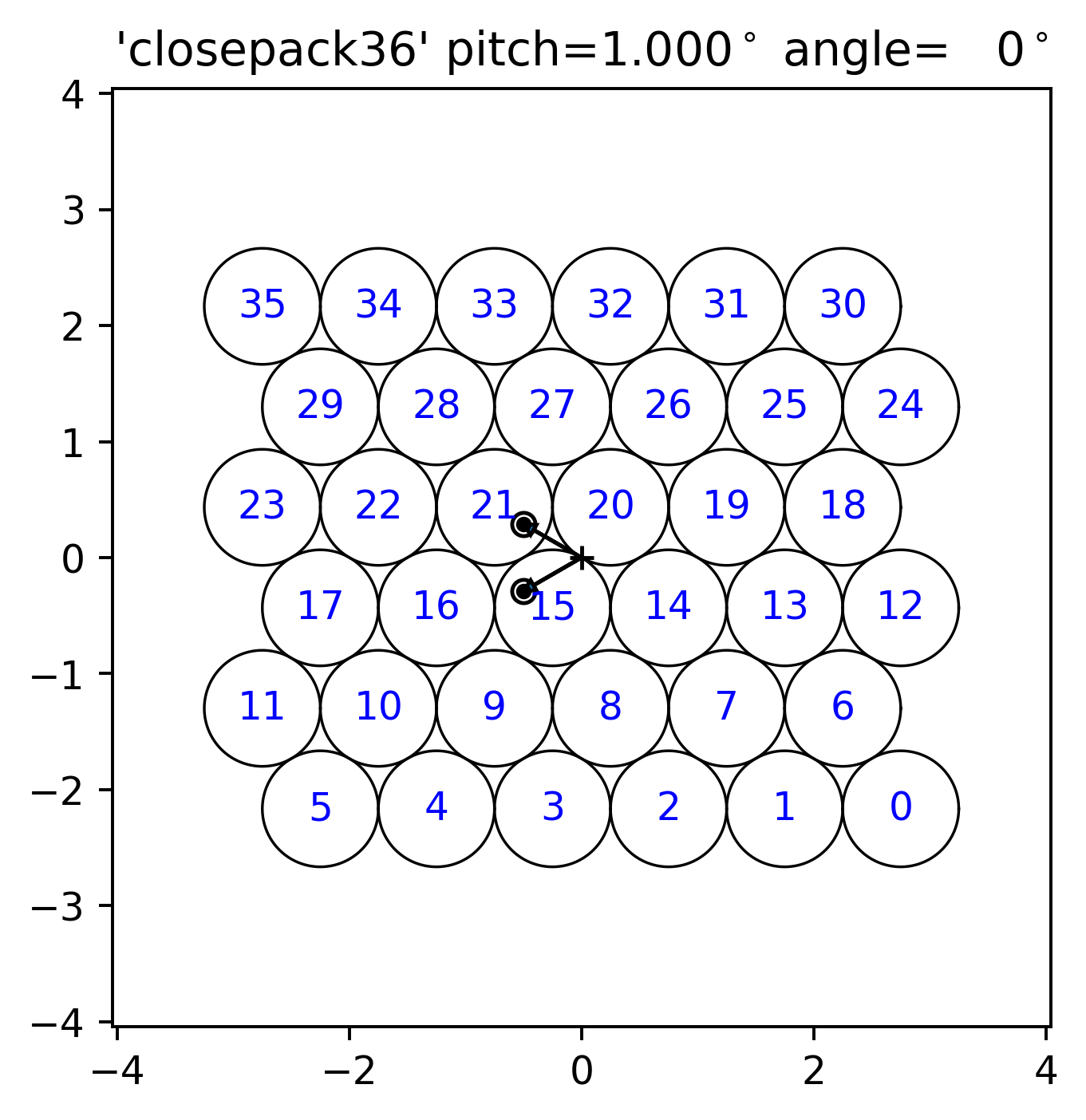}
\end{array}$
\end{center}
\caption{The two commonly-used footprints: {\tt square\_6x6} (left) and {\tt closepack36} (right). In this case both have beam spacings (pitch) of \ang{1.0} and a position angle of zero. The scales are in degrees, and celestial north (west) is to the top (right) of both diagrams. The + symbol represents the optical axis of the reflector and the lines extending from this position ending in filled circles represent the pointing shift required to optimally interleave the footprint by using other observations to fill in the least sensitive parts. The square arrangement has larger intrinsic sensitivity variation but requires only one interleaving position, while the closepack arrangement is more uniform initially but requires two interleaving positions for improvement.}
\label{fig:footprints}
\end{figure}

We call the arrangement of beams within the field of view the footprint.  Two footprints are in common use on ASKAP: {\tt square\_6x6} that follows a square grid and {\tt closepack36} that offsets alternating rows to provide more uniform sensitivity as shown in Figure~\ref{fig:footprints}.

Note that neither of these footprints has a beam located on the optical axis of the antenna, which would have been desirable for calibration purposes. Having an extra beam would have allowed the use of a hexagonal arrangement containing 37 beams, including one on the optical axis. One of the future enhancements we are considering would allow the exchange of bandwidth for additional beams. As part of this work, we may investigate adding one more beamformer engine to the standard full-bandwidth mode if resources on the FPGAs will allow it.

\subsection{Tiling footprints for surveys}
\begin{figure}
\begin{center}
\includegraphics[width=\columnwidth]{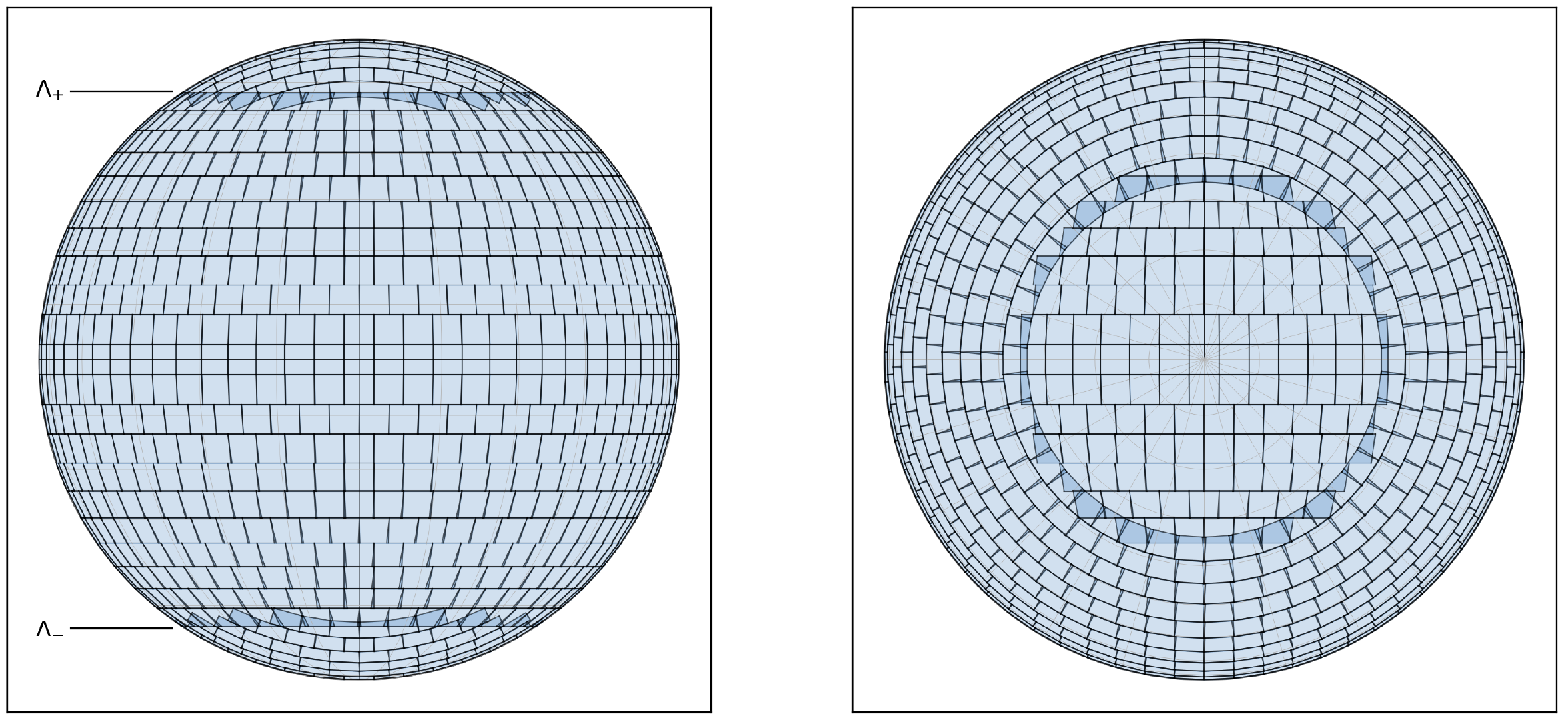}
\caption{Two views of the survey tiling scheme: equatorial (left) and polar (right). The boundaries of the two polar zones at $\Lambda_{-}$ and $\Lambda_{+}$ are indicated in the left-hand panel. The whole sphere is shown here; the tiling can be defined for any coordinate system (Equatorial, Galactic, etc) and then transformed into the natural operating frame.} 

\label{fig:twoviews}
\end{center}
\end{figure}

Since ASKAP will spend most of its time conducting all-sky surveys, it is useful to define a standard tiling scheme to cover the celestial sphere with the footprints described above. In consultation with ASKAP's survey science teams\footnote{\url{https://www.atnf.csiro.au/projects/askap/ssps.html}}, we have developed a tiling scheme in which most of the sky is covered by bands of footprints stacked end-to-end at constant declination (see Figure~\ref{fig:twoviews}). This becomes less efficient towards the celestial poles, so at 72 degrees above or below the equator we switch to polar caps consisting of a square grid that is truncated at the overlap with the outermost declination band. This approach allows a fixed beam footprint to be used for the entire survey. Methods such as HEALPix \citep{2005ApJ...622..759G} were considered, but deemed unsuitable. HEALPix was designed to efficiently partition a spherical surface by allowing the shape of each partition to vary. Matching the different partition shapes would greatly increase the operational overheads associated with beamforming.

The number of tiles required to cover the sky depends on the beam spacing within each tile, which can be adjusted to trade between survey speed and uniformity or mitigate widefield polarisation leakage. This optimisation is also a function of the observing frequency.

The first ASKAP all-sky survey, known as the Rapid ASKAP Continuum Survey (RACS), used a {\tt square\_6x6} footprint with beam spacing of \ang{1.05} at a centre frequency of 888\,MHz.  RACS consisted of 904 tiles observed for \SI{15}{\minute} each, covering the entire sky south of \ang[retain-explicit-plus]{+40} declination. A future publication will describe the forthcoming data release and source catalogue.

\subsection{Beamformer weight storage and management}

Beamformer weights are stored in a custom HDF5 \citep{hdf5} file format. Currently, each scheduling block specifies which weights file to use. We have developed, and are now commissioning, a database service that tracks weights files and can automatically retrieve the best available weights for a requested footprint and frequency setup. It is important to ensure that details of the beamformer weights are included in the metadata associated with an observation, to provide a complete record of the system configuration.

\section{RAW DATA INGEST}

Visibility data leaves the ASKAP correlator hardware as UDP packets which are transmitted over a standard Ethernet network. An EPICS IOC running at the observatory gathers data from each correlator block and transmits it to the Pawsey supercomputing centre. Pawsey hosts a dedicated ingest cluster for ASKAP that is isolated from general purpose activity to ensure predictable load.

Each UDP packet sent from the observatory contains one fine channel (packed in a non-contiguous way), one beam and a range of baselines (the whole baseline space is split into four groups). A specially-designed ingest pipeline software application receives the data using multiple threads, tags it with metadata from the telescope control system and reorders it into the form required for further processing. Data sent from each correlator hardware module are staggered in time to spread the network transmission load out over the full accumulation cycle. The ingest pipeline can also perform basic tasks such as channel averaging if required.

Data are written in parallel to a series of CASA {\tt MeasurementSet} \citep[version 2, ][]{MSv2} files on a Lustre\footnote{\url{http://lustre.org}} file system that is common to all ingest nodes. We need to split each of the PAF beams into its own measurement set to avoid writing to a single file at an excessive rate. For full spectral resolution mode, we must further divide each beam into six frequency channel groups. ASKAP's image processing pipeline builds a complete image by combining data from all of these independent measurement sets.

Although the performance specifications of the file system are adequate for capturing ASKAP data at full rate, we have run into practical problems where other processes operating on the same Lustre hardware can impact performance.

\section{SCIENCE DATA PROCESSING}
\label{sec:SDP}

The plan \citep{ASKAPSW0020} for ASKAP science data processing (SDP) was to have a near-real-time autonomous pipeline. This would handle the calibration, imaging, cataloguing and archiving of the data, in a way that would keep up with the observing rate. The plan hinged on the ability to forward-predict calibration parameters for the array using a sky model and apply the solution on-the-fly.

Through commissioning and early science, we were unable to bootstrap this method of calibration and instead reverted to the use of dedicated observations of a standard calibrator source in each beam. Calibration can therefore be applied offline and improved iteratively with additional flagging, which has been essential to meeting data quality requirements. Online calibration is now considered a possible future improvement, once a suitable sky model has been obtained from the first all-sky surveys calibrated using traditional methods.

Using dedicated calibration observations increases operational overheads. Due to the large number of beams, the current calibration procedure uses about 2.5 hours of observing time and must be done whenever the beamformer weights are changed or updated, which is typically once every day or two. This is feasible for survey observations, but adds scheduling complexity and reduces overall efficiency by about 10 percent. Although the time commitment for per-beam calibration using a reference source is significant, the storage requirements can be mitigated by keeping only the bandpass solution or discarding visibility data from beams when they are not pointing at the calibrator.

The batch-processing pipeline that has been developed provides the functionality required to process pilot survey data to a science-ready state. This section describes the high-level functionality of the pipelines, with more detail provided elsewhere \citep{Whiting2020}.

The pipeline and the software it runs are collectively known as ASKAPsoft. The calibration and imaging software of ASKAPsoft\footnote{The core ASKAPsoft imaging software is available as the \textit{Yandasoft} package \url{https://bitbucket.csiro.au/projects/ASKAPSDP/repos/yandasoft}. \textit{Yanda} is the Wajarri word for \textit{image}.} has been custom-written to address the particular imaging requirements of ASKAP, especially imaging over a wide field-of-view and in near-real-time. 

\subsection{Pipeline workflow}
The processing pipeline currently runs on \textit{Galaxy}, the Cray XC30 at the Pawsey Supercomputing Centre. It creates a workflow from a series of connected compute jobs, running under the Slurm workload manager, with appropriate dependencies to ensure completion in the correct order. 

The processing is mostly performed on individual beams independently, before combining images in a linear mosaic. The top-level workflow for a single beam can be expressed as follows:
\begin{enumerate}
    \item Bandpass calibration is done using a separate observation of PKS B1934$-$638, establishing the overall flux scale. Frequency-dependent complex gains are determined for each beam and later applied to the science observation. See section \ref{sdp:calibration}.
    \item Visibility data are prepared by splitting raw data measurement sets into single beam datasets (where necessary), then applying the bandpass calibration. The splitting may also create datasets for short time windows (typically one hour in length), which may be processed independently.
    \item RFI is flagged, before the data are averaged to 1\,MHz resolution (this will be used for the continuum imaging). Coarse-resolution data may be flagged again. If time-splitting was used, the time windows are recombined.
    \item The coarse-resolution measurement set is imaged and a sequence of self-calibration is performed. The self-calibration results in a fully-calibrated image and table of time-dependent (but frequency-independent) complex gains. See section \ref{sdp:calibration}.
    \item Both the coarse and full-resolution measurement sets are calibrated with the self-calibration gains.
    \item If necessary, continuum emission is subtracted from the full-resolution visibilities. This emission may be characterised by clean components, fitted components identified through source-finding, or through fitting to the visibility spectra. The optimal strategy for a given science case is being determined in conjunction with the Survey Science Teams, using the pilot survey processing.
    \item The calibrated visibilities are imaged, producing final continuum images, coarse-resolution spectral cubes or full-resolution spectral cubes as required.
    \item If necessary, any residual continuum emission remaining in spectral cubes (as a result of incomplete modelling of the continuum emission) is removed through an image-based continuum subtraction step.
    \item Diagnostic and quality control information is assembled into a set of reports that are archived along with the image data.
\end{enumerate}

Once all the beams have been imaged, full-field images and cubes are created through linear mosaicking. This corrects for the primary-beam attenuation by applying a two-dimensional Gaussian primary beam model (see Section~\ref{sec:holography} for efforts to develop more realistic models) centred on the position of each beam in the footprint. Once the mosaics have been created, source finding is run to create catalogues. The data products can then be uploaded to the archive (see Sec.~\ref{sec:CASDA}).

\subsection{Calibration}
\label{sdp:calibration}

Most ASKAP observations are calibrated in two steps. First, the flux scale and frequency-dependent complex gains (the ``bandpass'') are determined from a dedicated observation of PKS B1934$-$638. This is the primary southern-hemisphere flux calibrator, and has a well-determined spectral energy distribution \citep{reynolds1994}. The bandpass observation points each PAF beam in turn at PKS B1934$-$638 and observes for several minutes per beam. Although we currently assume a simple point source model for the calibrator source, \cite{2020MNRAS.494.5018H} have shown that slight improvements could be made using a model of the full field that includes confusing sources within the primary beam. The calibration parameters derived from this observation are applied to the raw science data prior to any other processing.

Since the bandpass gains are determined from a different observation, some refinement to them is often necessary due to time-variable factors such as atmospheric and ionospheric conditions that can impact the calibration solution at a low level. This is currently done through a self-calibration process, using the continuum-averaged dataset. A continuum image is made, then calibration is performed using a shallow sky model derived from that image. The complex gains are solved for within short ($\sim \SI{5}{\minute}$) intervals across the observation. The data are then re-imaged using this updated gain calibration to provide an improved result. This process can be iterated several times if required. We typically solve for the phases of the complex gains and leave the amplitudes fixed, in order to preserve the overall flux scale. Once complete, the gain solution can be applied to the full-spectral-resolution visibilities to facilitate spectral imaging.

\subsection{Wide-field imaging}
\label{sdp:imaging}

Wide-field imaging is enabled via the $W$-projection algorithm \citep{2005ASPC..347...86C}, allowing handling of image-plane effects due to non-coplanar baselines. This gives accurate imaging over the field of view of a single beam. Multiple beams could be combined via $AW$-projection \citep{2008A&A...487..419B,2013ApJ...770...91B}, taking into account the aperture illumination, although this approach is currently not used due to its large memory requirements. As described above, the aperture correction is instead applied when mosaicking the individual beams together.

The near-real-time imaging requirement has driven a number of design decisions within the code. The weighting of visibility data prior to imaging is traditionally done with an initial pass over the data to determine the distribution of visibilities as a function of baseline length and hence the weighting function, and then a subsequent pass for the imaging. To prevent repeated iteration over the large ASKAP data products, the ASKAPsoft code incorporates the \textit{preconditioning} approach \citep{rau2010}, where the re-weighting is performed after the construction of the dirty image and point spread function (PSF). ASKAPsoft imaging makes use of robust weighting, similar to that described in \citet{briggs1995}, allowing a tradeoff between the suppression of sidelobes and the overall sensitivity.

Continuum imaging is performed using Taylor-term multi-frequency synthesis \citep{rau2010, Rau2011}. This creates an image of the total intensity at a reference frequency, along with images of the spectral variation across the band, from which can be determined the spectral index (and curvature, if enough terms are used) of sources. For spectral imaging, each channel is imaged independently, using either the averaged (continuum) data to create ``continuum cubes'', or the full-resolution data to create spectral cubes. Continuum cubes can be made in all Stokes parameters, allowing polarisation imaging.

The imaging software has distributed processing designed in to the code at a low level. This allows the processing to be distributed over many compute nodes. The continuum imaging has the gridding distributed per coarse channel and Taylor term, with the deconvolution currently performed on the head node after combination of all channels. The spectral imaging is performed independently on each channel, and so is extremely parallel, allowing distribution over as many nodes as can be accommodated.

It should be noted that this extremely parallel case assumes a coarse form of Doppler correction done per beam and with only whole-channel offsets. This may be a limiting factor for some science applications. 

Images are currently stored in FITS format \citep{2010A&A...524A..42P} for broad compatibility and because the science data archive was designed around this format. ASKAP data cubes can be very large (tens of terabytes) and it is possible that future research and development may provide a more efficient data storage format for use in parallel processing environments with remote visualisation services. If a new standard is established, we will investigate upgrading ASKAP's software platforms to incorporate it.

\subsection{Source finding and cataloguing}

As part of the pipeline processing, we make catalogues of the sources in the images, using the Selavy source-finder \citep{whiting2012b}. Selavy uses the source-finding algorithms of Duchamp \citep{whiting2012a}, a source-finder developed for three-dimensional datasets. It can produce catalogues of spectral-line sources from the spectral cubes, largely following the Duchamp methodology, as well as catalogues of continuum sources.

The continuum source-finding produces two different catalogues. The first is the catalogue of \textit{islands} in the image. These are groupings of image pixels that are above the detection threshold. Each island is parameterised by its location and shape, determined solely from the detected pixels. To each island, we fit some number of two-dimensional Gaussians --- these are the \textit{components}, and the second catalogue reports their parameters. Each component will have a location, flux and shape, as well as information on the spectral index. This is obtained by fitting the Gaussian in question to the Taylor-1 image, which gives the product of the Taylor-0 flux and the spectral index.

Source finding is an active topic of research within the broader community and some of the ASKAP survey science teams will produce their own catalogues using different methods and tools. These are referred to as ``value-added'' data products. For example, the SoFiA spectral-line source-finding software developed by \cite{2015MNRAS.448.1922S} is being used on neutral hydrogen (HI) emission-line pilot survey data, and the FLASHfinder line-finding software developed by \cite{allison2012} is being used to search for HI and OH absorption lines on sightlines to bright radio continuum sources in ASKAP spectral-line cubes.

\section{DATA ARCHIVING AND ACCESS}
\label{sec:CASDA}

Science data products produced by the ASKAPsoft pipeline are archived and made available to the astronomy community by the CSIRO ASKAP Science Data Archive (CASDA\footnote{\url{https://research.csiro.au/casda}}). Data deposited into CASDA are made available to the survey science team (SST) members initially, and then made publicly available after quality control and validation.

CASDA is described in detail in \cite{Chapman2017} and \cite{Huynh2020}. In brief, CASDA is implemented across two data centres, Pawsey and the CSIRO data centre in Canberra. The CSIRO data centre runs the CSIRO data access portal, an enterprise-wide system that archives and provides access to data across many areas of CSIRO research, including ASKAP/CASDA. The ASKAP data are deposited on to tape and disk storage at Pawsey, with storage managed by the next-generation archive system (NGAS) \citep{Wu2013}. Functions such as metadata search and authentication are implemented at the Canberra data centre, while functions that need to be close to the data, such as data deposit and data access, are implemented at Pawsey.

At full operations the ingest data rate of ASKAP is expected to be approximately \SI{200}{\tera\byte\per\day}, or more than \SI{70}{\peta\byte\per\year}, for \SI{100}{\percent} duty cycle. Given this extremely high data rate, the full resolution uncalibrated visibilities are not stored. The data products archived and served by CASDA are:
\begin{itemize}
\item calibrated visibilities (continuum resolution);
\item images and cubes, including intermediate imaging products such as clean models;
\item catalogues; and
\item metadata from ASKAPsoft and quality assessment.
\end{itemize}
In addition, derived data products such as moment maps and spectra of detected sources will also be stored in CASDA, once they are produced by ASKAPsoft (planned in future development).  Given \SIrange{15}{20}{\tera\byte\per\day} of expected science data products from ASKAPsoft we expect to have more than \SI{5}{\peta\byte\per\year} of data flowing into CASDA during full operations. CASDA has a current allocation of \SI{10}{\peta\byte} of long term tape storage ($+\SI{10}{\peta\byte}$ for redundancy) at Pawsey.

CASDA data search and access is possible through both the data access portal web user interface\footnote{\url{https://data.csiro.au}} and virtual observatory (VO) services. CASDA implements a wide suite of VO services to maximise the usability and interoperability of ASKAP data products. For example, the table access protocol (TAP) can be used to search for specific observations via an application such as TOPCAT \citep{2005ASPC..347...29T}. Uploaded catalogues can be filtered for desired radio sources using the astronomical data query language (ADQL). Python scripts can be used for automated data discovery and retrieval, including the production and downloading of cube cutouts\footnote{\url{https://github.com/csiro-rds/casda-samples}}. A CASDA module has also recently been added to the astropy astroquery\footnote{\url{https://astroquery.readthedocs.io}} package.

\section{PERFORMANCE MEASUREMENTS}
\label{sec:performance}

In this section we provide a broad overview of measured key performance parameters. Detailed analysis of ASKAP pilot surveys will yield further performance information that will be summarised in future papers.

\subsection{Sensitivity}

The sensitivity of each ASKAP antenna depends on the physical area $A$ of the reflector aperture and the effective system temperature $T_\text{sys}/\eta$ achieved by the receiver as installed on the reflector.  Here  $T_\text{sys}$ is the system equivalent noise temperature and $\eta$ is the antenna efficiency. 

\begin{figure}
\begin{center}
\includegraphics[width=\columnwidth]{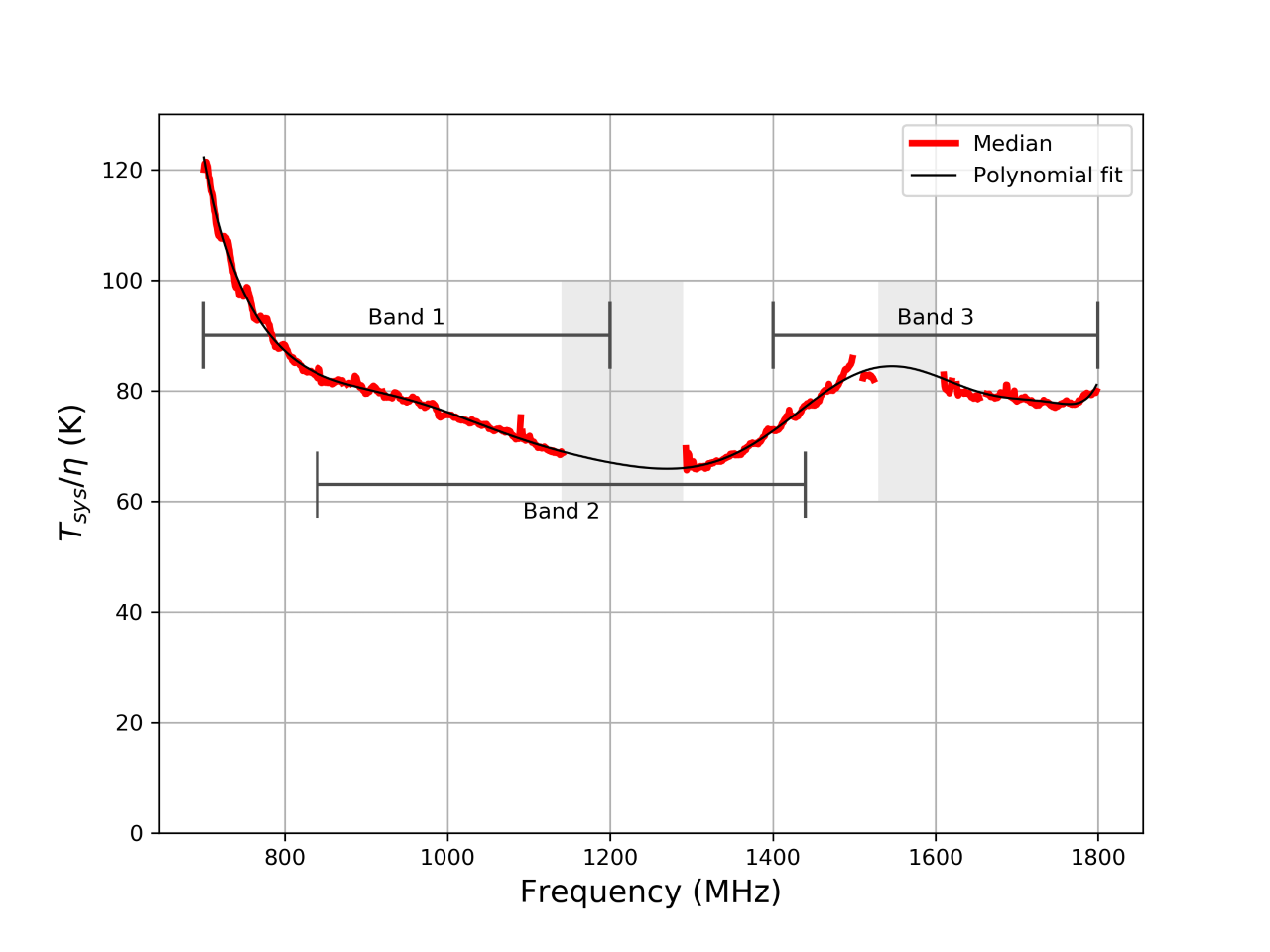}
\caption{Effective system temperature $T_\text{sys}/\eta$ across the ASKAP band. The median value over all antennas is plotted (red) for a beam close to the antennas' boresight. The black curve is a polynomial fitted to the spectrum.  Sensitivity estimates in the shaded frequency ranges are difficult because of persistent RFI.  The ASKAP operating bands are also shown.}
\label{fig:sefdSpectrum}
\end{center}
\end{figure}

Both $T_\text{sys}$ and $\eta$ vary with beamformer weights for a PAF receiver, so we report results for beamformer weights that maximise sensitivity (as used in operation; see Section~\ref{sec:beamforming}).  Figure~\ref{fig:sefdSpectrum} gives $T_\text{sys}/\eta$ across the ASKAP band measured with observations of PKS B1934$-$638 using the same technique as \citet{mcconnell_2016}. We use \SI{75}{\kelvin} as a nominal value of $T_\text{sys}/\eta$ when summarising ASKAP performance. 

The midband $T_\text{sys}$ of an ASKAP PAF beam, excluding spillover contribution from the ground, is \SI{50}{\kelvin} \citep{Chippendale2016}.  This measurement was made with the PAF on the ground pointing at the zenith, but using the same MaxSNR beamformer weights as are applied when the PAF is installed on the reflector. 

Astronomers characterise sensitivity of each antenna plus receiver system in terms of the system equivalent flux density (SEFD) defined by
\begin{equation}
\label{eq:sefd}
\text{SEFD} = \frac{2kT_\text{sys}}{\eta A}.
\end{equation}
This is the flux an unpolarised point source must have to yield a signal-to-noise ratio of one at the output of a given beam.  The factor of two appears in \eqref{eq:sefd} because a single-polarisation beam can only collect half of the total incident flux from an unpolarised source \citep{Wrobel1999}. ASKAP's nominal $T_\text{sys}/\eta$ of \SI{75}{\kelvin} corresponds to an SEFD of approximately \SI{1800}{\jansky} for a single ASKAP antenna.  

The sensitivity of the telescope can also be defined by the effective area divided by the system temperature
\begin{equation}
S(\Omega) = \frac{A_\text{e}}{T_\text{sys}}.
\end{equation}

This is generally a function of direction $\Omega$ within the field of view and antenna efficiency $\eta$ achieved together with the PAF receiver. The sensitivity of the full ASKAP array evaluates to \SI{54}{\meter\squared\per\kelvin} for a beam close to the boresight.

\subsection{Field of view}

ASKAP's field of view is set primarily by the size of the array of receptors at the focal plane and the size and focal ratio of the reflector.  It is also modified by the aperture illumination, which is in turn controlled by the beamformer weights.  

\begin{figure}
\begin{center}

\includegraphics[width=\columnwidth]{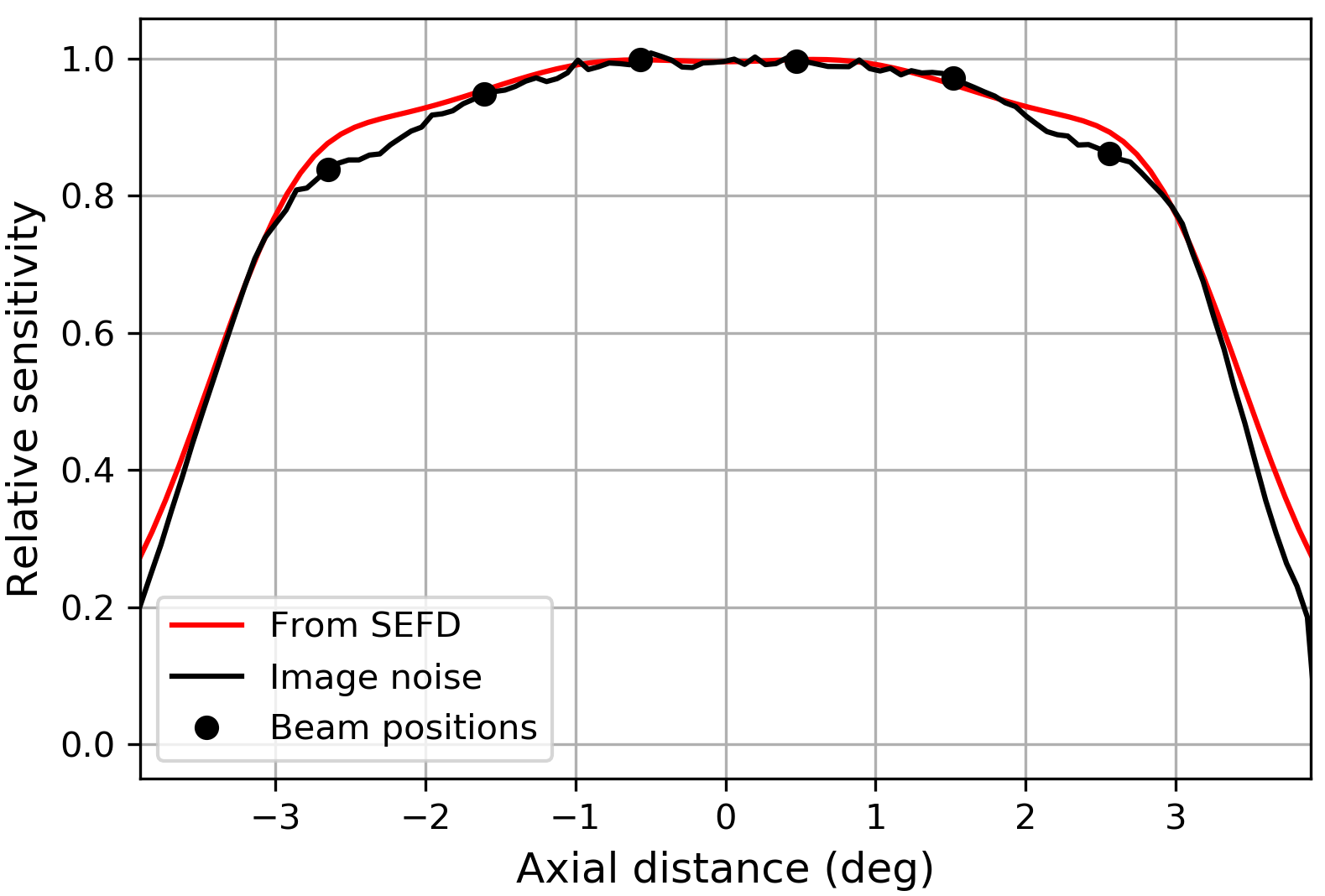}
\caption{The sensitivity profile over the field of view. The black line traces the observed sensitivity (see text), and the red line shows an analytic approximation of the sensitivity estimated from SEFD observations. These data were obtained from observations using the {\tt square\_6x6} footprint with a beam pitch of \ang{1.05} at a centre frequency of 888\,MHz.}

\label{fig:fovProfile}
\end{center}
\end{figure}

A measurement of the field of view  with maxSNR weights was made using the ASKAP-12 array and reported in \cite{McConnell_2017_Survey}. More recently, that measurement was compared with the observed variation of image noise over the field of view. Figure~\ref{fig:fovProfile} shows this comparison in one dimension as an east-west profile of image sensitivity across the median of a sample containing 44 36-beam fields. The observations were made with the {\tt square\_6x6} footprint using a beam spacing of \ang{1.05} at a centre frequency of \SI{888}{\mega\hertz}. Relative sensitivity is calculated as the reciprocal of the image rms, divided by its maximum. Sensitivity is maximum near the centre of the field of view and is reduced by less than \SI{20}{\percent} at the edge. 

These measurements show that ASKAP achieves its design specification of a \SI{30}{\deg\squared} field of view at \SI{850}{\mega\hertz}.  The field of view reduces to \SI{15}{\deg\squared} at \SI{1700}{\mega\hertz} due to the current limit of 36 dual-polarisation beams.  The field of view at the high end of the band may be increased to \SI{30}{\deg\squared} in the future by adding or reconfiguring digital hardware to process more than 36 beams. 

\subsection{Survey speed}

As seen in Figure~\ref{fig:fovProfile} the shape of ASKAP's field of view is very different from that of a dish with a single feed, which is approximately Gaussian. The traditional survey speed figure of merit \eqref{eq:ssfom} should not be calculated using the half power field of view.  When the sensitivity varies significantly over the field of view, \eqref{eq:ssfom} should be viewed as a point estimate that is integrated over the field of view to yield the survey speed \citep{5651120}
\begin{equation}
\label{eq:ss}
\text{SS} = \int S^2(\Omega)d\Omega.
\end{equation}
Equivalently, \eqref{eq:ssfom} may be used with the maximum sensitivity $S_\text{max}$ and a survey speed weighted field of view 
\begin{equation}
\label{eq:foveff}
\Omega_{\text{FoV}} = \frac{1}{S_\text{max}^2}\int S^2(\Omega)d\Omega.
\end{equation}
All ASKAP survey speed and field of view numbers are calculated by \eqref{eq:ss} and \eqref{eq:foveff} in this paper.

Using this definition the survey speed of ASKAP is currently \SI[per-mode=reciprocal]{91400}{\meter\tothe{4}\deg\squared\per\kelvin\squared} at \SI{800}{\mega\hertz} with a smooth reduction to \SI[per-mode=reciprocal]{44200}{\meter\tothe{4}\deg\squared\per\kelvin\squared} at \SI{1700}{\mega\hertz}. It may be possible to double survey speed at the high end of the band in future by processing more than 36 dual-polarisation beams. However, with current hardware the extra beams would come at the expense of bandwidth.

As a point of comparison, evaluating \eqref{eq:ss} for a Gaussian beam yields a survey speed of $0.721\Omega_{G}S_\text{max}^2$, where $\Omega_{G}$ is the half power field of view of the Gaussian beam. Using this measure, the survey speed of ASKAP at \SI{1.4}{\giga\hertz} is 245 times faster than the Parkes \SI{64}{\metre} dish with a cryogenic single-pixel receiver having $T_\text{sys}/\eta$ of \SI{40}{\kelvin} \citep{2020PASA...37...12H}.

\section{RADIO FREQUENCY INTERFERENCE}
\label{sec:rfi}

ASKAP's location on a protected radio quiet site makes it possible to observe in parts of the spectrum that are no longer usable at other radio telescopes. In particular, the band from \SIrange{700}{1080}{\mega\hertz} is almost entirely free of persistent RFI. Occasionally, atmospheric conditions can lead to ducting of distant mobile communications signals, but this can be predicted to some extent based on meteorological data \citep{2018SPIE10704E..2SI} and impacts relatively few observations.

Satellites and aircraft have been the most common sources of RFI in observations to date \citep{RFI2016}. Aircraft automatic dependent surveillance broadcast (ADS-B) transponders transmit at 1090\,MHz and are visible when flights pass over the site. This signal is strong, but has a relatively low duty cycle so future real-time interference mitigation methods may be able to remove it.

Satellite navigation systems are visible between approximately \SIrange{1150}{1300}{\mega\hertz}. Occupancy of interference approaches \SI{100}{\percent} at several places within this range and typical flagging algorithms leave very little usable data. The band from \SIrange{1500}{1620}{\mega\hertz} also contains significant satellite interference, with up to \SI{70}{\percent} occupancy. 

In future, adaptive beamforming could be used as a form of RFI mitigation on ASKAP, by placing nulls at the locations of transmitting satellites \citep{Black2015, 7032330, Hellbourg2016}. This would involve changing the beams during an observation, which would need main-beam constraints and corrections for pattern rumble bias \citep{Jeffs2008_Signal} to avoid invalidating array calibration solutions. 

\section{TELESCOPE OPERATIONS}
\label{sec:ops}

In a departure from other Australia Telescope National Facility (ATNF) telescopes, \SI{75}{\percent} of observing time on ASKAP has been pre-allocated to several international survey science teams with five-year observing plans. The remaining \SI{25}{\percent} of available time will be allocated to guest science projects, subject to peer review through the ATNF time allocation committee (TAC). ASKAP will be operated by CSIRO staff and astronomers will not interact directly with the telescope. Survey plans and guest science proposals will be converted into scheduling blocks that are observed, calibrated and imaged as part of operations.  Science-ready images will be made available on CASDA. This means that calibration, imaging, archiving and initial quality control all fall within the domain of the observatory.

Multi-year surveys are commonly considered legacy projects at other observatories and are usually done after the telescope has been operational for many years. Since ASKAP is conducting large-scale survey projects as one of its first activities, we devised a sequence of steps to test the telescope's readiness. See Table \ref{tab:timeline} for a timeline showing significant milestones on the road to full survey operations. Descriptions of the various phases can be found below.

\begin{table}
\caption{ASKAP science operations timeline}
\label{tab:timeline}
\begin{tabular}{ l|l }
Dec 2016 & Early science begins \\
July 2018 & Early science ends \\
Feb 2019 & Full array operational \\
July 2019 & Pilot Surveys Phase I observing begins \\
May 2020 & Pilot Surveys Phase I observing ends \\
Q1 2021 & Pilot Surveys Phase II planned start \\
Q3 2021 & Pilot Surveys Phase II planned end \\
Q4 2021 & Full Surveys planned start \\
\end{tabular}
\end{table}

One of the most important lessons from the early science program was the need to maintain close engagement with the science community. Taking on the responsibility of providing science-ready data products means that the observatory must know exactly what is required by its users. We developed a commissioning team that included representatives from the observatory and the survey science teams working together.

Due to the high data rate, we do not plan to keep all visibilities long term, although we will archive visibilities that have been calibrated and averaged to 1\,MHz frequency resolution. We have also found it useful to keep raw calibration data, which accumulates at a more modest rate than the science observations. Outputs from the imaging pipeline are tested for quality and reprocessed if necessary, with raw data being deleted to make way for the next project once the image products have been archived.

\subsection{Early science}

The ASKAP early science program started with 12 antennas equipped with Mk~II PAFs in 2016. We planned to run two survey projects on behalf of all the survey science teams, but found that this was not feasible so early in the life of the instrument. Instead we devoted time to smaller test observations for each science team, using the data to test ASKAP's image processing software and identify problems with the visibility data. These test observations were separated by commissioning periods of a few weeks duration, where antennas were added to the array and features necessary for full operations were implemented.

During this time, various existing software tools were used to cross-check and supplement ASKAP's custom processing software as it was developed. Eventually we were able to complete a modest continuum survey known as the cosmology survey, which is available on CASDA.

Many publications arose from the early science program, as a result of identifying projects that the telescope could excel at with only a third of its collecting area. These include a continuum survey and detection of cold gas outflows from the small magellanic cloud \citep{2018NatAs...2..901M,2019MNRAS.490.1202J}, studies of neutral hydrogen in nearby galaxy groups \citep{2019MNRAS.482.3591R,2019MNRAS.tmp...27L,elagali2019,2019MNRAS.488.5352K,for2019}, studies of absorption lines \citep{2019MNRAS.489.4926G,allison2020}, continuum observations of the galaxy and mass assembly (GAMA) G23 field \citep{2019PASA...36...24L}, searches for transient and variable sources \citep{bhandari2018}, and a few targeted studies of individual radio galaxies  \citep[e.g.][]{2020PASA...37...13S}, including the polarisation characteristics of Centaurus A \citep{Anderson2018}. ASKAP was also the first telescope to localise a non-repeating fast radio burst to its host galaxy \citep{2019Sci...365..565B}.

\subsection{Pilot surveys}
Integration of all 36 antennas with Mk~II PAFs and digital backends was completed in February 2019. Soon after, we commenced a series of test observations designed to verify all the modes required for survey operations with the full array. This involved relatively simple imaging of representative fields. On 15 July 2019, we began a pilot survey program to demonstrate readiness for extended science operations, and to highlight places of improvement for sustainable full operations. Pilot surveys involved observing small parts (100 hours per project) of the larger survey plans developed by the survey science teams. The pilot survey concept arose as a natural extension of the early science program. Reaching peak operational efficiency is a process that takes time and these pilot surveys provide useful science data while improvements and updates to the telescope's core systems are ongoing.

There are several goals associated with pilot surveys:
\begin{itemize}
\item Provide representative data to each SST
\item Test SST survey strategies
\item Push operational limits to find pressure points 
\item Develop operational plans for full surveys
\item Assess and improve processing pipelines
\end{itemize}

Observations for phase I of these pilot surveys were completed on 14 May 2020, although processing is expected to take several additional months. Upon completion of pilot surveys phase I observations we entered a 6-month consolidation period, in which the priority for telescope access shifted back to development, testing and maintenance. This provided time to consider lessons learned and deploy various software and firmware updates, while processing the data backlog. The primary goal of pilot surveys phase I was to demonstrate production of science-ready data products. Phase II of pilot surveys will focus on optimising commensality and overall data processing efficiency. We are not currently planning a third phase of pilot surveys, but this may be considered if significant issues arise from phase II.

Full survey operations are expected to begin by the end of 2021 and ramp up in efficiency by 10 percent each year, reaching a limit of 70 percent efficiency in 2024.

\subsection{Specifying observations}
The ASKAP control system implements a flexible observation management system built around the concept of observing procedures and scheduling blocks. Procedures are implemented as Python scripts using a library of functions that provide access to the hardware. The simplest (and most commonly used) procedure simply tracks a source position specified in equatorial coordinates. More complicated procedures have been developed for bandpass calibration and holography observations where on-the-fly calculation of offset tracking positions is required.

A scheduling block is specified as a list of parameters and an associated procedure. The list of parameters currently includes source coordinates and system configuration options. Eventually, we plan to expand this list to include data processing parameters, which will allow automated execution of the image processing pipeline upon completion of a scheduling block. All-sky surveys require roughly \num{1000} individual fields, but we have found in practice that scheduling each one individually (rather than specifying multiple fields in a single block) is the most effective approach, since it provides a level of fault tolerance and allows for more predictable timing.

\subsection{Observation scheduling}
Over the course of the pilot survey period, various scheduling requirements have been incorporated to meet the needs of different survey science cases. Developing a working model for efficient and effective scheduling is an ongoing process that depends on the level of system stability, robustness and automation. For now, scheduling is carried out in one of two modes: (i) transit-centric, or (ii) target of opportunity. The first approach applies largely to long tracks, where the requirement is to schedule a long observation while the target is above the horizon and pair it automatically with a bandpass calibration observation. The second approach is more suitable for short tracks such as the \SI{15}{\minute} observations used for the Rapid ASKAP Continuum Survey. It selects an appropriate target from a specified list based on a series of constraints (e.g. above the horizon, solar distance, lunar distance, hour angle, etc.). The scheduling algorithms are being refined and evolved as we gain experience with the requirements of survey projects and ensure the system is able to handle these in a way that is as automated and efficient as possible. 

Our current intention is to aim for autonomous scheduling, in that the scheduling tool will detect system status and environment and have access to the pool of survey observations as well as survey constraints and historical observational data. It would then use this information to decide on the next observation to schedule and ensure the telescope is in the right state to complete this observation. The scheduling tool would also provide a forecast of the coming observations if the system status remained the same, but it would maintain the flexibility to adapt as necessary based on the changing system and environment. We intend to use the next period of pilot survey observations to prototype this approach and determine its feasibility for full surveys, ultimately maximising automation as much as possible which will help us to maximise efficiency gains. Recent developments by the ASKAP team to enable automated array start-ups based on the input parset have been a key milestone towards this vision for scheduling, and encourage us that our goal of autonomous scheduling may be realised for ASKAP.

\begin{figure*}[t!]
    \centering
    \includegraphics[height=0.41\textwidth]{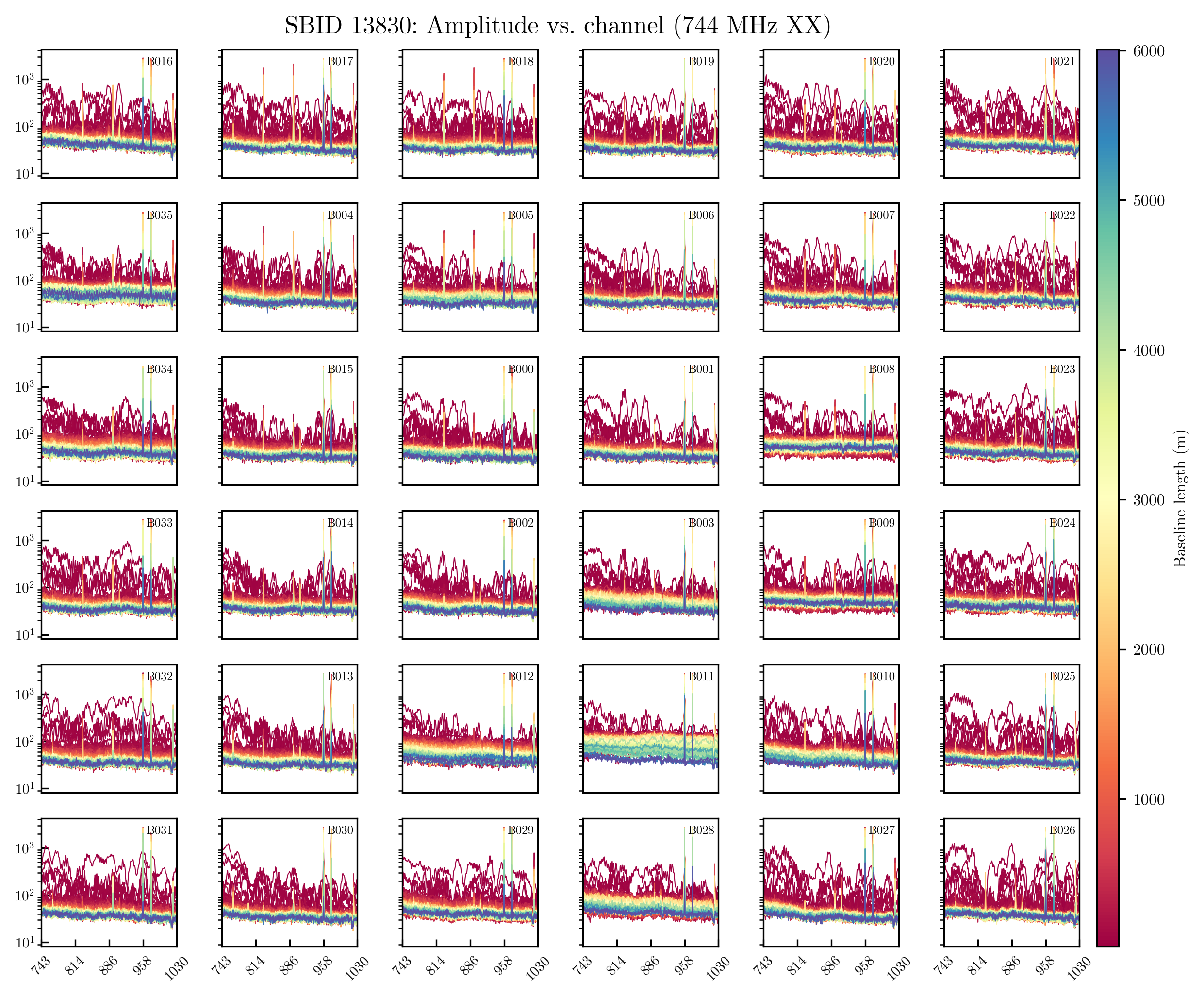}
    \includegraphics[height=0.41\textwidth]{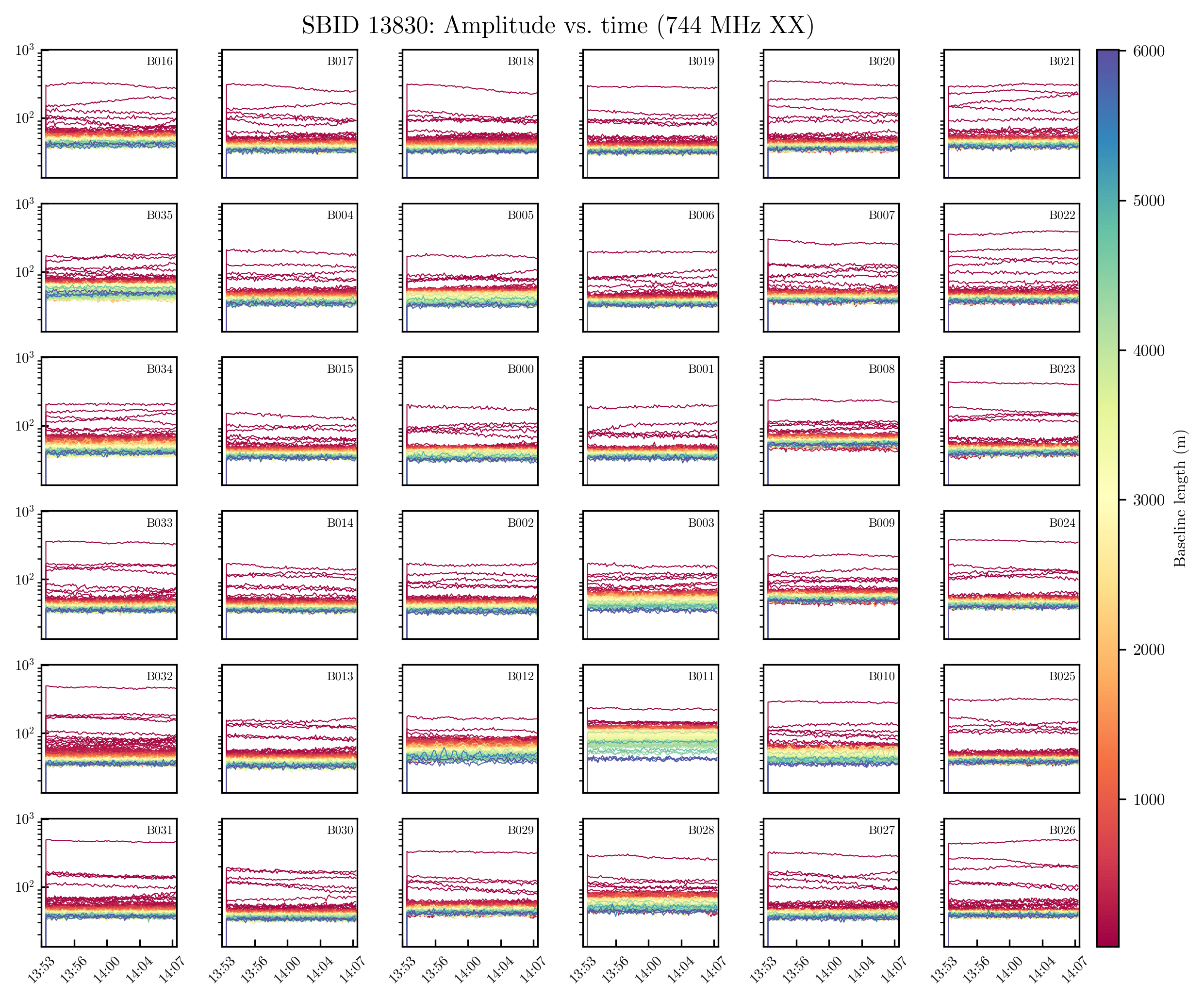}
    \includegraphics[height=0.45\textwidth]{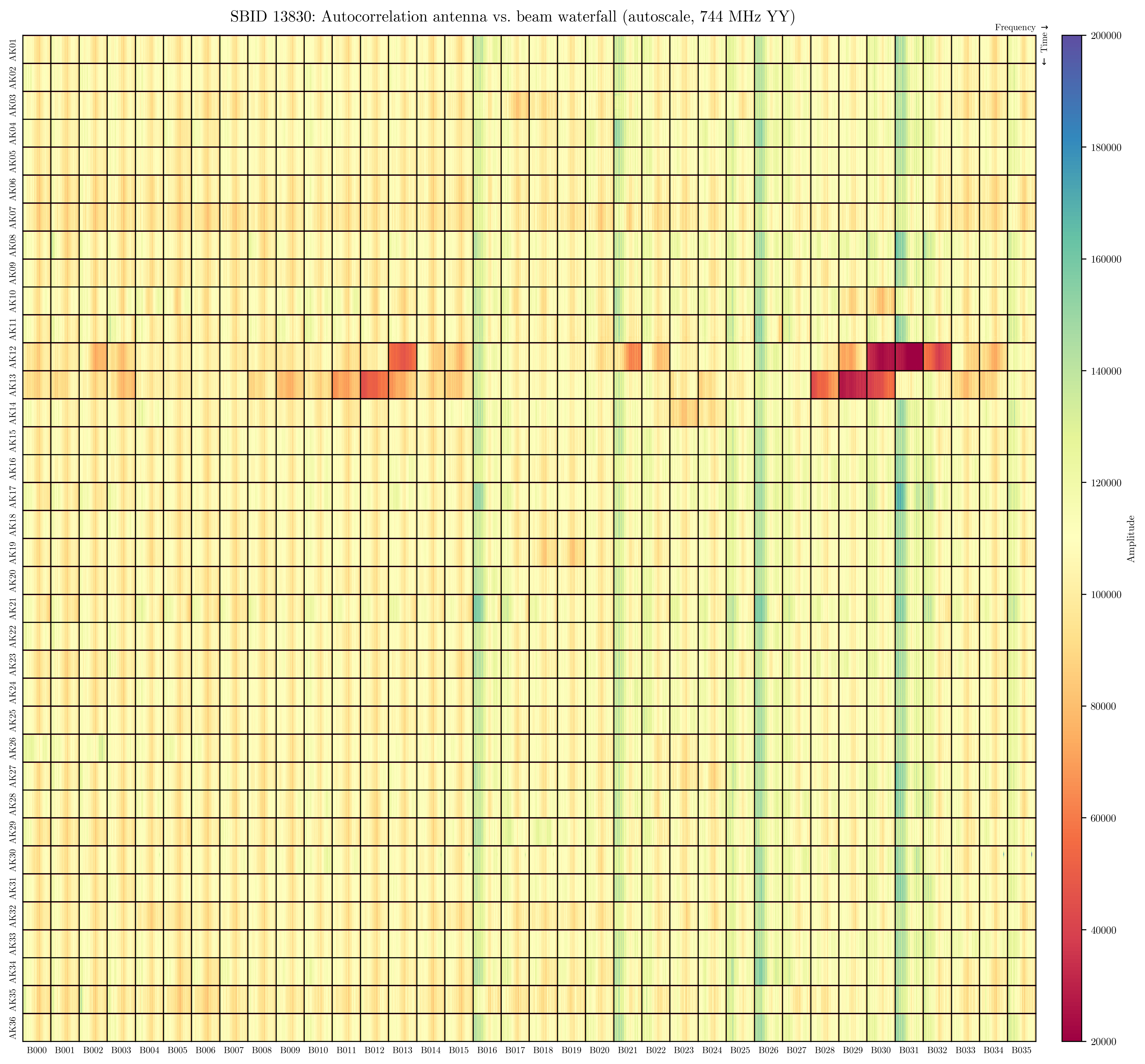}
    \includegraphics[height=0.45\textwidth]{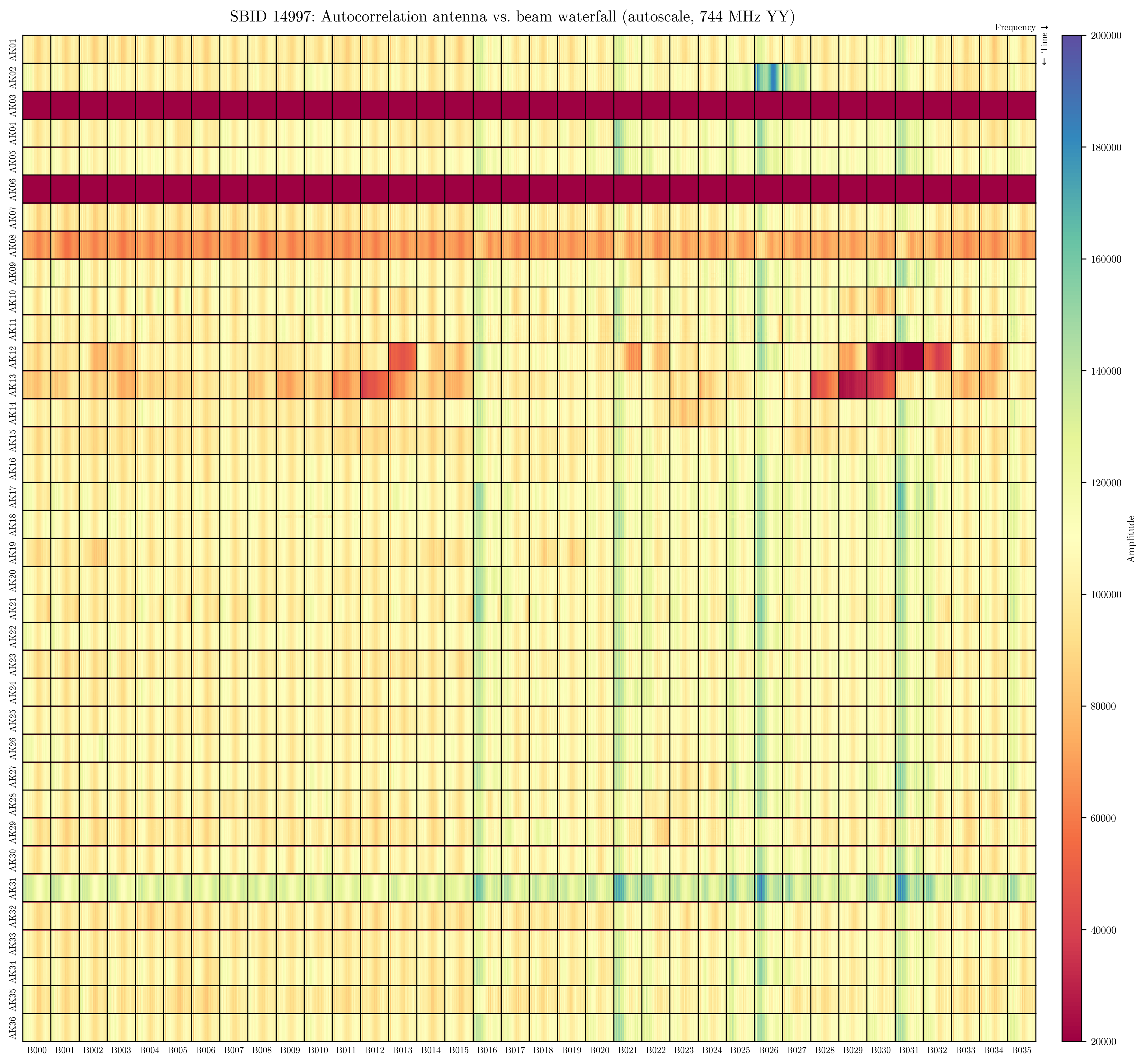}    
    \caption{Example ASKAP raw data diagnostic plots for Scheduling Block IDs 13830 and 14997, demonstrating visualisations of correlation amplitude vs. frequency (\emph{top-left}), correlation amplitude vs. time (\emph{top-right}), and time-frequency waterfall of autocorrelation for all beams and antennas (\emph{bottom-left}) for SBID 13830. For comparison, we show the same diagnostic plot for SBID 14997. These diagnostics give us a direct insight into the recorded data prior to processing, and are used to inform overall data quality assessment. In the top two plots, each line represents a baseline colour-coded by baseline length (where red is shortest and purple is longest). In the autocorrelation waterfall plots, orange and red sections indicate beams on particular antennas that may be affected by faulty PAF dominos (in this case, the same bad YY elements are seen on ak12/ak13 for both observations), while the four vertical green stripes across entire beams trace the less-sensitive beams at the corners of the field of view. Dark red stripes indicate a missing antenna, and any antennas with visually-outlying amplitudes are likely to have scaling issues. The vertical spikes seen in amplitude vs. channel and the baseline-average waterfall plot are RFI.}
    \label{fig:diagnostics}
\end{figure*}

\subsection{Data quality diagnostics}
ASKAP presents an incredible big-data challenge, marking a noticeable shift in the ways astronomical data can be managed and interacted with. It is quite difficult to gauge data quality in meaningful ways when data sizes are tens of terabytes, including complex multi-dimensional axes across time, frequency, antennas, beams and polarisations. While working within the constraints of no longer being able to visualise all data, effort has been invested to ensure that certain aspects of the data can be visualised and used for diagnostic purposes, particularly to ensure that the system is working as expected in a timely way. 

For the raw data, a series of plots have been designed and implemented to allow insight into the recorded data regardless of their size. Though currently the amount of time for processing does scale with data size (improving this is the subject of ongoing investigation), we are able to obtain these raw data diagnostic plots automatically in close to real time for short observations, which has been important for determining system status. These plots are extensions of standard astronomical visualisations of the visibilities, but necessarily more complex due to the amount of data needing to be represented. The exact form of ASKAP diagnostics is under development as part of the continuing transition between commissioning and operations. The ultimate goal is to assess data quality in an automated way, making use of anomaly detection algorithms and other machine learning techniques where feasible, and this is the subject of further investigation. Currently, diagnostic plots are visually inspected where necessary and allow human pattern recognition to spot anomalous conditions. 

Some examples of diagnostic plots for a continuum dataset (SBID 13830 and 14997) are shown in Figure~\ref{fig:diagnostics}, demonstrating visualisation of (correlation) amplitude vs. time, amplitude vs. frequency channel, and time-frequency autocorrelation waterfall visualisations that combine the two. In particular, the last two plots are an attempt to visualise the contents of the entire dataset at once, showing a time-frequency waterfall plot of autocorrelation amplitude for each antenna and beam within the array. This is extremely effective at highlighting particular hardware issues, but less so for issues that depend on cross-correlation baselines. Although in this visualisation each sub-plot is too small to be used for quantitative analysis, broad trends or hardware faults stand out clearly against a background of relatively uniform behaviour. Conversely, the data used to generate these visualisations is extremely rich in diagnostic information, and will likely form a useful input dataset for either classification or machine-learning algorithms.  

Extensive quality control is also done post-imaging. Diagnostics are being developed by the science teams and will be incorporated into the standard observatory-driven processing pipelines over time. CASDA includes a validation stage that must occur prior to data release. Currently this requires human inspection of diagnostic information and selection of a three-tier quality grade. As diagnostic algorithms improve, we will seek to automate this process as much as possible.

\section{Future enhancements}
\label{sec:future}
One of the first upgrades to ASKAP's systems could be a coherent transient detection module that improves the sensitivity of fast radio burst searches. This would provide access to an alternative search data stream from the correlator rather than the beamformers, increasing the sensitivity to fast transients by a factor of six and providing near-instantaneous arcsecond localisation.

Another possible improvement is the addition of a tied-array module to create phased array beams from all 36 antennas. This would allow the full array to be used for very long baseline interferometry, pulsar timing experiments, and the search for extraterrestrial intelligence.

Satellite interference breaks up ASKAP bands 2 and 3 into clean sections smaller than the telescope's instantaneous bandwidth, so it could be beneficial to split the observing band itself into two or more disconnected spectral windows. Although this is not currently possible, we are planning to develop such a split-band mode through software and control system upgrades. This would only be possible within one of the band-limiting filters (see Table~\ref{tab:drxbands}). Split-band mode would also allow simultaneous observations of widely-separated spectral lines and may also be beneficial for determining spectral indices and rotation measures.

Another possible improvement would be allowing the exchange of bandwidth for additional PAF beams. This would be particularly beneficial at the high-frequency end of ASKAP's frequency range since, above \SI{850}{\mega\hertz}, 36 dual-polarisation beams cannot fully sample the \SI{30}{\deg\squared} field of view of the PAF \citep{5651120, McConnell_2017_Survey}.  At \SI{1700}{\mega\hertz}, the field of view is currently limited to \SI{15}{\deg\squared}.  Processing more beams may therefore double the current field of view at \SI{1700}{\mega\hertz}.  Additional beams would also reduce the need for interleaved observations, but further analysis of the effects of correlated noise between closely spaced beams may be needed to understand the exact benefit \citep{Serra2015, McConnell_2017_Estimating, McConnell_2017_Survey}.

An additional 48\,MHz correlator block could be installed to bring the instantaneous bandwidth to 336\,MHz. To achieve the full 384\,MHz bandwidth a full complement of Redback modules for the beamformer and correlator are needed. The additional data rate would be difficult to support on the current supercomputing platform and would diminish the amount of spare hardware modules, so any such upgrade needs careful consideration.  Alternatively, upgrading the correlator with new hardware would release sufficient hardware to expand the beamformer (since the same physical boards are currently used in both subsystems).  An option for a new correlator is the Gemini \citep{8104976} board (currently in development) which is a good match to the current Redback modules. An upgraded correlator would allow increased frequency resolution across the band to \SI{9.2}{\kilo\hertz} or \SI{4.6}{\kilo\hertz}. Implementing tied-array beams would also be easier with new correlator hardware, and the extra resources would likely make an increased number of tied array beams possible. The upgrade would also allow full bandwidth ``standard'' resolution observing with simultaneous zoom bands, increased DM range for transient searches and integration of transient searches within the correlator hardware.

Increased sensitivity could be obtained with an increase in antenna diameter. The diameter of the existing antennas could be increased to 15\,m or more while still maintaining ground clearance at minimum elevation. Initial work shows the existing counterweights would be insufficient to support a heavier reflector and a lightweight design would be needed. The implications of additional wind loading on the drives would have to be carefully considered as well. At 15\,m, the ASKAP dish has an $f/D$ similar to Parkes, which has been tested successfully with an ASKAP PAF \citep{PAFonParkes, Chippendale2017, Deng2017, Reynolds2017}. With the additional collecting area, the instantaneous sensitivity of ASKAP increases by \SI{56}{\percent}. Survey speed is also increased, but more beams may be required to compensate for the reduced primary beam size.

Sensitivity could also be improved significantly by updating the room-temperature LNA design with new transistors \citep{Shaw2015, Weinreb2019} or by scaling up the manufacturability and affordability of cryogenic PAF technology like that under development for the Parkes 64\,m telescope \citep{Dunning2016, Dunning2019}, as mentioned earlier in Section~\ref{sec:surveytelescope}.

\section{CONCLUSION}
\label{sec:conclusion}

ASKAP is one of the first radio telescopes to employ PAF technology, giving the telescope a wide field of view, rapid survey speed and excellent polarisation characteristics. This has been demonstrated in several pilot surveys and observatory projects that will be published elsewhere. We have shown that PAF technology is a powerful solution for wide-field radio astronomy. So far we have used ASKAP's PAFs in familiar ways, creating beams that are treated independently and kept fixed for the duration of an observation. PAFs offer much more power and flexibility that we hope to explore in future, including adaptive interference mitigation.

ASKAP's survey science projects will occupy most of the next few years of observing time and all data will be released to the public after quality control. At the present time, ASKAP pushes the boundaries of signal processing technology and requires a combination of techniques to convert the torrent of raw data into calibrated, science-ready images that can be archived permanently. As computing technology improves, we expect that it will be possible to extract even more information from the data. One important design consideration for other radio telescopes intending to deliver science-ready output is to budget for reprocessing. Radio astronomy imaging often requires iteration to perfect, which can be very resource-intensive.

As a precursor telescope for the Square Kilometre Array, ASKAP has demonstrated several key lessons. It is important to consider hardware, firmware and software development as connected activities and integrate subsystems as early as possible. Design aspects may need to be changed on the basis of experience in the field, so it is important to begin science-oriented test observations as soon as possible. Commissioning a new radio telescope begins during subsystem tests and continues well into operations. Automation has played a key role in smoothing the transition from commissioning to operations, and will continue to be a core value applied to ASKAP operations. Based on our experience, automation acts as a catalyst for improving all aspects of system behaviour, identifying issues and fixing where possible or building robust work-around procedures otherwise. This is especially important for a telescope designed to spend a majority of its time conducting systematic surveys.

We have found with ASKAP that effective ongoing communications between the design, construction, operations, science and management teams is critical to achieving the goal of delivering a productive radio telescope, and expect this to be one of the biggest challenges to solve for the next generation of large-scale science facilities to come.

\begin{acknowledgements}
The Australian SKA Pathfinder is part of the Australia Telescope National Facility which is managed by CSIRO. Operation of ASKAP is funded by the Australian Government with support from the National Collaborative Research Infrastructure Strategy. ASKAP uses the resources of the Pawsey Supercomputing Centre. Establishment of ASKAP, the Murchison Radio-astronomy Observatory and the Pawsey Supercomputing Centre are initiatives of the Australian Government, with support from the Government of Western Australia and the Science and Industry Endowment Fund. We acknowledge the Wajarri Yamatji as the traditional owners of the Observatory site.
\end{acknowledgements}

\bibliographystyle{pasa-mnras}
\bibliography{IEEEabrv,askap}

\begin{thebibliography}{}
\makeatletter
\relax
\def\mn@urlcharsother{\let\do\@makeother \do\$\do\&\do\#\do\^\do\_\do\%\do\~}
\definecolor{darkblue}{rgb}{0,0,0.597656}
\def\mndoi{\begingroup\mn@urlcharsother \@ifnextchar [ {\mndoi@} {\mndoi@[]}}
\def\mndoi@[#1]#2{\def\@tempa{#1}\ifx\@tempa\@empty \href
  {http://dx.doi.org/#2} {\textcolor{darkblue}{doi:#2}}\else \href
  {http://dx.doi.org/#2} {\textcolor{darkblue}{#1}}\fi \endgroup}
\def\mn@eprint#1#2{\mn@eprint@#1:#2::\@nil}
\def\mn@eprint@arXiv#1{\href {http://arxiv.org/abs/#1} {{\tt arXiv:#1}}}
\def\mn@eprint@dblp#1{\href {http://dblp.uni-trier.de/rec/bibtex/#1.xml}
  {dblp:#1}}
\def\mn@eprint@#1:#2:#3:#4\@nil{\def\@tempa {#1}\def\@tempb {#2}\def\@tempc
  {#3}\ifx \@tempc \@empty \let \@tempc \@tempb \let \@tempb \@tempa \fi \ifx
  \@tempb \@empty \def\@tempb {arXiv}\fi \@ifundefined
  {mn@eprint@\@tempb}{\@tempb:\@tempc}{\expandafter \expandafter \csname
  mn@eprint@\@tempb\endcsname \expandafter{\@tempc}}}

\bibitem[\protect\citeauthoryear{ACMA}{ACMA}{2014}]{acma2007}
ACMA 2014, Radiocommunications assignment and licensing instruction MS 32,
  Coordination of Apparatus Licensed Services Within the Mid West Radio Quiet
  Zone Western Australia,
  \url{https://www.acma.gov.au/publications/2019-08/publication/rali-ms32-mid-west-radio-quiet-zone}

\bibitem[\protect\citeauthoryear{{Abeywickrema}, {Allen}, {Ardern},
  {Schinckel}, {Leitch}, {Wilson}  \& {Beresford}}{{Abeywickrema}
  et~al.}{2013}]{Abeywickrema2013}
{Abeywickrema} S.,  {Allen} G.,  {Ardern} K.,  {Schinckel} A.,  {Leitch} A.,
  {Wilson} C.,   {Beresford} R.,  2013, in 2013 Asia-Pacific Symposium on
  Electromagnetic Compatibility (APEMC). pp~1--4,
  \mndoi{10.1109/APEMC.2013.7360649}

\bibitem[\protect\citeauthoryear{{Allison}, {Sadler}  \& {Whiting}}{{Allison}
  et~al.}{2012}]{allison2012}
{Allison} J.~R.,  {Sadler} E.~M.,   {Whiting} M.~T.,  2012, \mndoi [\pasa]
  {10.1071/AS11040}, \href
  {https://ui.adsabs.harvard.edu/abs/2012PASA...29..221A} {29, 221}

\bibitem[\protect\citeauthoryear{{Allison} et~al.,}{{Allison}
  et~al.}{2015}]{10.1093/mnras/stv1532}
{Allison} J.~R.,  et~al., 2015, \mndoi [\mnras] {10.1093/mnras/stv1532}, \href
  {https://ui.adsabs.harvard.edu/abs/2015MNRAS.453.1249A} {453, 1249}

\bibitem[\protect\citeauthoryear{{Allison} et~al.,}{{Allison}
  et~al.}{2017}]{2017MNRAS.465.4450A}
{Allison} J.~R.,  et~al., 2017, \mndoi [\mnras] {10.1093/mnras/stw2860}, \href
  {https://ui.adsabs.harvard.edu/abs/2017MNRAS.465.4450A} {465, 4450}

\bibitem[\protect\citeauthoryear{{Allison} et~al.,}{{Allison}
  et~al.}{2020}]{allison2020}
{Allison} J.~R.,  et~al., 2020, \mndoi [\mnras] {10.1093/mnras/staa949}, \href
  {https://ui.adsabs.harvard.edu/abs/2020MNRAS.494.3627A} {494, 3627}

\bibitem[\protect\citeauthoryear{{Anderson} et~al.,}{{Anderson}
  et~al.}{2018}]{Anderson2018}
{Anderson} C.,  et~al., 2018, \mndoi [Galaxies] {10.3390/galaxies6040127},
  \href {https://ui.adsabs.harvard.edu/abs/2018Galax...6..127A} {6, 127}

\bibitem[\protect\citeauthoryear{{Applebaum}}{{Applebaum}}{1976}]{Applebaum1976}
{Applebaum} S.~P.,  1976, \mndoi [{IEEE} Trans. Antennas Propag.]
  {10.1109/TAP.1976.1141417}, \href
  {https://ui.adsabs.harvard.edu/abs/1976ITAP...24..585A} {24, 585}

\bibitem[\protect\citeauthoryear{{Bannister}, {Zackay}, {Qiu}, {James}  \&
  {Shannon}}{{Bannister} et~al.}{2019a}]{2019ascl.soft06003B}
{Bannister} K.,  {Zackay} B.,  {Qiu} H.,  {James} C.,   {Shannon} R.,  2019a,
  {{FREDDA}: A fast, real-time engine for de-dispersing amplitudes,
  astrophysics source code library, record
  \href{http://www.ascl.net/1906.003}{{\tt ascl:1906.003}}}

\bibitem[\protect\citeauthoryear{{Bannister} et~al.,}{{Bannister}
  et~al.}{2019b}]{2019Sci...365..565B}
{Bannister} K.~W.,  et~al., 2019b, \mndoi [Science] {10.1126/science.aaw5903},
  \href {https://ui.adsabs.harvard.edu/abs/2019Sci...365..565B} {365, 565}

\bibitem[\protect\citeauthoryear{{Beresford} \& {Bunton}}{{Beresford} \&
  {Bunton}}{2013}]{PAF_shielding}
{Beresford} R.,  {Bunton} J.,  2013, in 2013 Asia-Pacific Symposium on
  Electromagnetic Compatibility (APEMC). pp~1--4,
  \mndoi{10.1109/APEMC.2013.7360638}

\bibitem[\protect\citeauthoryear{{Beresford} \& {Li}}{{Beresford} \&
  {Li}}{2013}]{Beresford2013}
{Beresford} R.,  {Li} L.,  2013, in 2013 Asia-Pacific Symposium on
  Electromagnetic Compatibility (APEMC). pp~1--4,
  \mndoi{10.1109/APEMC.2013.7360637}

\bibitem[\protect\citeauthoryear{{Beresford}, {Cheng}  \&
  {Roberts}}{{Beresford} et~al.}{2017a}]{Beresford2017_low_cost}
{Beresford} R.,  {Cheng} W.,   {Roberts} P.,  2017a, in 2017 XXXIInd General
  Assembly and Scientific Symposium of the International Union of Radio Science
  (URSI GASS). pp~1--4, \mndoi{10.23919/URSIGASS.2017.8105422}

\bibitem[\protect\citeauthoryear{{Beresford} et~al.,}{{Beresford}
  et~al.}{2017b}]{Beresford2017}
{Beresford} R.,  et~al., 2017b, in 2017 XXXIInd General Assembly and Scientific
  Symposium of the International Union of Radio Science (URSI GASS). pp~1--4,
  \mndoi{10.23919/URSIGASS.2017.8105423}

\bibitem[\protect\citeauthoryear{{Beresford}, {Ferris}, {Cheng}, {Hampson},
  {Bunton}, {Chippendale}  \& {Kanapathippillai}}{{Beresford}
  et~al.}{2017c}]{RFoF}
{Beresford} R.,  {Ferris} D.,  {Cheng} W.,  {Hampson} G.,  {Bunton} J.,
  {Chippendale} A.,   {Kanapathippillai} J.,  2017c, in 2017 International
  Topical Meeting on Microwave Photonics (MWP). pp~1--4,
  \mndoi{10.1109/MWP.2017.8168641}

\bibitem[\protect\citeauthoryear{{Bhandari} et~al.,}{{Bhandari}
  et~al.}{2018}]{bhandari2018}
{Bhandari} S.,  et~al., 2018, \mndoi [\mnras] {10.1093/mnras/sty1157}, \href
  {https://ui.adsabs.harvard.edu/abs/2018MNRAS.478.1784B} {478, 1784}

\bibitem[\protect\citeauthoryear{{Bhatnagar}, {Cornwell}, {Golap}  \&
  {Uson}}{{Bhatnagar} et~al.}{2008}]{2008A&A...487..419B}
{Bhatnagar} S.,  {Cornwell} T.~J.,  {Golap} K.,   {Uson} J.~M.,  2008, \mndoi
  [\aap] {10.1051/0004-6361:20079284}, \href
  {https://ui.adsabs.harvard.edu/abs/2008A&A...487..419B} {487, 419}

\bibitem[\protect\citeauthoryear{{Bhatnagar}, {Rau}  \& {Golap}}{{Bhatnagar}
  et~al.}{2013}]{2013ApJ...770...91B}
{Bhatnagar} S.,  {Rau} U.,   {Golap} K.,  2013, \mndoi [\apj]
  {10.1088/0004-637X/770/2/91}, \href
  {https://ui.adsabs.harvard.edu/abs/2013ApJ...770...91B} {770, 91}

\bibitem[\protect\citeauthoryear{{Black}, {Jeffs}, {Warnick}, {Hellbourg}  \&
  {Chippendale}}{{Black} et~al.}{2015}]{Black2015}
{Black} R.~A.,  {Jeffs} B.~D.,  {Warnick} K.~F.,  {Hellbourg} G.,
  {Chippendale} A.,  2015, in 2015 IEEE Signal Processing and Signal Processing
  Education Workshop (SP/SPE). pp 261--266,
  \mndoi{10.1109/DSP-SPE.2015.7369563}

\bibitem[\protect\citeauthoryear{{Bowman}, {Rogers}, {Monsalve}, {Mozdzen}  \&
  {Mahesh}}{{Bowman} et~al.}{2018}]{2018Natur.555...67B}
{Bowman} J.~D.,  {Rogers} A. E.~E.,  {Monsalve} R.~A.,  {Mozdzen} T.~J.,
  {Mahesh} N.,  2018, \mndoi [Nature] {10.1038/nature25792}, \href
  {https://ui.adsabs.harvard.edu/abs/2018Natur.555...67B} {555, 67}

\bibitem[\protect\citeauthoryear{Briggs}{Briggs}{1995}]{briggs1995}
Briggs D.~S.,  1995, PhD thesis, New Mexico Institute of Mining and Technology,
  \url{http://www.aoc.nrao.edu/dissertations/dbriggs/}

\bibitem[\protect\citeauthoryear{{Brown} et~al.,}{{Brown}
  et~al.}{2014}]{Dragonfly}
{Brown} A.~J.,  et~al., 2014, in 2014 International Conference on
  Electromagnetics in Advanced Applications (ICEAA). pp 268--271,
  \mndoi{10.1109/ICEAA.2014.6903860}

\bibitem[\protect\citeauthoryear{Bunton}{Bunton}{2003a}]{Bunton2003}
Bunton J.~D.,  2003a, {SKA} Memo~40, Figure of merit for {SKA} survey speed.
\url{https://www.skatelescope.org/uploaded/51368_40_memo_Bunton.pdf}

\bibitem[\protect\citeauthoryear{{Bunton}}{{Bunton}}{2003b}]{2003_Bunton_ALMAmemo447}
{Bunton} J.~D.,  2003b, ALMA Memo~447, {Multi-resolution {FX} correlator}.
\url{http://library.nrao.edu/public/memos/alma/main/memo447.pdf}

\bibitem[\protect\citeauthoryear{{Bunton} \& {Hay}}{{Bunton} \&
  {Hay}}{2010}]{5651120}
{Bunton} J.~D.,  {Hay} S.~G.,  2010, in 2010 International Conference on
  Electromagnetics in Advanced Applications. pp 728--730,
  \mndoi{10.1109/ICEAA.2010.5651120}

\bibitem[\protect\citeauthoryear{{CSIRO}}{{CSIRO}}{2008}]{Jackson2008}
{CSIRO} 2008, Public specification, {ASKAP} antenna specification and operating
  parameters.
CSIRO,
  \url{https://www.atnf.csiro.au/projects/askap/ASKAP_Antenna_public_specification_Nov08_v0.0.pdf}

\bibitem[\protect\citeauthoryear{{Chapman}, {Dempsey}, {Miller}, {Heywood},
  {Pritchard}, {Sangster}, {Whiting}  \& {Dart}}{{Chapman}
  et~al.}{2017}]{Chapman2017}
{Chapman} J.~M.,  {Dempsey} J.,  {Miller} D.,  {Heywood} I.,  {Pritchard} J.,
  {Sangster} E.,  {Whiting} M.,   {Dart} M.,  2017, in {Lorente} N.~P.~F.,
  {Shortridge} K.,   {Wayth} R.,  eds,  Astronomical Society of the Pacific
  Conference Series Vol. 512, Astronomical Data Analysis Software and Systems
  XXV. pp 73,
  \url{http://aspbooks.org/custom/publications/paper/512--0073.html}

\bibitem[\protect\citeauthoryear{Chippendale \& Anderson}{Chippendale \&
  Anderson}{2019}]{Ch_An_2019_XYphase}
Chippendale A.~P.,  Anderson C.,  2019, ACES Memo~19, On-dish calibration of
  {XY} phase for {ASKAP} phased array feeds.
{CSIRO}, \url{http://www.atnf.csiro.au/projects/askap/ACES-memos}

\bibitem[\protect\citeauthoryear{{Chippendale} \& {Hellbourg}}{{Chippendale} \&
  {Hellbourg}}{2017}]{Chippendale2017}
{Chippendale} A.~P.,  {Hellbourg} G.,  2017, in 2017 International Conference
  on Electromagnetics in Advanced Applications (ICEAA). pp 948--951,
  \mndoi{10.1109/ICEAA.2017.8065413}

\bibitem[\protect\citeauthoryear{{Chippendale} \& {Wormnes}}{{Chippendale} \&
  {Wormnes}}{2013}]{Chippendale2013}
{Chippendale} A.,  {Wormnes} K.,  2013, in 2013 Asia-Pacific Symposium on
  Electromagnetic Compatibility (APEMC). pp~1--4,
  \mndoi{10.1109/APEMC.2013.7360640}

\bibitem[\protect\citeauthoryear{Chippendale, Colegate  \&
  O'Sullivan}{Chippendale et~al.}{2007}]{Chippendale2007}
Chippendale A.~P.,  Colegate T.~M.,   O'Sullivan J.~D.,  2007, {SKA} Memo~92,
  {SKAcost}: a tool for {SKA} cost and performance estimation.
\url{https://www.skatelescope.org/uploaded/33760_memo_92.pdf}

\bibitem[\protect\citeauthoryear{{Chippendale}, {O'Sullivan}, {Reynolds},
  {Gough}, {Hayman}  \& {Hay}}{{Chippendale} et~al.}{2010}]{5613298}
{Chippendale} A.,  {O'Sullivan} J.,  {Reynolds} J.,  {Gough} R.,  {Hayman} D.,
   {Hay} S.,  2010, in 2010 IEEE International Symposium on Phased Array
  Systems and Technology. pp 648--652, \mndoi{10.1109/ARRAY.2010.5613298}

\bibitem[\protect\citeauthoryear{{Chippendale}, {Hayman}  \&
  {Hay}}{{Chippendale} et~al.}{2014}]{2014PASA...31...19C}
{Chippendale} A.~P.,  {Hayman} D.~B.,   {Hay} S.~G.,  2014, \mndoi [\pasa]
  {10.1017/pasa.2014.11}, \href
  {https://ui.adsabs.harvard.edu/abs/2014PASA...31...19C} {31, e019}

\bibitem[\protect\citeauthoryear{{Chippendale} et~al.,}{{Chippendale}
  et~al.}{2015}]{PAF_T}
{Chippendale} A.~P.,  et~al., 2015, in 2015 International Symposium on Antennas
  and Propagation (ISAP). pp~1--4 (\mn@eprint {arXiv} {1509.05489})

\bibitem[\protect\citeauthoryear{{Chippendale} et~al.,}{{Chippendale}
  et~al.}{2016a}]{Chippendale2016}
{Chippendale} A.~P.,  et~al., 2016a, in 2016 10th European Conference on
  Antennas and Propagation (EuCAP). pp~1--5, \mndoi{10.1109/EuCAP.2016.7481741}

\bibitem[\protect\citeauthoryear{{Chippendale}, {Beresford}, {Deng}, {Leach},
  {Reynolds}, {Kramer}  \& {Tzioumis}}{{Chippendale}
  et~al.}{2016b}]{PAFonParkes}
{Chippendale} A.~P.,  {Beresford} R.~J.,  {Deng} X.,  {Leach} M.,  {Reynolds}
  J.~E.,  {Kramer} M.,   {Tzioumis} T.,  2016b, in 2016 International
  Conference on Electromagnetics in Advanced Applications (ICEAA). pp 909--912,
  \mndoi{10.1109/ICEAA.2016.7731550}

\bibitem[\protect\citeauthoryear{Chippendale, Button  \& Louren{\c
  c}o}{Chippendale et~al.}{2018}]{Chippendale_2018_ODC}
Chippendale A.~P.,  Button C.,   Louren{\c c}o L.,  2018, ACES Memo~18,
  Measuring {ASKAP}'s on-dish calibration signal level and its impact on beam
  sensitivity.
{CSIRO}, \url{http://www.atnf.csiro.au/projects/askap/ACES-memos}

\bibitem[\protect\citeauthoryear{{Cornwell}, {Golap}  \&
  {Bhatnagar}}{{Cornwell} et~al.}{2005}]{2005ASPC..347...86C}
{Cornwell} T.~J.,  {Golap} K.,   {Bhatnagar} S.,  2005, in {Shopbell} P.,
  {Britton} M.,   {Ebert} R.,  eds,  Astronomical Society of the Pacific
  Conference Series Vol. 347, Astronomical Data Analysis Software and Systems
  XIV. pp 86,
  \url{http://aspbooks.org/custom/publications/paper/347--0086.html}

\bibitem[\protect\citeauthoryear{Cornwell, Humphreys, Lenc, Voronkov, Whiting,
  Mitchell, Ord  \& Collins}{Cornwell et~al.}{2016}]{ASKAPSW0020}
Cornwell T.,  Humphreys B.,  Lenc E.,  Voronkov M.,  Whiting M.,  Mitchell D.,
  Ord S.,   Collins D.,  2016, Technical report, {{ASKAP} science processing}.
CSIRO, \url{https://www.atnf.csiro.au/projects/askap/ASKAP-SW-0020.pdf}

\bibitem[\protect\citeauthoryear{{DeBoer} et~al.,}{{DeBoer}
  et~al.}{2009}]{DeBoer_2009}
{DeBoer} D.~R.,  et~al., 2009, \mndoi [Proc. {IEEE}]
  {10.1109/JPROC.2009.2016516}, \href
  {https://ui.adsabs.harvard.edu/abs/2009IEEEP..97.1507D} {97, 1507}

\bibitem[\protect\citeauthoryear{{Deller} et~al.,}{{Deller}
  et~al.}{2011}]{2011PASP..123..275D}
{Deller} A.~T.,  et~al., 2011, \mndoi [\pasp] {10.1086/658907}, \href
  {https://ui.adsabs.harvard.edu/abs/2011PASP..123..275D} {123, 275}

\bibitem[\protect\citeauthoryear{{Deng} et~al.,}{{Deng}
  et~al.}{2017}]{Deng2017}
{Deng} X.,  et~al., 2017, \mndoi [\pasa] {10.1017/pasa.2017.20}, \href
  {https://ui.adsabs.harvard.edu/abs/2017PASA...34...26D} {34, e026}

\bibitem[\protect\citeauthoryear{{Dunning}, {Bowen}, {Hayman},
  {Kanapathippillai}, {Kanoniuk}, {Shaw}  \& {Severs}}{{Dunning}
  et~al.}{2016}]{Dunning2016}
{Dunning} A.,  {Bowen} M.~A.,  {Hayman} D.~B.,  {Kanapathippillai} J.,
  {Kanoniuk} H.,  {Shaw} R.~D.,   {Severs} S.,  2016, in 2016 46th European
  Microwave Conference (EuMC). pp 1568--1571, \mndoi{10.1109/EuMC.2016.7824657}

\bibitem[\protect\citeauthoryear{{Dunning} et~al.,}{{Dunning}
  et~al.}{2019}]{Dunning2019}
{Dunning} A.,  et~al., 2019, in PAF Workshop 2019.

\bibitem[\protect\citeauthoryear{{Elagali} et~al.,}{{Elagali}
  et~al.}{2019}]{elagali2019}
{Elagali} A.,  et~al., 2019, \mndoi [\mnras] {10.1093/mnras/stz1448}, \href
  {https://ui.adsabs.harvard.edu/abs/2019MNRAS.487.2797E} {487, 2797}

\bibitem[\protect\citeauthoryear{{Elmer}, {Jeffs}  \& {Warnick}}{{Elmer}
  et~al.}{2014}]{Elmer2014}
{Elmer} M.,  {Jeffs} B.~D.,   {Warnick} K.~F.,  2014, \mndoi [{IEEE} Trans.
  Antennas Propag.] {10.1109/TAP.2014.2359473}, 62, 6067

\bibitem[\protect\citeauthoryear{{Feng}, {Li}  \& {Li}}{{Feng}
  et~al.}{2010}]{Feng2010}
{Feng} Q.-Q.,  {Li} Z.-C.,   {Li} G.-Y.,  2010, in \procspie. p. 782022,
  \mndoi{10.1117/12.866765}

\bibitem[\protect\citeauthoryear{{For} et~al.,}{{For} et~al.}{2019}]{for2019}
{For} B.~Q.,  et~al., 2019, \mndoi [\mnras] {10.1093/mnras/stz2501}, \href
  {https://ui.adsabs.harvard.edu/abs/2019MNRAS.489.5723F} {489, 5723}

\bibitem[\protect\citeauthoryear{{Frost}}{{Frost}}{1972}]{frost1972}
{Frost} O.~L.,  1972, \mndoi [Proc. {IEEE}] {10.1109/PROC.1972.8817}, 60, 926

\bibitem[\protect\citeauthoryear{{Glowacki} et~al.,}{{Glowacki}
  et~al.}{2019}]{2019MNRAS.489.4926G}
{Glowacki} M.,  et~al., 2019, \mndoi [\mnras] {10.1093/mnras/stz2452}, \href
  {https://ui.adsabs.harvard.edu/abs/2019MNRAS.489.4926G} {489, 4926}

\bibitem[\protect\citeauthoryear{{G{\'o}rski}, {Hivon}, {Banday}, {Wand elt},
  {Hansen}, {Reinecke}  \& {Bartelmann}}{{G{\'o}rski}
  et~al.}{2005}]{2005ApJ...622..759G}
{G{\'o}rski} K.~M.,  {Hivon} E.,  {Banday} A.~J.,  {Wand elt} B.~D.,  {Hansen}
  F.~K.,  {Reinecke} M.,   {Bartelmann} M.,  2005, \mndoi [\apj]
  {10.1086/427976}, \href
  {https://ui.adsabs.harvard.edu/abs/2005ApJ...622..759G} {622, 759}

\bibitem[\protect\citeauthoryear{Gupta, Johnston, Feain  \& Cornwell}{Gupta
  et~al.}{2008}]{ASKAPconfig}
Gupta N.,  Johnston S.,  Feain I.,   Cornwell T.,  2008, Memo~1, The initial
  array configuration for {ASKAP}.
{CSIRO}, \url{https://www.atnf.csiro.au/projects/askap/newdocs/configs-3.pdf}

\bibitem[\protect\citeauthoryear{{Guzman} \& {Humphreys}}{{Guzman} \&
  {Humphreys}}{2010}]{2010SPIE.7740E..1JG}
{Guzman} J.~C.,  {Humphreys} B.,  2010, in \procspie. p. 77401J,
  \mndoi{10.1117/12.856962}

\bibitem[\protect\citeauthoryear{{Guzman} et~al.,}{{Guzman}
  et~al.}{2019}]{2019ascl.soft12003G}
{Guzman} J.,  et~al., 2019, {{ASKAPsoft: ASKAP} science data processor
  software, astrophysics source code library, record
  \href{https://ascl.net/1912.003}{{\tt ascl:1912.003}}}

\bibitem[\protect\citeauthoryear{Hampson et~al.,}{Hampson
  et~al.}{2012}]{Hampson2012}
Hampson G.,  et~al., 2012. pp 807--809, \mndoi{10.1109/ICEAA.2012.6328742}

\bibitem[\protect\citeauthoryear{{Hampson}, {Brown}, {Bunton}, {Neuhold},
  {Chekkala}, {Bateman}  \& {Tuthill}}{{Hampson}
  et~al.}{2014}]{Hampson_2014_ASKAP}
{Hampson} G.~A.,  {Brown} A.,  {Bunton} J.~D.,  {Neuhold} S.,  {Chekkala} R.,
  {Bateman} T.,   {Tuthill} J.,  2014, in Proc. XXXIth URSI General Assembly
  and Scientific Symposium. pp~1--4, \mndoi{10.1109/URSIGASS.2014.6930062}

\bibitem[\protect\citeauthoryear{{Harvey-Smith} et~al.,}{{Harvey-Smith}
  et~al.}{2016}]{2016MNRAS.460.2180H}
{Harvey-Smith} L.,  et~al., 2016, \mndoi [\mnras] {10.1093/mnras/stw974}, \href
  {https://ui.adsabs.harvard.edu/abs/2016MNRAS.460.2180H} {460, 2180}

\bibitem[\protect\citeauthoryear{Hay}{Hay}{2010}]{Hay2010}
Hay S.,  2010,
  \href{https://www.ijmot.com/ijmot/uploaded/trmzswogu.pdf}{\textcolor{darkblue}{International
  Journal of Microwave and Optical Technology}}, 5, 375

\bibitem[\protect\citeauthoryear{Hay \& O'Sullivan}{Hay \&
  O'Sullivan}{2008}]{Hay2008}
Hay S.~G.,  O'Sullivan J.~D.,  2008, \mndoi [Radio Science]
  {10.1029/2007RS003798}, 43

\bibitem[\protect\citeauthoryear{{Hay}, {O'Sullivan}, {Kot}, {Granet},
  {Grancea}, {Forsyth}  \& {Hayman}}{{Hay} et~al.}{2007}]{Hay2007}
{Hay} S.~G.,  {O'Sullivan} J.~D.,  {Kot} J.~S.,  {Granet} C.,  {Grancea} A.,
  {Forsyth} A.~R.,   {Hayman} D.~H.,  2007, in The Second European Conference
  on Antennas and Propagation, EuCAP 2007. pp~1--5,
  \mndoi{10.1049/ic.2007.0899}

\bibitem[\protect\citeauthoryear{{Hay}, {O'Sullivan}  \& {Mittra}}{{Hay}
  et~al.}{2011}]{Hay2011}
{Hay} S.~G.,  {O'Sullivan} J.~D.,   {Mittra} R.,  2011, \mndoi [{IEEE} Trans.
  Antennas Propag.] {10.1109/TAP.2011.2123867}, 59, 1828

\bibitem[\protect\citeauthoryear{{Hayman}, {Chippendale}, {Qiao}, {Bunton},
  {Beresford}, {Roberts}  \& {Axtens}}{{Hayman} et~al.}{2010}]{Hayman2010}
{Hayman} D.~B.,  {Chippendale} A.,  {Qiao} R.,  {Bunton} J.~D.,  {Beresford}
  R.~J.,  {Roberts} P.,   {Axtens} P.,  2010, in 2010 International Conference
  on Electromagnetics in Advanced Applications. pp 418--421,
  \mndoi{10.1109/ICEAA.2010.5653177}

\bibitem[\protect\citeauthoryear{{Hellbourg}, {Chippendale}, {Kesteven}  \&
  {Jeffs}}{{Hellbourg} et~al.}{2014}]{7032330}
{Hellbourg} G.,  {Chippendale} A.~P.,  {Kesteven} M.~J.,   {Jeffs} B.~D.,
  2014, in 2014 IEEE Global Conference on Signal and Information Processing
  (GlobalSIP). pp 1286--1290, \mndoi{10.1109/GlobalSIP.2014.7032330}

\bibitem[\protect\citeauthoryear{{Hellbourg}, {Bannister}  \&
  {Hotan}}{{Hellbourg} et~al.}{2016}]{Hellbourg2016}
{Hellbourg} G.,  {Bannister} K.,   {Hotan} A.,  2016, in 2016 Radio Frequency
  Interference (RFI). IEEE Conference Series.
pp 37--42, \mndoi{10.1109/RFINT.2016.7833528}

\bibitem[\protect\citeauthoryear{{Heywood} et~al.,}{{Heywood}
  et~al.}{2016}]{Heywood2016}
{Heywood} I.,  et~al., 2016, \mndoi [\mnras] {10.1093/mnras/stw186}, \href
  {https://ui.adsabs.harvard.edu/abs/2016MNRAS.457.4160H} {457, 4160}

\bibitem[\protect\citeauthoryear{{Heywood} et~al.,}{{Heywood}
  et~al.}{2020}]{2020MNRAS.494.5018H}
{Heywood} I.,  et~al., 2020, \mndoi [\mnras] {10.1093/mnras/staa941}, \href
  {https://ui.adsabs.harvard.edu/abs/2020MNRAS.494.5018H} {494, 5018}

\bibitem[\protect\citeauthoryear{{Hobbs} et~al.,}{{Hobbs}
  et~al.}{2016}]{Hobbs2016}
{Hobbs} G.,  et~al., 2016, \mndoi [\mnras] {10.1093/mnras/stv2893}, \href
  {https://ui.adsabs.harvard.edu/abs/2016MNRAS.456.3948H} {456, 3948}

\bibitem[\protect\citeauthoryear{{Hobbs} et~al.,}{{Hobbs}
  et~al.}{2020}]{2020PASA...37...12H}
{Hobbs} G.,  et~al., 2020, \mndoi [\pasa] {10.1017/pasa.2020.2}, \href
  {https://ui.adsabs.harvard.edu/abs/2020PASA...37...12H} {37, e012}

\bibitem[\protect\citeauthoryear{Hotan}{Hotan}{2016}]{Hotan_2016_Holography}
Hotan A.~W.,  2016, ACES Memo~11, Holographic Measurement of {ASKAP} Primary
  Beams.
{CSIRO}, \url{http://www.atnf.csiro.au/projects/askap/ACES-memos}

\bibitem[\protect\citeauthoryear{{Hotan} et~al.,}{{Hotan}
  et~al.}{2014}]{hotan_2014}
{Hotan} A.~W.,  et~al., 2014, \mndoi [\pasa] {10.1017/pasa.2014.36}, \href
  {https://ui.adsabs.harvard.edu/abs/2014PASA...31...41H} {31, e041}

\bibitem[\protect\citeauthoryear{Hoyle \& Mirtschin}{Hoyle \&
  Mirtschin}{2015}]{Hoyle2015}
Hoyle S.~A.,  Mirtschin P.~L.,  2015, in Proc. 15th Int. Conf. on Accelerator
  and Large Experimental Physics Control Systems ({ICALEPCS'15}). pp 660--663,
  \mndoi{doi:10.18429/JACoW-ICALEPCS2015-WEM301}

\bibitem[\protect\citeauthoryear{{Huynh}, {Dempsey}, {Whiting}  \&
  {Ophel}}{{Huynh} et~al.}{2020}]{Huynh2020}
{Huynh} M.,  {Dempsey} J.,  {Whiting} M.~T.,   {Ophel} M.,  2020, in
  {Ballester} P.,  {Ibsen} J.,  {Solar} M.,   {Shortridge} K.,  eds,
  Astronomical Society of the Pacific Conference Series Vol. 522, Astronomical
  Data Analysis Software and Systems XXVII. pp 263,
  \url{http://aspbooks.org/custom/publications/paper/522--0263.html}

\bibitem[\protect\citeauthoryear{{Indermuehle}, {Harvey-Smith}, {Wilson}  \&
  {Chow}}{{Indermuehle} et~al.}{2016}]{RFI2016}
{Indermuehle} B.,  {Harvey-Smith} L.,  {Wilson} C.,   {Chow} K.,  2016, in 2016
  Radio Frequency Interference (RFI) workshop. IEEE Conference Series.
pp 43--48, \mndoi{10.1109/RFINT.2016.7833529}

\bibitem[\protect\citeauthoryear{{Indermuehle}, {Harvey-Smith}, {Marquarding}
  \& {Reynolds}}{{Indermuehle} et~al.}{2018a}]{2018SPIE10704E..1WI}
{Indermuehle} B.~T.,  {Harvey-Smith} L.,  {Marquarding} M.,   {Reynolds} J.,
  2018a, in \procspie. p. 107041W, \mndoi{10.1117/12.2311926}

\bibitem[\protect\citeauthoryear{{Indermuehle}, {Harvey-Smith}, {Marquarding}
  \& {Reynolds}}{{Indermuehle} et~al.}{2018b}]{2018SPIE10704E..2SI}
{Indermuehle} B.~T.,  {Harvey-Smith} L.,  {Marquarding} M.,   {Reynolds} J.,
  2018b, in \procspie. p. 107042S, \mndoi{10.1117/12.2311917}

\bibitem[\protect\citeauthoryear{{Jeffs}, {Warnick}, {Landon}, {Waldron},
  {Jones}, {Fisher}  \& {Norrod}}{{Jeffs} et~al.}{2008}]{Jeffs2008_Signal}
{Jeffs} B.~D.,  {Warnick} K.~F.,  {Landon} J.,  {Waldron} J.,  {Jones} D.,
  {Fisher} J.~R.,   {Norrod} R.~D.,  2008, \mndoi [{IEEE} J. Sel. Topics Signal
  Process.] {10.1109/JSTSP.2008.2005023}, \href
  {https://ui.adsabs.harvard.edu/abs/2008ISTSP...2..635J} {2, 635}

\bibitem[\protect\citeauthoryear{Johnston \& Grey}{Johnston \&
  Grey}{2006}]{Johnston2006}
Johnston S.,  Grey A.,  2006, {SKA} Memo~72, Surveys with the {xNTD} and
  {CLAR}.
\url{https://www.skatelescope.org/uploaded/24448_72_Johnston.pdf}

\bibitem[\protect\citeauthoryear{{Johnston} et~al.,}{{Johnston}
  et~al.}{2007}]{2007PASA...24..174J}
{Johnston} S.,  et~al., 2007, \mndoi [\pasa] {10.1071/AS07033}, \href
  {https://ui.adsabs.harvard.edu/abs/2007PASA...24..174J} {24, 174}

\bibitem[\protect\citeauthoryear{{Joseph} et~al.,}{{Joseph}
  et~al.}{2019}]{2019MNRAS.490.1202J}
{Joseph} T.~D.,  et~al., 2019, \mndoi [\mnras] {10.1093/mnras/stz2650}, \href
  {https://ui.adsabs.harvard.edu/abs/2019MNRAS.490.1202J} {490, 1202}

\bibitem[\protect\citeauthoryear{{Kadler} et~al.,}{{Kadler}
  et~al.}{2016}]{Kadler2016}
{Kadler} M.,  et~al., 2016, \mndoi [Nature Physics] {10.1038/nphys3715}, \href
  {https://ui.adsabs.harvard.edu/abs/2016NatPh..12..807K} {12, 807}

\bibitem[\protect\citeauthoryear{Kemball \& Wieringa}{Kemball \&
  Wieringa}{2000}]{MSv2}
Kemball A.,  Wieringa M.,  2000, AIPS++ Memo~229, {{\tt MeasurementSet}
  definition version 2.0 }.
NRAO, \url{https://casa.nrao.edu/Memos/229.html}

\bibitem[\protect\citeauthoryear{{Kleiner} et~al.,}{{Kleiner}
  et~al.}{2019}]{2019MNRAS.488.5352K}
{Kleiner} D.,  et~al., 2019, \mndoi [\mnras] {10.1093/mnras/stz2063}, \href
  {https://ui.adsabs.harvard.edu/abs/2019MNRAS.488.5352K} {488, 5352}

\bibitem[\protect\citeauthoryear{{Kooistra}, {Hampson}, {Gunst}, {Bunton},
  {Schoonderbeek}  \& {Brown}}{{Kooistra} et~al.}{2017}]{8104976}
{Kooistra} E.,  {Hampson} G.~A.,  {Gunst} A.~W.,  {Bunton} J.~D.,
  {Schoonderbeek} G.~W.,   {Brown} A.,  2017, in 2017 XXXIInd General Assembly
  and Scientific Symposium of the International Union of Radio Science (URSI
  GASS). pp~1--4, \mndoi{10.23919/URSIGASS.2017.8104976}

\bibitem[\protect\citeauthoryear{{Leahy} et~al.,}{{Leahy}
  et~al.}{2019}]{2019PASA...36...24L}
{Leahy} D.~A.,  et~al., 2019, \mndoi [\pasa] {10.1017/pasa.2019.16}, \href
  {https://ui.adsabs.harvard.edu/abs/2019PASA...36...24L} {36, e024}

\bibitem[\protect\citeauthoryear{{Lee-Waddell} et~al.,}{{Lee-Waddell}
  et~al.}{2019}]{2019MNRAS.tmp...27L}
{Lee-Waddell} K.,  et~al., 2019, \mndoi [\mnras] {10.1093/mnras/stz017}, \href
  {https://ui.adsabs.harvard.edu/abs/2019MNRAS.487.5248L} {487, 5248}

\bibitem[\protect\citeauthoryear{{Lo}, {Lee}  \& {Lee}}{{Lo}
  et~al.}{1966}]{Lo1966}
{Lo} Y.~T.,  {Lee} S.~W.,   {Lee} Q.~H.,  1966, \mndoi [Proc. {IEEE}]
  {10.1109/PROC.1966.4988}, 54, 1033

\bibitem[\protect\citeauthoryear{{McClure-Griffiths}
  et~al.,}{{McClure-Griffiths} et~al.}{2018}]{2018NatAs...2..901M}
{McClure-Griffiths} N.~M.,  et~al., 2018, \mndoi [Nature Astronomy]
  {10.1038/s41550-018-0608-8}, \href
  {https://ui.adsabs.harvard.edu/abs/2018NatAs...2..901M} {2, 901}

\bibitem[\protect\citeauthoryear{McConnell}{McConnell}{2016}]{McConnell_2016_Field}
McConnell D.,  2016, ACES Memo~10, Field structure at the focus of paraboloidal
  reflectors and comparison with {ASKAP} beams.
{CSIRO}, \url{http://www.atnf.csiro.au/projects/askap/ACES-memos}

\bibitem[\protect\citeauthoryear{McConnell}{McConnell}{2017a}]{McConnell_2017_Estimating}
McConnell D.,  2017a, ACES Memo~14, Estimating {ASAKP} beam to beam
  correlation.
{CSIRO}, \url{http://www.atnf.csiro.au/projects/askap/ACES-memos}

\bibitem[\protect\citeauthoryear{McConnell}{McConnell}{2017b}]{McConnell_2017_Survey}
McConnell D.,  2017b, ACES Memo~15, Observing with {ASKAP}: Optimisation for
  survey speed.
{CSIRO}, \url{http://www.atnf.csiro.au/projects/askap/ACES-memos}

\bibitem[\protect\citeauthoryear{{McConnell} et~al.,}{{McConnell}
  et~al.}{2016}]{mcconnell_2016}
{McConnell} D.,  et~al., 2016, \mndoi [\pasa] {10.1017/pasa.2016.37}, \href
  {https://ui.adsabs.harvard.edu/abs/2016PASA...33...42M} {33, e042}

\bibitem[\protect\citeauthoryear{{Moss} et~al.,}{{Moss}
  et~al.}{2017}]{2017MNRAS.471.2952M}
{Moss} V.~A.,  et~al., 2017, \mndoi [\mnras] {10.1093/mnras/stx1679}, \href
  {https://ui.adsabs.harvard.edu/abs/2017MNRAS.471.2952M} {471, 2952}

\bibitem[\protect\citeauthoryear{{Pence}, {Chiappetti}, {Page}, {Shaw}  \&
  {Stobie}}{{Pence} et~al.}{2010}]{2010A&A...524A..42P}
{Pence} W.~D.,  {Chiappetti} L.,  {Page} C.~G.,  {Shaw} R.~A.,   {Stobie} E.,
  2010, \mndoi [\aap] {10.1051/0004-6361/201015362}, \href
  {https://ui.adsabs.harvard.edu/abs/2010A&A...524A..42P} {524, A42}

\bibitem[\protect\citeauthoryear{{Rau}}{{Rau}}{2010}]{rau2010}
{Rau} U.,  2010, PhD thesis, New Mexico Institute of Mining and Technology,
  \url{http://www.aoc.nrao.edu/~rurvashi/DataFiles/UrvashiRV_PhdThesis.pdf}

\bibitem[\protect\citeauthoryear{{Rau} \& {Cornwell}}{{Rau} \&
  {Cornwell}}{2011}]{Rau2011}
{Rau} U.,  {Cornwell} T.~J.,  2011, \mndoi [\aap]
  {10.1051/0004-6361/201117104}, \href
  {https://ui.adsabs.harvard.edu/abs/2011A&A...532A..71R} {532, A71}

\bibitem[\protect\citeauthoryear{{Reynolds}}{{Reynolds}}{1994}]{reynolds1994}
{Reynolds} J.~E.,  1994, Technical Report 39.3/040, A revised flux scale for
  the {AT} compact array.
Australia Telescope National Facility,
  \url{https://www.atnf.csiro.au/observers/memos/d96783~1.pdf}

\bibitem[\protect\citeauthoryear{{Reynolds}, {Staveley-Smith}, {Rhee},
  {Westmeier}, {Chippendale}, {Deng}, {Ekers}  \& {Kramer}}{{Reynolds}
  et~al.}{2017}]{Reynolds2017}
{Reynolds} T.~N.,  {Staveley-Smith} L.,  {Rhee} J.,  {Westmeier} T.,
  {Chippendale} A.~P.,  {Deng} X.,  {Ekers} R.~D.,   {Kramer} M.,  2017, \mndoi
  [\pasa] {10.1017/pasa.2017.45}, \href
  {https://ui.adsabs.harvard.edu/abs/2017PASA...34...51R} {34, e051}

\bibitem[\protect\citeauthoryear{{Reynolds} et~al.,}{{Reynolds}
  et~al.}{2019}]{2019MNRAS.482.3591R}
{Reynolds} T.~N.,  et~al., 2019, \mndoi [\mnras] {10.1093/mnras/sty2930}, \href
  {https://ui.adsabs.harvard.edu/abs/2019MNRAS.482.3591R} {482, 3591}

\bibitem[\protect\citeauthoryear{{Rice}}{{Rice}}{1982}]{1982_Sig_Env_Rice}
{Rice} S.~O.,  1982, \mndoi [Proc. {IEEE}]
  {https://doi.org/10.1109/PROC.1982.12376}, 70, 692

\bibitem[\protect\citeauthoryear{{Sarkissian}, {Reynolds}, {Hobbs}  \&
  {Harvey-Smith}}{{Sarkissian} et~al.}{2017}]{Sarkissian2017}
{Sarkissian} J.~M.,  {Reynolds} J.~E.,  {Hobbs} G.,   {Harvey-Smith} L.,  2017,
  \mndoi [\pasa] {10.1017/pasa.2017.19}, \href
  {https://ui.adsabs.harvard.edu/abs/2017PASA...34...27S} {34, e027}

\bibitem[\protect\citeauthoryear{Sault}{Sault}{2014}]{Sault_2014_Initial}
Sault B.,  2014, ACES Memo~2, Initial characterisation of {BETA} polarimetric
  response.
{CSIRO}, \url{http://www.atnf.csiro.au/projects/askap/ACES-memos}

\bibitem[\protect\citeauthoryear{Sault}{Sault}{2015}]{Sault_2015_Widefield}
Sault B.,  2015, ACES Memo~7, Widefield polarimetric considerations for
  {ASKAP}.
{CSIRO}, \url{http://www.atnf.csiro.au/projects/askap/ACES-memos}

\bibitem[\protect\citeauthoryear{{Schinckel} et~al.,}{{Schinckel}
  et~al.}{2011}]{Schinckel2011}
{Schinckel} A.~E.,  et~al., 2011, in Asia-Pacific Microwave Conference 2011. pp
  1178--1181

\bibitem[\protect\citeauthoryear{{Serra} et~al.,}{{Serra}
  et~al.}{2015a}]{2015MNRAS.448.1922S}
{Serra} P.,  et~al., 2015a, \mndoi [\mnras] {10.1093/mnras/stv079}, \href
  {https://ui.adsabs.harvard.edu/abs/2015MNRAS.448.1922S} {448, 1922}

\bibitem[\protect\citeauthoryear{{Serra} et~al.,}{{Serra}
  et~al.}{2015b}]{Serra2015}
{Serra} P.,  et~al., 2015b, \mndoi [\mnras] {10.1093/mnras/stv1326}, \href
  {https://ui.adsabs.harvard.edu/abs/2015MNRAS.452.2680S} {452, 2680}

\bibitem[\protect\citeauthoryear{{Seymour} et~al.,}{{Seymour}
  et~al.}{2020}]{2020PASA...37...13S}
{Seymour} N.,  et~al., 2020, \mndoi [\pasa] {10.1017/pasa.2019.49}, \href
  {https://ui.adsabs.harvard.edu/abs/2020PASA...37...13S} {37, e013}

\bibitem[\protect\citeauthoryear{{Shaw} \& {Hay}}{{Shaw} \&
  {Hay}}{2015}]{Shaw2015}
{Shaw} R.~D.,  {Hay} S.~G.,  2015, in 2015 9th European Conference on Antennas
  and Propagation ({EuCAP}). pp 1--4,
  \url{https://ieeexplore.ieee.org/document/7228365}

\bibitem[\protect\citeauthoryear{{Shaw}, {Hay}  \& {Ranga}}{{Shaw}
  et~al.}{2012}]{Shaw2012}
{Shaw} R.~D.,  {Hay} S.~G.,   {Ranga} Y.,  2012, in 2012 International
  Conference on Electromagnetics in Advanced Applications. pp 438--441,
  \mndoi{10.1109/ICEAA.2012.6328666}

\bibitem[\protect\citeauthoryear{{Taylor}}{{Taylor}}{2005}]{2005ASPC..347...29T}
{Taylor} M.~B.,  2005, in {Shopbell} P.,  {Britton} M.,   {Ebert} R.,  eds,
  Astronomical Society of the Pacific Conference Series Vol. 347, Astronomical
  Data Analysis Software and Systems XIV. p.~29

\bibitem[\protect\citeauthoryear{{The HDF Group}}{{The HDF Group}}{2020}]{hdf5}
{The HDF Group} 1997-2020, {Hierarchical data format, version 5,
  \url{https://www.hdfgroup.org/HDF5/}}

\bibitem[\protect\citeauthoryear{{Tingay} et~al.,}{{Tingay}
  et~al.}{2013}]{2013PASA...30....7T}
{Tingay} S.~J.,  et~al., 2013, \mndoi [\pasa] {10.1017/pasa.2012.007}, \href
  {https://ui.adsabs.harvard.edu/abs/2013PASA...30....7T} {30, e007}

\bibitem[\protect\citeauthoryear{{Tuthill}, {Hampson}, {Bunton}, {Brown},
  {Neuhold}, {Bateman}, {de Souza}  \& {Joseph}}{{Tuthill}
  et~al.}{2012}]{Tuthill2012}
{Tuthill} J.,  {Hampson} G.,  {Bunton} J.,  {Brown} A.,  {Neuhold} S.,
  {Bateman} T.,  {de Souza} L.,   {Joseph} J.,  2012, in 2012 International
  Conference on Electromagnetics in Advanced Applications. pp 1067--1070,
  \mndoi{10.1109/ICEAA.2012.6328788}

\bibitem[\protect\citeauthoryear{{Tuthill}, {Hampson}, {Bunton}, {Harris},
  {Brown}, {Ferris}  \& {Bateman}}{{Tuthill} et~al.}{2015}]{Tuthill2015}
{Tuthill} J.,  {Hampson} G.,  {Bunton} J.~D.,  {Harris} F.~J.,  {Brown} A.,
  {Ferris} R.,   {Bateman} T.,  2015, in 2015 IEEE Signal Processing and Signal
  Processing Education Workshop (SP/SPE). pp 255--260,
  \mndoi{10.1109/DSP-SPE.2015.7369562}

\bibitem[\protect\citeauthoryear{{Van Veen} \& {Buckley}}{{Van Veen} \&
  {Buckley}}{1988}]{1988_Beamforming_VanVeen}
{Van Veen} B.~D.,  {Buckley} K.~M.,  1988, \mndoi [IEEE ASSP Magazine]
  {10.1109/53.665}, \href
  {https://ui.adsabs.harvard.edu/abs/1988IASSP...5....4V} {5, 4}

\bibitem[\protect\citeauthoryear{{Wayth} et~al.,}{{Wayth}
  et~al.}{2018}]{2018PASA...35...33W}
{Wayth} R.~B.,  et~al., 2018, \mndoi [\pasa] {10.1017/pasa.2018.37}, \href
  {http://adsabs.harvard.edu/abs/2018PASA...35...33W} {35}

\bibitem[\protect\citeauthoryear{{Weinreb} \& {Shi}}{{Weinreb} \&
  {Shi}}{2019}]{Weinreb2019}
{Weinreb} S.,  {Shi} J.,  2019, in PAF Workshop 2019. \url
  {https://events.mpifr-bonn.mpg.de/indico/event/108/session/17/contribution/27}

\bibitem[\protect\citeauthoryear{{Whiting}}{{Whiting}}{2012}]{whiting2012a}
{Whiting} M.~T.,  2012, \mndoi [\mnras] {10.1111/j.1365-2966.2012.20548.x},
  \href {http://adsabs.harvard.edu/abs/2012MNRAS.421.3242W} {421, 3242}

\bibitem[\protect\citeauthoryear{{Whiting}}{{Whiting}}{2020}]{Whiting2020}
{Whiting} M.~T.,  2020, in {Ballester} P.,  {Ibsen} J.,  {Solar} M.,
  {Shortridge} K.,  eds,  Astronomical Society of the Pacific Conference Series
  Vol. 522, Astronomical Data Analysis Software and Systems XXVII. p.~469, \url
  {http://aspbooks.org/custom/publications/paper/522-0469.html}

\bibitem[\protect\citeauthoryear{{Whiting} \& {Humphreys}}{{Whiting} \&
  {Humphreys}}{2012}]{whiting2012b}
{Whiting} M.,  {Humphreys} B.,  2012, \mndoi [\pasa] {10.1071/AS12028}, \href
  {http://adsabs.harvard.edu/abs/2012PASA...29..371W} {29, 371}

\bibitem[\protect\citeauthoryear{{Wilson}, {Storey}  \& {Tzioumis}}{{Wilson}
  et~al.}{2013}]{Wilson2013}
{Wilson} C.,  {Storey} M.,   {Tzioumis} T.,  2013, in 2013 Asia-Pacific
  Symposium on Electromagnetic Compatibility (APEMC). pp~1--4,
  \mndoi{10.1109/APEMC.2013.7360651}

\bibitem[\protect\citeauthoryear{{Wilson}, {Chow}, {Harvey-Smith},
  {Indermuehle}, {Sokolowski}  \& {Wayth}}{{Wilson} et~al.}{2016}]{Wilson2016}
{Wilson} C.,  {Chow} K.,  {Harvey-Smith} L.,  {Indermuehle} B.,  {Sokolowski}
  M.,   {Wayth} R.,  2016. Proc. 2016 International Conference on
  Electromagnetics in Advanced Applications (ICEAA).
pp 922--923, \mndoi{10.1109/ICEAA.2016.7731554}

\bibitem[\protect\citeauthoryear{{Wrobel} \& {Walker}}{{Wrobel} \&
  {Walker}}{1999}]{Wrobel1999}
{Wrobel} J.~M.,  {Walker} R.~C.,  1999, in {Taylor} G.~B.,  {Carilli} C.~L.,
  {Perley} R.~A.,  eds,  Astronomical Society of the Pacific Conference Series
  Vol. 180, Synthesis Imaging in Radio Astronomy II. p.~171, \url
  {https://ui.adsabs.harvard.edu/abs/1999ASPC..180..171W}

\bibitem[\protect\citeauthoryear{{Wu}, {Wicenec}, {Pallot}  \&
  {Checcucci}}{{Wu} et~al.}{2013}]{Wu2013}
{Wu} C.,  {Wicenec} A.,  {Pallot} D.,   {Checcucci} A.,  2013, \mndoi
  [Experimental Astronomy] {10.1007/s10686-013-9354-1}, \href
  {https://ui.adsabs.harvard.edu/abs/2013ExA....36..679W} {36, 679}

\makeatother
\end{thebibliography}

\end{document}